\documentclass[a4paper,11pt]{article}
\usepackage{jheppub} 
\usepackage{lineno}

\usepackage{graphicx} 
\usepackage{dcolumn} 
\usepackage{bm} 
\usepackage{slashed}
\usepackage[dvipsnames]{xcolor}
\usepackage{xspace}
\usepackage{amsmath}
\usepackage{amsfonts}
\usepackage{lmodern}
\usepackage{tabulary}
\usepackage{dsfont}
\usepackage{hyperref}
\usepackage{placeins}
\usepackage{subcaption}

\usepackage{tabularx,array}

\newcolumntype{C}{>{\centering\arraybackslash}X}
\usepackage{multirow}

\usepackage[colorlinks=true,
urlcolor=blue,
linkcolor=blue,
citecolor=blue,
linktocpage=true,
pdfproducer=medialab,
pdfa=true,
anchorcolor=blue]{hyperref}

\usepackage{float}
\usepackage{xcolor}

\usepackage{defs}

\leftline{MITHIG-MOD-NOTE-25-002}
\leftline{Version 2 as of \today}

\title{Analysis note: measurement of thrust in \boldmath $\ee$ collisions at $\sqrt{s}$ = 91 GeV with archived ALEPH data}

\author[a]{Anthony Badea}
\author[i]{Austin Baty}
\author[f]{Hannah Bossi}
\author[f]{Yu-Chen Chen}
\author[h]{Yi Chen}
\author[h]{Jingyu Zhang}
\author[f]{Gian Michele Innocenti}
\author[g]{Marcello Maggi}
\author[f]{Chris McGinn}
\author[f]{Michael Peters} 
\author[f]{Tzu-An Sheng}
\author[d,e]{Vinicius Mikuni} 
\author[d,e]{Matthew Avaylon} 
\author[f]{Patrick Komiske} 
\author[f]{Eric Metodiev} 
\author[f]{Jesse Thaler}
\author[b,c]{Benjamin Nachman}
\author[f]{Yen-Jie Lee}

\affiliation[a]{Enrico Fermi Institute, University of Chicago, Chicago, IL 60637, USA}
\affiliation[b]{Department of Particle Physics and Astrophysics, Stanford University, Stanford, CA 94305, USA}
\affiliation[c]{Fundamental Physics Directorate, SLAC National Accelerator Laboratory, Menlo Park, CA 94025, USA}
\affiliation[d]{Physics Division, Lawrence Berkeley National Laboratory, Berkeley, CA 94720, USA}
\affiliation[e]{Berkeley Institute for Data Science, University of California, Berkeley, CA 94720, USA}
\affiliation[f]{Laboratory for Nuclear Science, Massachusetts Institute of Technology, Cambridge, MA 02139, USA}%
\affiliation[g]{Istituto Nazionale di Fisica Nucleare, Bari Division, BA 70126, Italy}%
\affiliation[h]{Department of Physics and Astronomy, Vanderbilt University, Nashville, TN 37235, USA}
\affiliation[i]{Department of Physics, University of Illinois Chicago, Chicago, IL 60607, USA}%

\emailAdd{anthony.badea@cern.ch}
\emailAdd{abaty@uic.edu}
\emailAdd{hannah.bossi@cern.ch}
\emailAdd{janice\_c@mit.edu}
\emailAdd{luna.chen@vanderbilt.edu}
\emailAdd{jingyu.zhang@cern.ch}
\emailAdd{ginnocen@mit.edu}
\emailAdd{cfmcginn@mit.edu}
\emailAdd{jthaler@mit.edu}
\emailAdd{nachman@stanford.edu}
\emailAdd{yenjie@mit.edu}

\abstract{
A measurement of the thrust distribution in $e^{+}e^{-}$ collisions at $\sqrt{s} = 91.2$ GeV with archived data from the ALEPH experiment at the Large Electron-Positron Collider is presented. The thrust distribution is reconstructed from charged and neutral particles resulting from hadronic $Z$-boson decays. For the first time with $e^{+}e^{-}$ data, detector effects are corrected using a machine learning based method for unbinned unfolding. The measurement provides new input for resolving current discrepancies between theoretical calculations and experimental determinations of $\alpha_{s}$, constraining non-perturbative effects through logarithmic moments, developing differential hadronization models, and enabling new precision studies using the archived data.
}


\begin{document}
\maketitle

\flushbottom

\section{Introduction}
\label{sec:intro}

Precise measurements of high energy electron-positron (\ee) collisions are essential for constructing a comprehensive and global picture of particle physics. These collisions offer a clean environment free from hadronic initial states, making them an ideal testing ground for fundamental aspects of the Standard Model and beyond. Despite the fact that the Large Electron-Positron (LEP) collider ceased operations more than two decades ago, the data collected then remains indispensable. LEP results continue to inform both experimental analyses and theoretical developments, particularly through their role in tuning phenomenological models used in modern parton shower Monte Carlo (MC) simulations~\cite{ParticleDataGroup:2022pth} and in comparisons with modern theoretical calculations.

One of the most important classes of observables in \ee collisions are event shape variables, which measure the collision event topology and are highly sensitive to the underlying Quantum Chromodynamics (QCD) dynamics. Those observables are crucial for probing both perturbative and non-perturbative physics, making them valuable tools in precision studies. A widely studied event shape variable is thrust~\cite{PhysRevLett.39.1587, Heister:2003aj}, a variable that quantifies how compatible an event is with two back-to-back objects. The thrust distribution has been extensively used to extract the strong coupling constant $\alpha_{s}(m_Z)$ with high precision, to constrain non-perturbative shape functions, and to test and refine models of hadronization~\cite{Abbate:2010xh, Davison:2009wzs, Hartgring:2013jma}. These capabilities make thrust a fundamental observable in both theoretical and experimental analyses of \ee data.


Recently, new theoretical results have re-emphasized the importance of the LEP thrust measurements for multiple research directions. Efforts to determine $\alpha_{s}(m_{Z})$ from dijets in \ee\ thrust have yielded precise theoretical predictions that lie several standard deviations below the world average~\cite{Benitez:2024nav}. The analysis reports $\alpha_{s}(m_{Z}) = 0.1136 \pm 0.0012$, a value significantly lower and incompatible with the 2023 PDG world average of $\alpha_{s}(m_{Z}) = 0.1180 \pm 0.0009$. The experimental world average has also evolved, with the 2023 re-calculation excluding \ee\ event-shape observables due to MC estimated hadronization corrections. In parallel, recent studies have highlighted the importance of logarithmic moments of event-shape variables, such as thrust, in constraining both perturbative and non-perturbative QCD effects~\cite{Assi:2025ibi}. Based on those results, in addition to others, it is well motivated to re-examine the $e^{+}e^{-}$ event shape variables, in particular the impact of the systematic experimental uncertainties.










The possibility to re-analyze the thrust distribution from \ee collisions using new experimental techniques has recently been made possible by the successful analysis of archived data from the ALEPH experiment. Our collaboration has utilized the archived data to measure two particle correlation functions of charged particles~\cite{Badea:2019vey} as well as jet energy spectrum and substructure~\cite{Chen:2021uws}. Those works established an experimental handle on the charged and neutral particle components of the archived dataset, enabling a new research direction into precision measurements of global observables such as the thrust distribution and energy-energy correlators~\cite{Bossi:2025xsi}.

In this note, we present a re-analysis of the $\log(\tau)$ observable, where $\tau = 1 - T$, using archived $e^{+}e^{-}$ collision data at $\sqrt{s} = 91.2$ GeV from the ALEPH experiment at LEP1~\cite{Decamp:1990jra}. We report the measured distribution along with a complete breakdown of the associated systematic uncertainties. An unbinned unfolding algorithm is employed to fully-correct the distribution for detector effects and new MC simulations are used to assess theoretical uncertainties. However, due to the limited availability of MC samples processed through the ALEPH detector simulation, the new simulations cannot yet be fully utilized in the unfolding. As a result, the overall uncertainty remains comparable to that of the original ALEPH measurement~\cite{ALEPH:2003obs}. With a robust experimental framework now established, however, the theory uncertainty, including the MC hadronization uncertainties, can be the focus of future studies. This work represents a key step toward leveraging the ALEPH dataset for precision measurements and addressing the existing discrepancies in $\alpha_{S}$ determinations.



As mentioned, we correct the $\log\tau$ distribution for detector effects using an unbinned unfolding algorithm. Recent advances in machine learning have enabled a new class of unbinned unfolding techniques, which have been successfully applied in several particle physics analyses~\cite{Andreassen:2019cjw, ATLAS:2024xxl, LHCb:2022rky, CMS-PAS-SMP-23-008, Song:2023sxb, H1:2023fzk, H1:2024mox, canelli2025practicalguideunbinnedunfolding}. Our work represents the first application of such methods to $e^{+}e^{-}$ data. We adopt an unbinned approach as opposed to traditional binned methods such as Iterative Bayesian Unfolding (IBU) for two main reasons. First, we investigate whether unbinned unfolding can reduce experimental uncertainties by better capturing higher-order detector effects. In this analysis, we do not observe a significant advantage in that respect. Second, we aim to increase the robustness and flexibility of the result for downstream applications. For instance, while a binned unfolded distribution may be sufficient for extracting $\alpha_{s}$, it may be suboptimal for computing logarithmic moments or training machine learning–based hadronization models that rely on access to the full, high-dimensional event structure. Our final measurement consists of weighted events, which can be re-binned with systematics recomputed for each binning choice, or used directly in non-histogram–based analyses. Although binning can, in principle, be made arbitrarily fine, practical constraints from detector resolution and data statistics still apply.

This document is organized as follows: Section~\ref{sec:aleph} describes the ALEPH detector, Section~\ref{sec:datasets} covers the datasets, Section~\ref{sec:eventReconAndSel} details event reconstruction and selection, Section~\ref{sec:analysismethod} outlines the analysis method, Section~\ref{sec:systematics} discusses the systematic uncertainties, Section~\ref{sec:results} presents the results, and Section~\ref{sec:conclusion} discusses takeaways and next steps. Additional supporting material is given in Appendices~\ref{sec:app_supplemental}-\ref{sec:app_unifoldAsIBU} as referenced throughout the note.

\section{ALEPH detector}
\label{sec:aleph}

The ALEPH detector~\cite{Decamp:1990jra, ALEPH:1994ayc} at the LEP covered nearly the entire solid angle around the collision point. The detector consisted of an inner tracker surrounded by a superconducting solenoid, electromagnetic and hadronic calorimeters, and outer muon detection system.

The central tracking system was immersed in a 1.5 T solenoidal magnetic field and provided tracking in the pseudorapidity range $|\eta| < 1.74$. A two-layer silicon strip vertex detector (VDET) at radii of 6.3~cm and 11~cm provided two pairs of coordinates per track for precise vertex reconstruction. A cylindrical multi-wire drift chamber, known as the inner track chamber (ITC), with inner and outer diameters of 13~cm and 29~cm delivered an additional eight coordinates per track. Finally, a large time projection chamber (TPC), 4.4~m long and 3.6~m in diameter, offered up to 330 ionization measurements per track. Together, these detectors provided a transverse momentum resolution for charged particles of $\delta p_{\mathrm{T}}/p_{\mathrm{T}} = 0.6 \times 10^{-3} p_{\mathrm{T}} \oplus 0.005$ ($p_{\mathrm{T}}$ in GeV/$c$).

The calorimeter system was comprised of an electromagnetic calorimeter (ECAL) and a hadronic calorimeter (HCAL). The ECAL, positioned between the TPC and the solenoid coil, was a sampling calorimeter of lead and proportional wire chambers segmented into 
projective towers covering $0.9^\circ \times 0.9^\circ$ each. The ECAL had a thickness of 22 radiation lengths and provided a relative energy resolution of $\sigma(E)/E = 0.18/\sqrt{E} + 0.009$ ($E$ in GeV) for isolated electrons. The HCAL was formed by instrumenting the 1.2~m thick iron return yoke of the solenoid magnet with 23 layers of plastic streamer tubes separated by a 5 cm thick iron slabs, giving a total of 7.2 interaction lengths at 90$^\circ$. The HCAL was read out in 
projective towers and delivered a relative energy 
resolution of $\sigma(E)/E = 0.85/\sqrt{E}$ ($E$ in GeV) for pions at normal incidence.

The muon system was located outside of the HCAL and consisted of two double-layers of streamer tube chambers. The muons were distinguished from hadrons by their penetration patterns in the HCAL and muon chambers. The delivered luminosity by the LEP was measured using small-angle Bhabha scattering events, with systematic errors of 0.6\% (experimental) and 0.3\% (theoretical)~\cite{Neugebauer:804776}. The measurements were done by luminosity monitors near the inner detectors and end-caps. A two-level trigger system selected events from collisions. The first-level trigger used information from the calorimeters and ITC track candidates. The second-level trigger incorporated TPC track information to enhance event selection. The system was nearly fully efficient for collecting hadronic Z-boson events. An extensive software suite supported the data simulation, reconstruction, analysis, detector operations, trigger systems, and data acquisition. 

\section{Data and simulated event samples}
\label{sec:datasets}


The analysis uses data from \ee~collisions at
$\sqrt{s} = 91.2$ GeV recorded with the ALEPH detector during 1994, corresponding to an integrated luminosity of approximately 40~\ipb after data quality
requirements. Only the 1994 dataset is used because full detector simulation is only available in the archived sample for 1994 conditions. Events were recorded using the trigger system with an efficiency of 99.99\% for hadronic Z decays and an uncertainty of less than 0.01\%~\cite{ALEPH:1994ayc}. 

MC simulations are used to correct the reported measurement for detector effects. Alongside the archived ALEPH data, a dedicated tune of $\textsc{Pythia}$ 6.1 MC was archived as well~\cite{Sjostrand:2000wi}. The simulation includes generator level particles and matched events processed through a simulation of the ALEPH detector, known as GALEPH. 
The events were generated at leading order (LO) in QCD and tuned with a dedicated ALEPH configuration. Electromagnetic initial state radiation (ISR) and final state radiation (FSR) were included in the generation. Those contributions are corrected for through selections to facilitate comparisons with modern theory predictions as explained in Section~\ref{sec:analysismethod_corrections}. The archived data and MC were converted to an MIT Open Data (MOD) format~\cite{Tripathee:2017ybi}. Additional samples are generated using modern $\textsc{Pythia}$ 8~\cite{bierlich2022comprehensiveguidephysicsusage}, $\textsc{Herwig}$~\cite{B_hr_2008, bewick2024herwig73releasenote}, and $\textsc{Sherpa}$~\cite{Bothmann2019} MC simulations and used to assess the theoretical uncertainty arising from the choice of prior in the unfolding. The alternative MC samples are only produced at generator level since GALEPH is not currently accessible.

\section{Event reconstruction and selection}
\label{sec:eventReconAndSel}

The information from the tracking detectors and the calorimeters were combined in an energy-flow algorithm~\cite{ALEPH:1994ayc}. For each event, the algorithm provided a set of charged and neutral reconstructed particles, called \textit{energy-flow objects}, with measured momentum vectors and information on particle type. \textit{Good tracks} are defined as charged particle tracks reconstructed with at least four hits in the TPC, originating from within a cylinder of length 20 cm 
and radius 2 cm coaxial with the beam and centered at the nominal collision point. The charged energy-flow objects are required to have a polar angle $\theta$ with respect to the beam such that $|\cos\theta| < 0.94$ and a minimum transverse momentum (\pT) of 200 MeV/$c$. The neutral energy-flow objects are required to have $|\cos\theta| < 0.98$ and a minimum energy of 400 MeV/$c$. A missing momentum vector is constructed using the charged and neutral objects $\Vec{p}_{\rm MET} = -\sum\nolimits_{\rm neu, chg} \Vec{p}$ to account for invisible objects, such as neutrinos. 


The selection of hadronic events collected at the Z resonance follows the ALEPH standard choices as described in~\cite{Barate:1996fi, Badea:2019vey}. 
Hadronic events are selected by requiring 
at least 5 good tracks, 
at least 13 reconstructed particles (charged or neutral),  
and a total charged energy $E_{\mathrm{ch}} = \sum_{i} E_{i} = \sum_{i} \sqrt{p_{i}^{2} + m_{\pi}^{2}}$ in excess of 15~GeV. 
To ensure that an event was well contained within the detector, the polar angle of the sphericity axis is determined from all selected particles and required to satisfy $|\cos\theta_{\text{sph}}| < 0.82$~\cite{Barate:1996fi}.  
The selections are summarized in Table~\ref{tab:selections}. 

\begin{table}[ht]
\caption{Selections to identify high-quality charged and neutral particles and hadronic Z decays events, following ALEPH standard choices as described in~\cite{Barate:1996fi, Badea:2019vey}.}


\begin{center}
\begin{tabularx}{0.55\textwidth}{ll}
\hline
\multicolumn{2}{l}{\textit{Charged Particle Selections}}  \\ 
\hline
Acceptance              & $|\cos\theta| \le 0.94$ \\
Quality                 & $p_{\rm T} \ge 0.2$ GeV\\
                        & TPC hits $\ge 4$\\
                        & $d_0 \le 2$~cm \\
                        & $z_0 \le 10$~cm \\
\hline
\multicolumn{2}{l}{\textit{Neutral Particle Selections}}  \\ 
\hline
Acceptance              & $|\cos\theta| \le 0.98$ \\
Quality                 & $E \ge 0.4$~GeV \\
\hline 
\multicolumn{2}{l}{\textit{Event Selections}}  \\
\hline
Acceptance              & $|\cos\theta_{\text{sph}}| \le 0.82$ \\
Hadronic events         & N$_{\rm Trk} \ge 5$ \\ 
                        & E$_{\rm ch} \ge 15$~GeV \\
                        & N$_{\rm Trk + Neu} \ge 13$ \\ 
\hline
\end{tabularx}
\label{tab:selections}
\end{center}
\end{table}





\subsection{Data and simulated event comparisons}

To assess the application of the selections to the data and agreement with the archived MC a number of key variables are compared below, with additional checks shown in Appendix~\ref{sec:app_supplemental}. The distributions are shown without corrections for detector effects, as those are accounted for in the unfolding procedure outlined in Section~\ref{sec:unfolding}. The particle and track level distributions are shown after all selections. In the archived samples, particles are categorized as charged tracks, charged lepton 1 (electron) candidates, charged lepton 2 (muon) candidates, neutral long-lived particles that decay to two charged tracks ($V^{0}$), photons, and neutrals based on the reconstructed properties utilizing different detector subsystems. The charged and neutral particle selections described in the previous section are applied to the corresponding particle categories.

\begin{figure}[t!]
\centering
\begin{subfigure}[b]{0.32\textwidth}
    \includegraphics[width=\textwidth,angle=0]{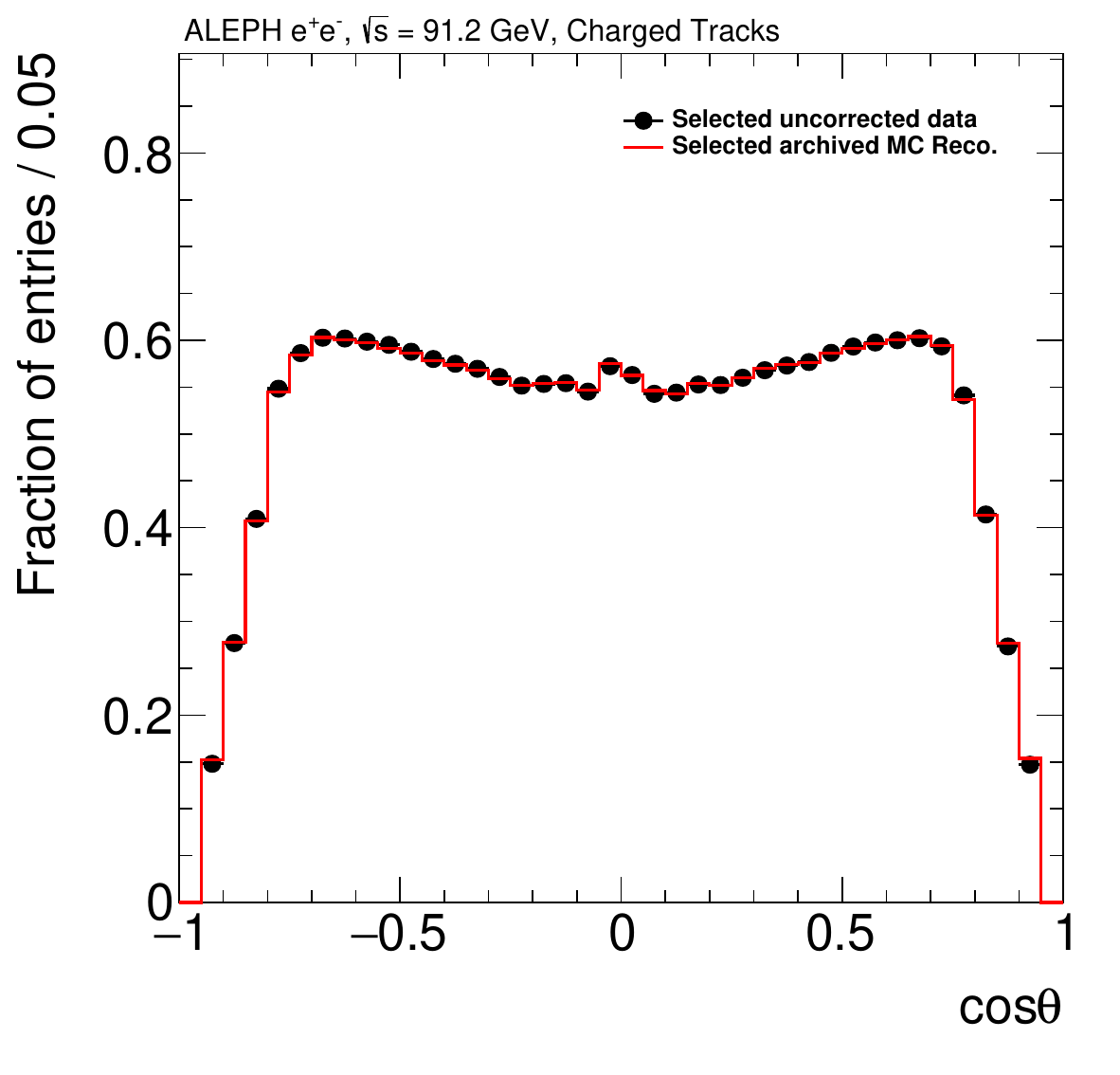}
    \caption{}
    \label{fig:kinem_pwflag0_mainbody_a}
\end{subfigure}
\begin{subfigure}[b]{0.32\textwidth}
    \includegraphics[width=\textwidth,angle=0]{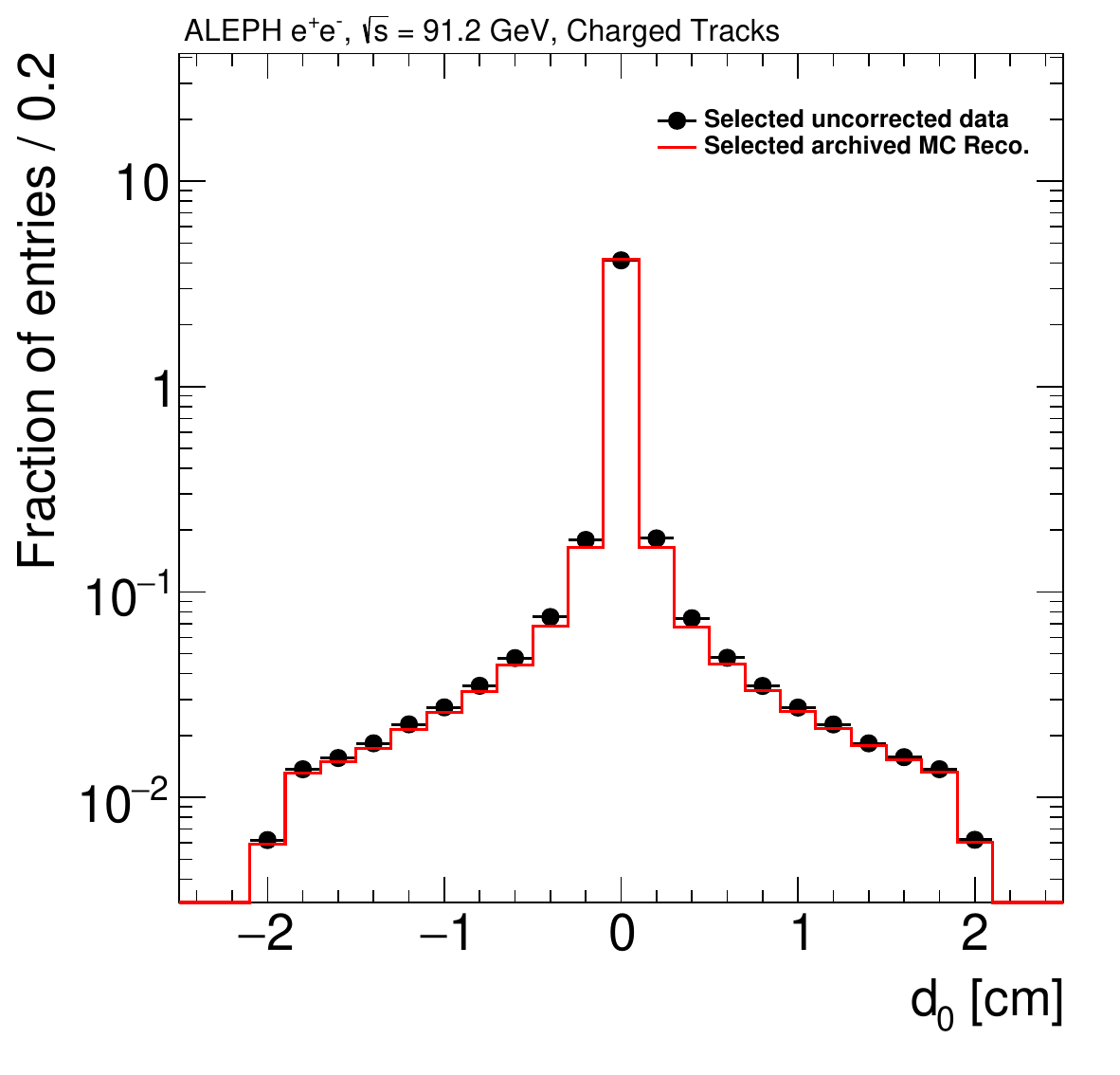}
    \caption{}
    \label{fig:kinem_pwflag0_mainbody_b}
\end{subfigure}
\begin{subfigure}[b]{0.32\textwidth}
    \includegraphics[width=\textwidth,angle=0]{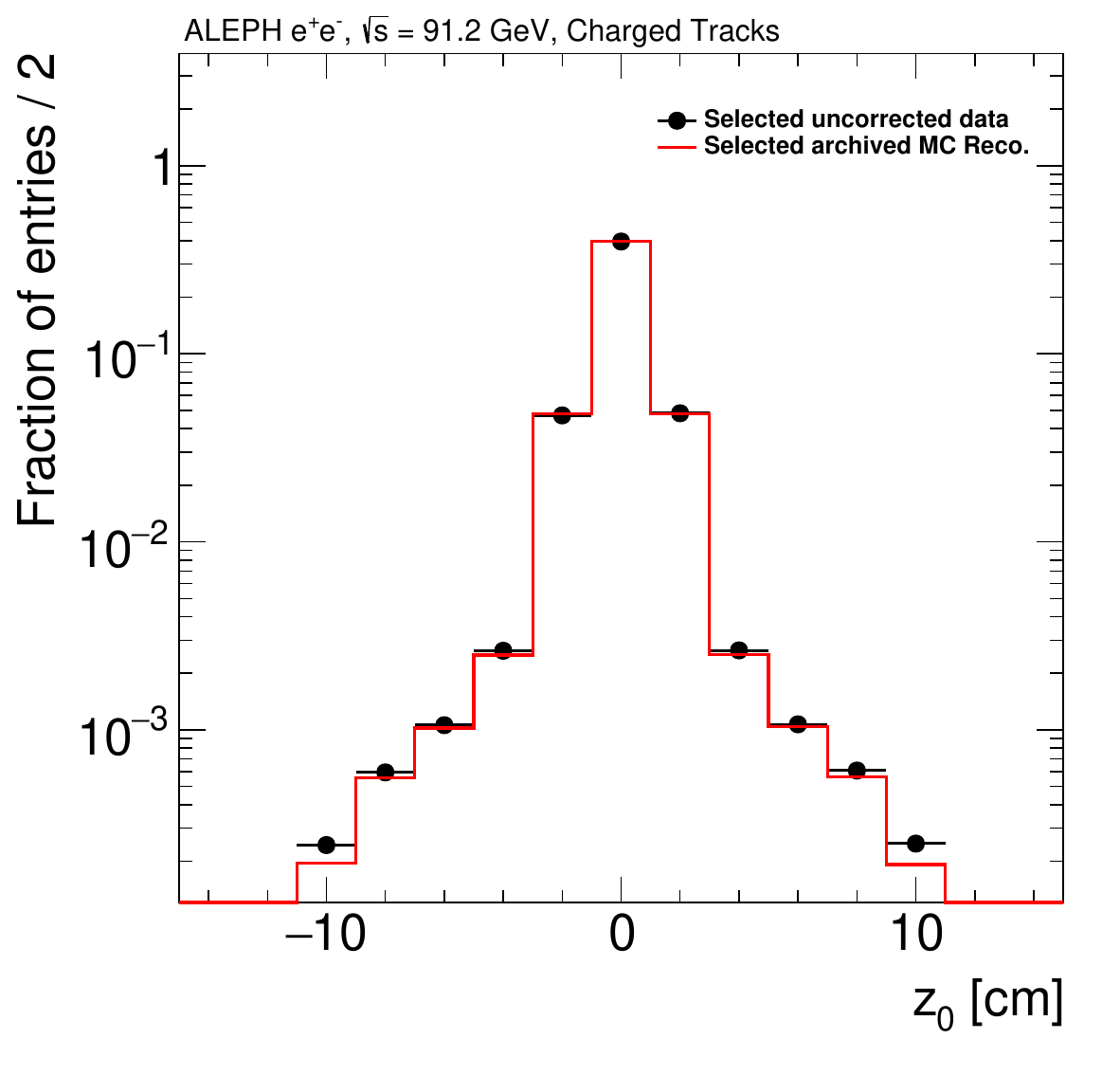}
    \caption{}
    \label{fig:kinem_pwflag0_mainbody_c}
\end{subfigure}
\begin{subfigure}[b]{0.32\textwidth}
    \includegraphics[width=\textwidth,angle=0]{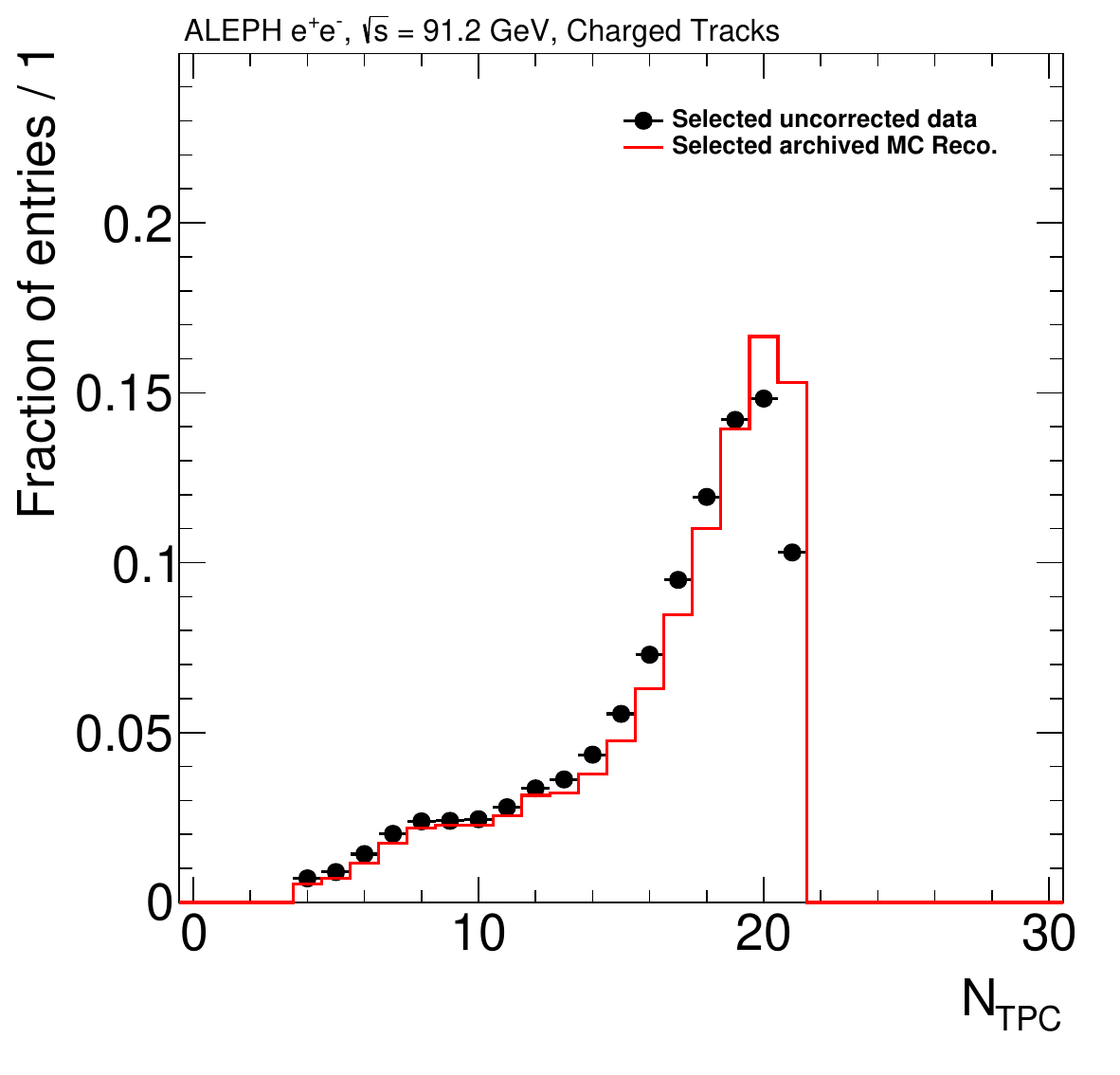}
    \caption{}
    \label{fig:kinem_pwflag0_mainbody_d}
\end{subfigure}
\begin{subfigure}[b]{0.32\textwidth}
    \includegraphics[width=\textwidth,angle=0]{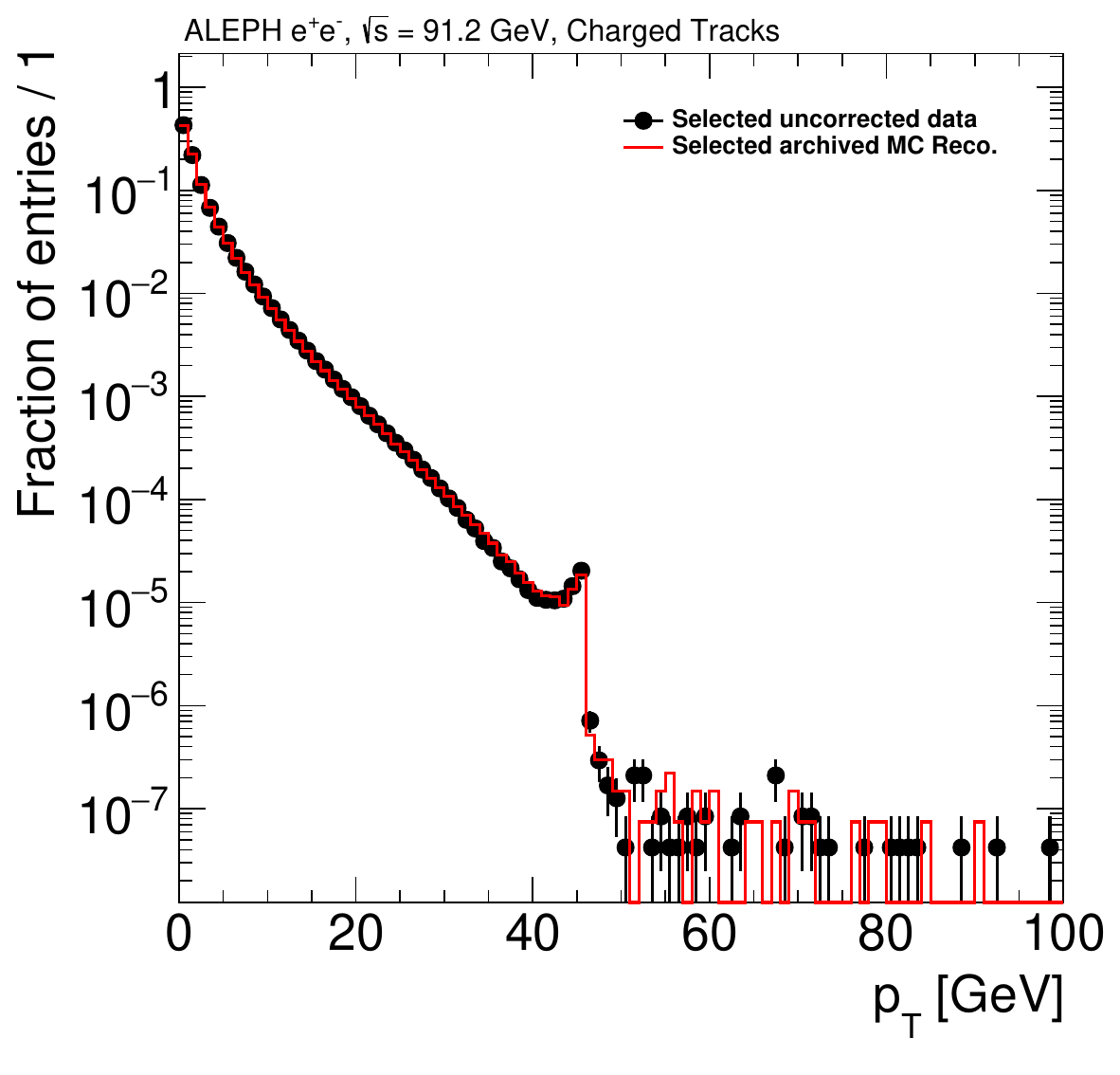}
    \caption{}
    \label{fig:kinem_pwflag0_mainbody_e}
\end{subfigure}
\begin{subfigure}[b]{0.32\textwidth}
    \includegraphics[width=\textwidth,angle=0]{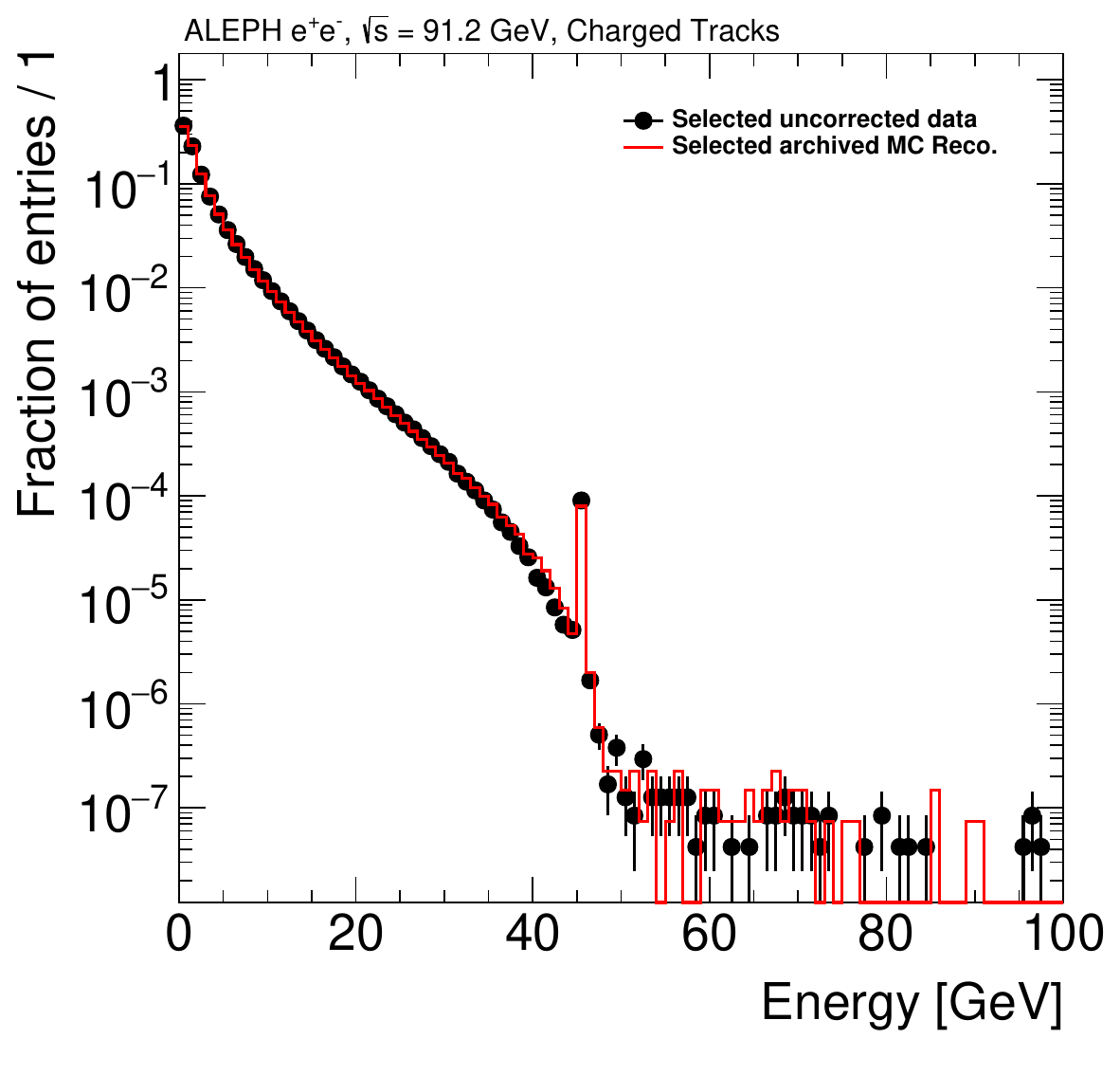}
    \caption{}
    \label{fig:kinem_pwflag0_mainbody_f}
\end{subfigure}
\caption{Observables used for selecting high-quality charged tracks include: (a) the cosine of the polar angle $\theta$, (b) the transverse impact parameter $d_{0}$, (c) the longitudinal impact parameter $z_{0}$, (d) the number of hits in the time projection chamber ($N_{\mathrm{TPC}}$), (e) the transverse momentum~\pT, and (f) the energy (E).}
\label{fig:kinem_pwflag0_mainbody}
\end{figure}

For the charged tracks, comparisons of the polar angle ($\theta$), transverse impact parameters ($d_{0}$), longitudinal impact parameters ($z_{0}$), number of hits in the TPC ($\mathrm{N}_\mathrm{TPC}$), \pT, and energy are shown in Figure~\ref{fig:kinem_pwflag0_mainbody}. There are slight differences in the $\mathrm{N}_\mathrm{TPC}$ distribution, likely arising from an imperfect modeling of the TPC. In both the \pT~and energy distributions there is a peak around $\sim$45 GeV that occurs at the reconstruction level, but not the generator level. The impact on the overall thrust distribution with and without this contribution was found to be negligible, and therefore no explicit cut was applied for this contribution. The $\cos\theta$ spectrum shows the characteristic enhancement in the forward regions from the first term in the $Z \rightarrow$ hadrons cross section $\frac{d\sigma}{d\cos\theta} \propto (1 + \cos^2\theta) + 2A_{\text{FB}} \cos\theta$. The drop in yields near $\cos\theta$ of 1 are from detector acceptance. The distributions generally show good agreement. 

\begin{figure}[t!]
\centering
\begin{subfigure}[b]{0.32\textwidth}
    \includegraphics[width=\textwidth,angle=0]{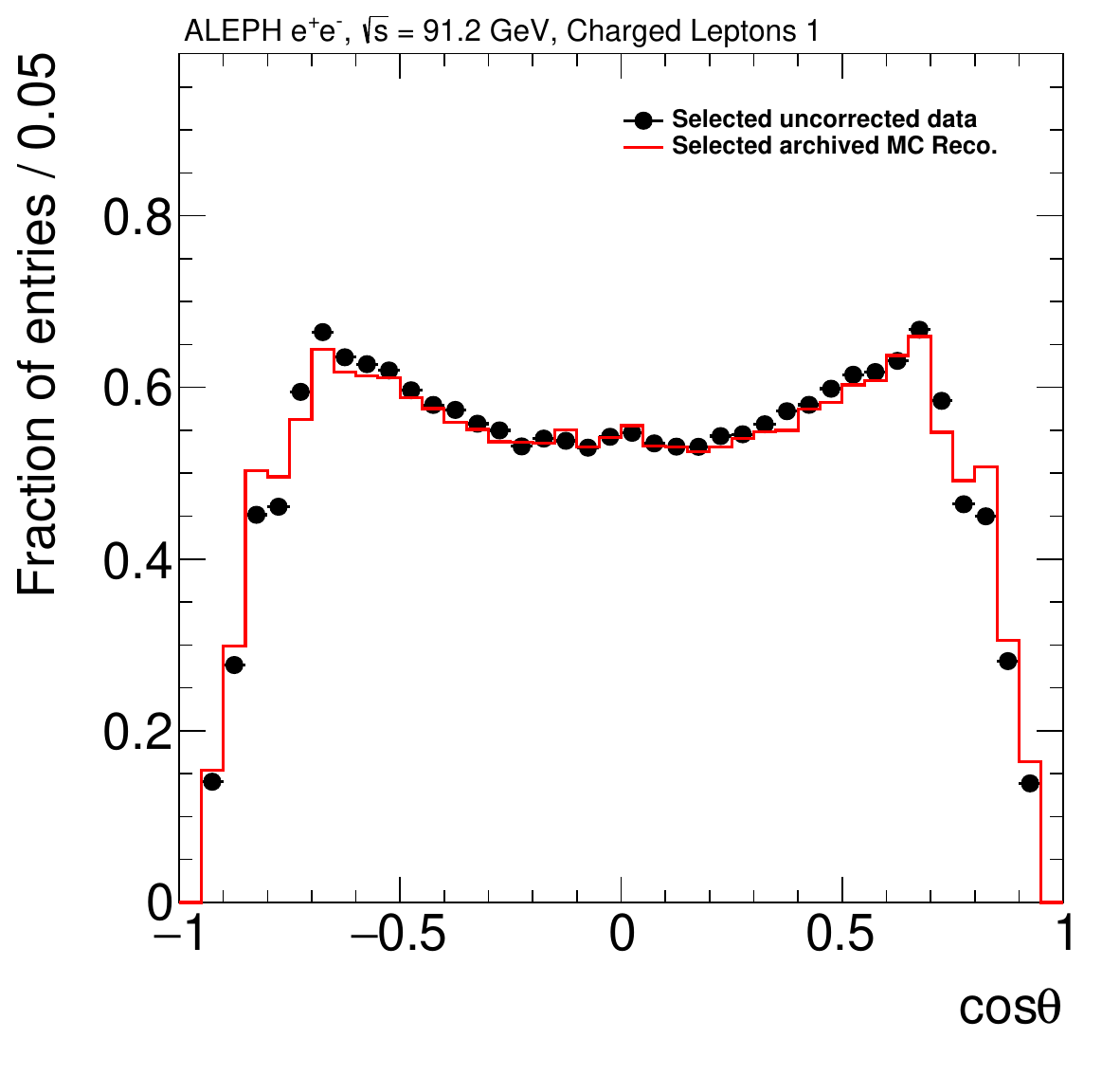}
    \caption{}
    \label{fig:kinem_pwflag1_mainbody_a}
\end{subfigure}
\begin{subfigure}[b]{0.32\textwidth}
    \includegraphics[width=\textwidth,angle=0]{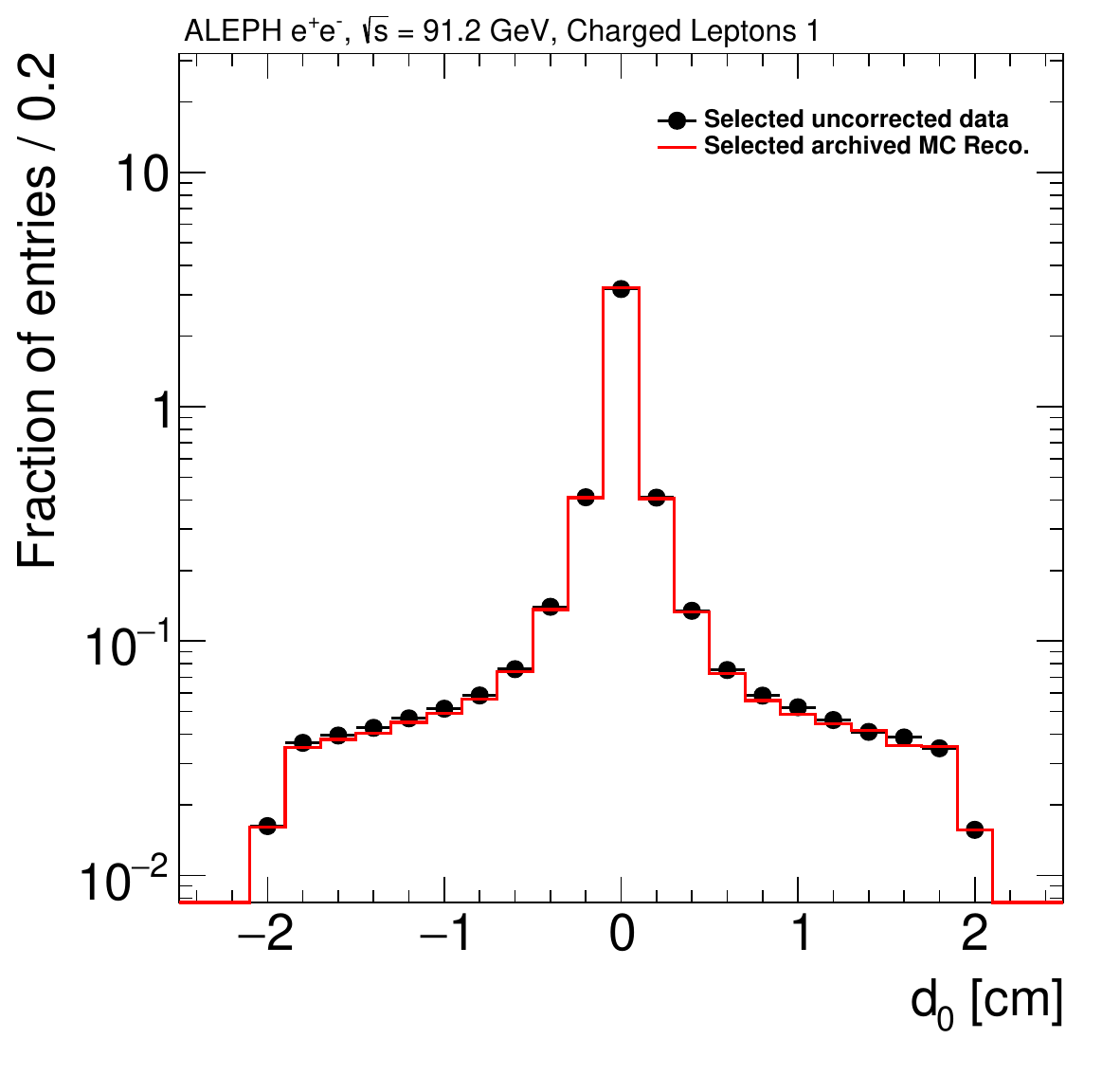}
    \caption{}
    \label{fig:kinem_pwflag1_mainbody_b}
\end{subfigure}
\begin{subfigure}[b]{0.32\textwidth}
    \includegraphics[width=\textwidth,angle=0]{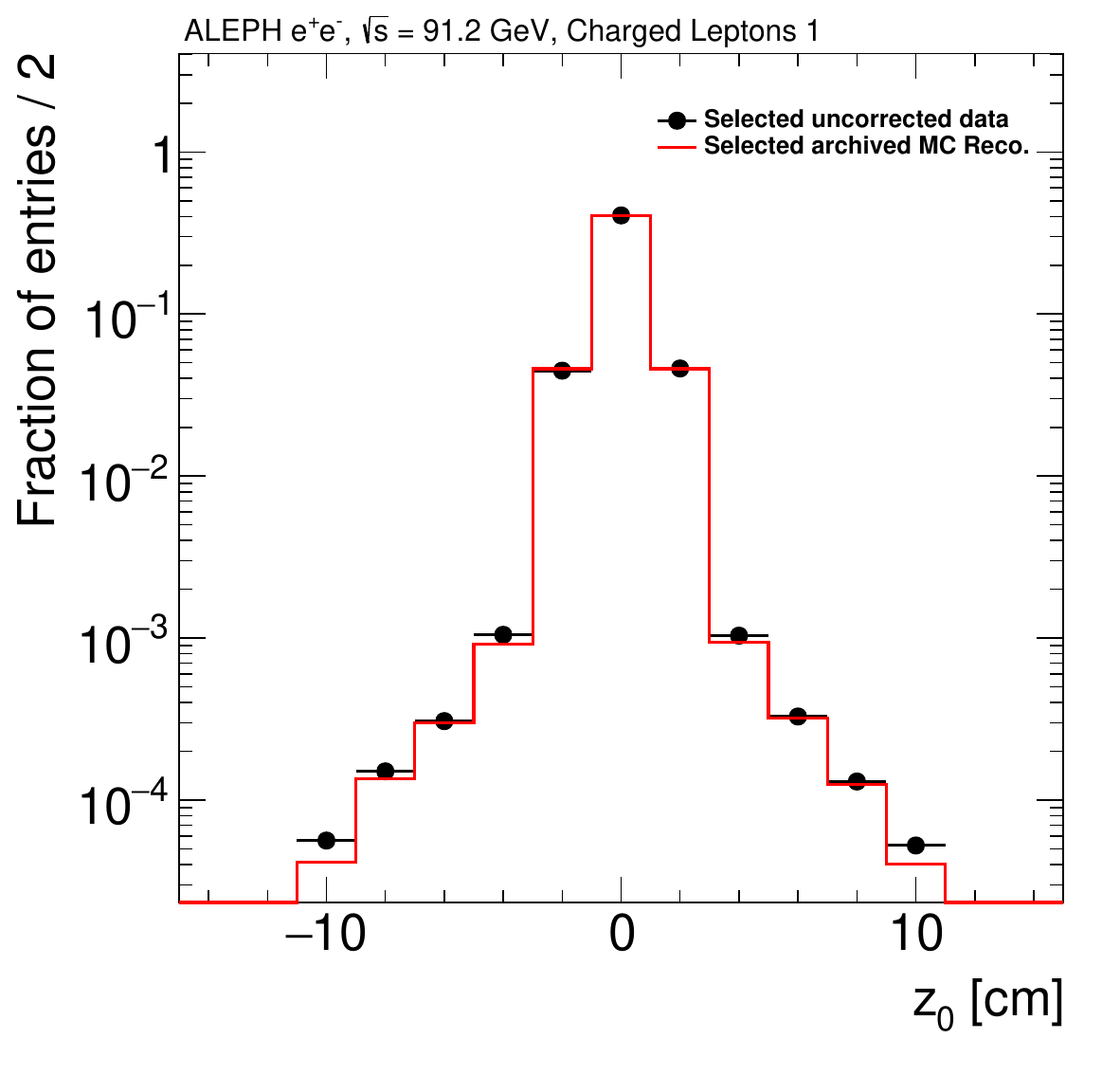}
    \caption{}
    \label{fig:kinem_pwflag1_mainbody_c}
\end{subfigure}
\begin{subfigure}[b]{0.32\textwidth}
    \includegraphics[width=\textwidth,angle=0]{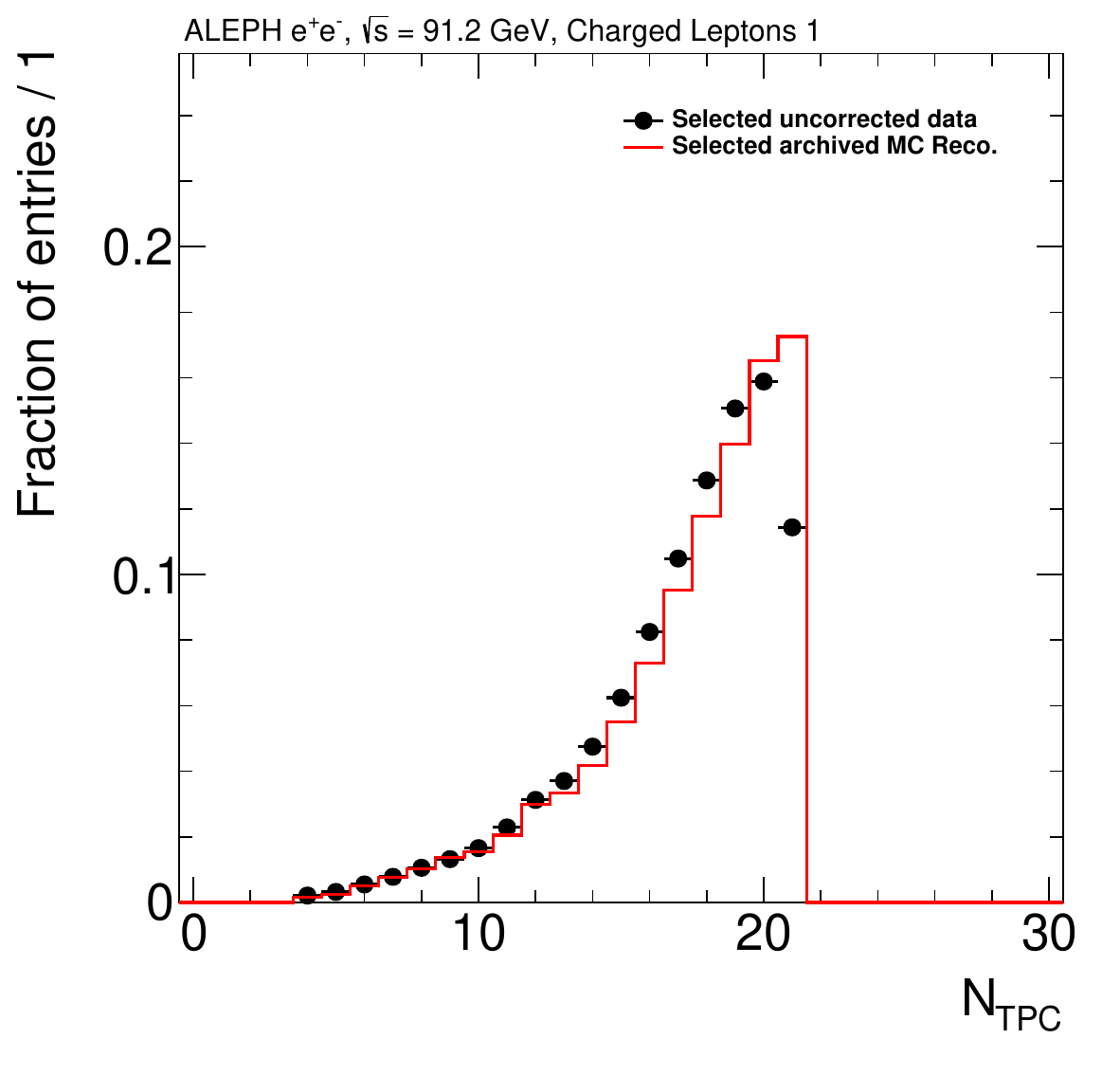}
    \caption{}
    \label{fig:kinem_pwflag1_mainbody_d}
\end{subfigure}
\begin{subfigure}[b]{0.32\textwidth}
    \includegraphics[width=\textwidth,angle=0]{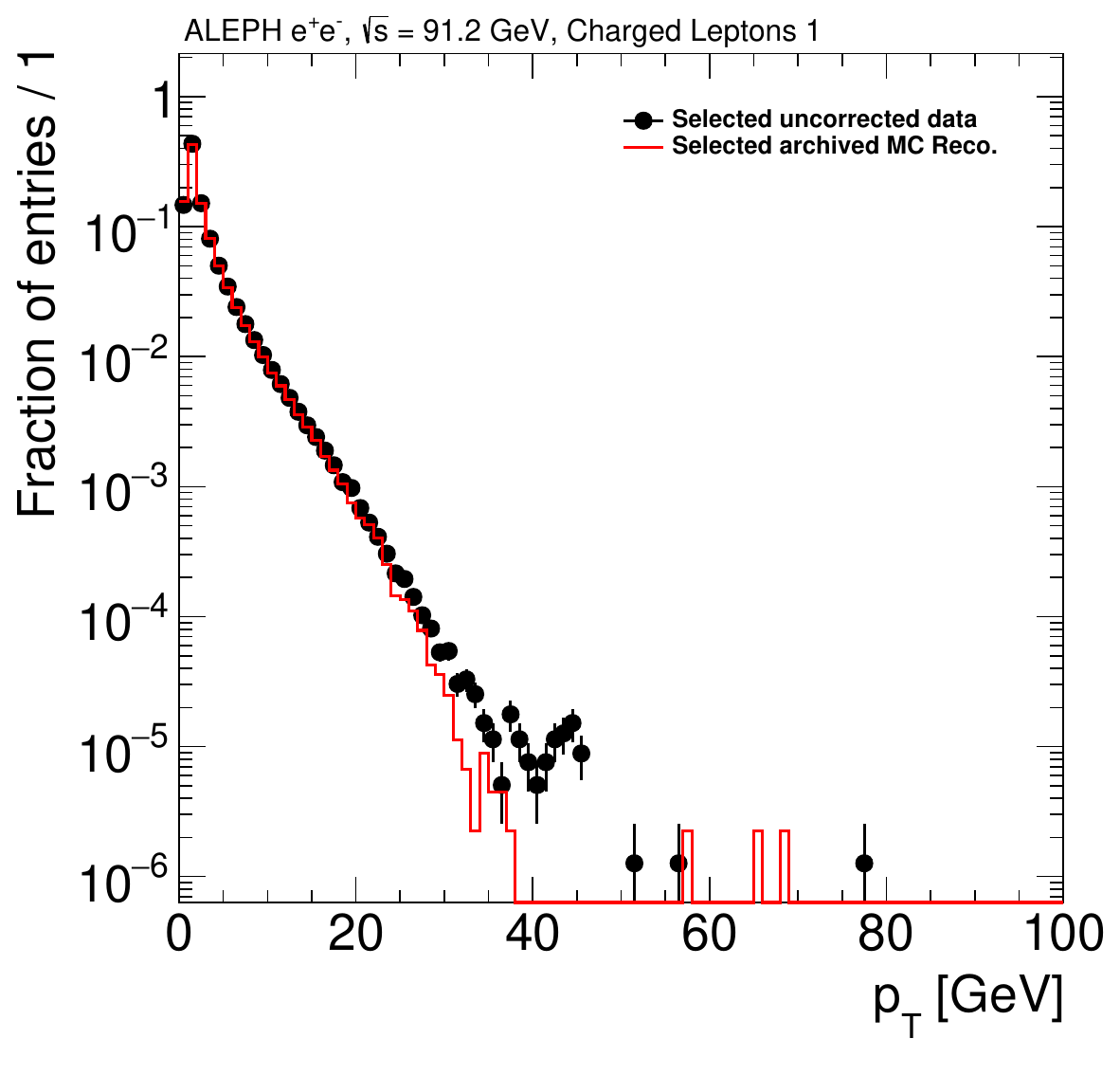}
    \caption{}
    \label{fig:kinem_pwflag1_mainbody_e}
\end{subfigure}
\begin{subfigure}[b]{0.32\textwidth}
    \includegraphics[width=\textwidth,angle=0]{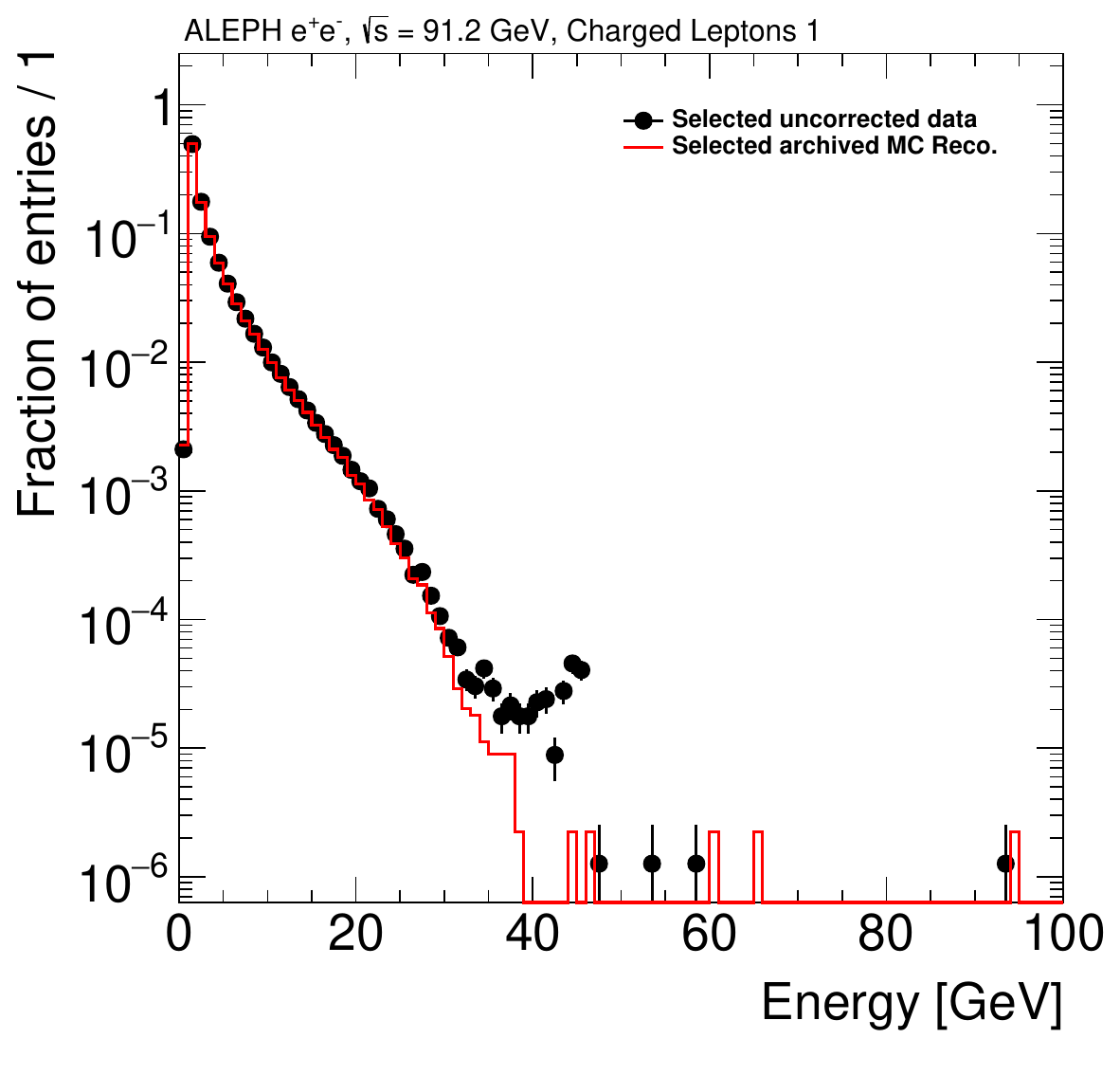}
    \caption{}
    \label{fig:kinem_pwflag1_mainbody_f}
\end{subfigure}
\caption{Observables used for selecting high-quality charged leptons with tracker and calorimeter hits: (a) the cosine of the polar angle $\theta$, (b) the transverse impact parameter $d_{0}$, (c) the longitudinal impact parameter $z_{0}$, (d) the number of hits in the time projection chamber ($N_{\mathrm{TPC}}$), (e) the transverse momentum~\pT, and (f) the energy (E).}
\label{fig:kinem_pwflag1_mainbody}
\end{figure}

\begin{figure}[t!]
\centering
\begin{subfigure}[b]{0.32\textwidth}
    \includegraphics[width=\textwidth,angle=0]{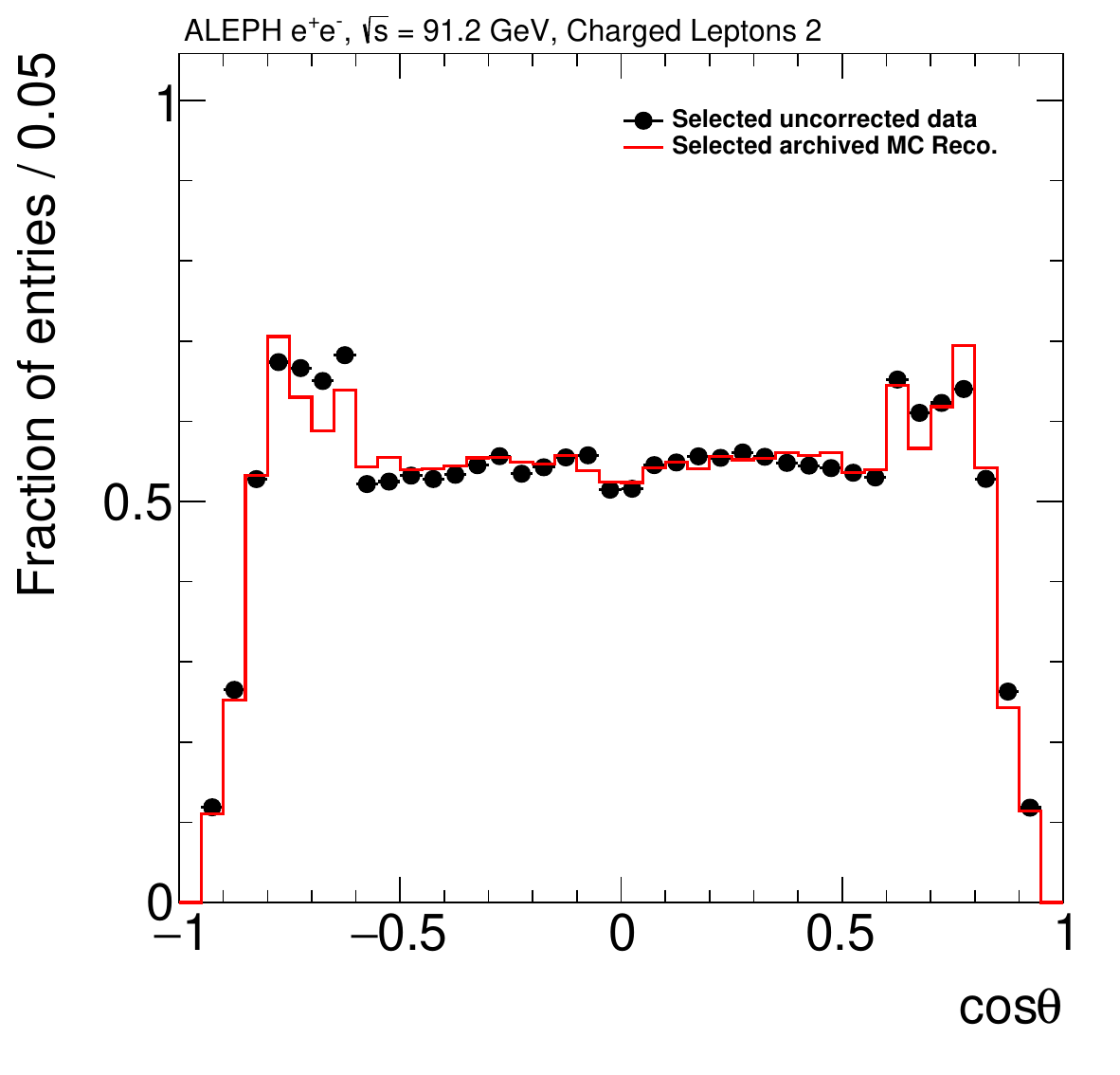}
    \caption{}
    \label{fig:kinem_pwflag2_mainbody_a}
\end{subfigure}
\begin{subfigure}[b]{0.32\textwidth}
    \includegraphics[width=\textwidth,angle=0]{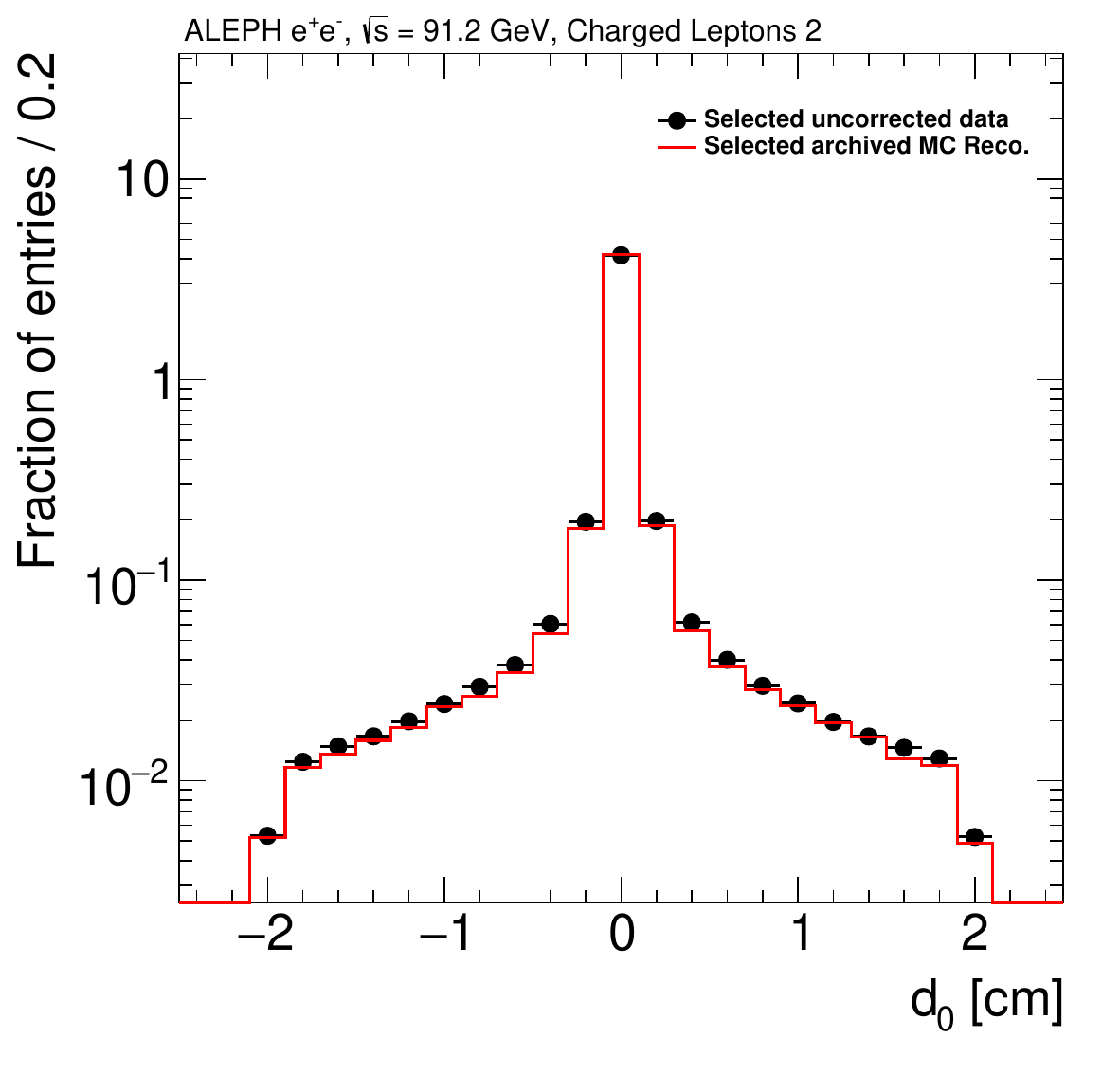}
    \caption{}
    \label{fig:kinem_pwflag2_mainbody_b}
\end{subfigure}
\begin{subfigure}[b]{0.32\textwidth}
    \includegraphics[width=\textwidth,angle=0]{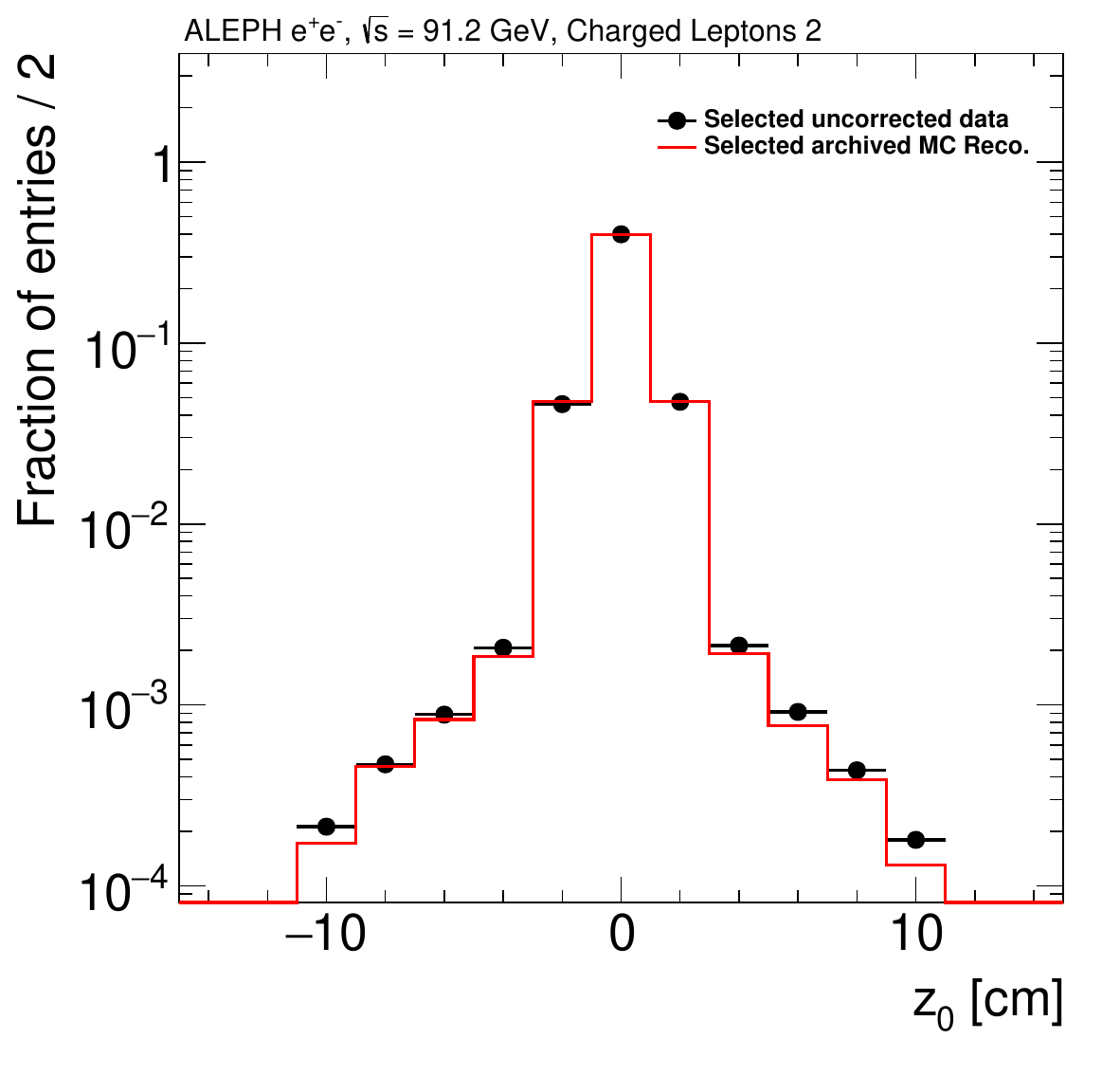}
    \caption{}
    \label{fig:kinem_pwflag2_mainbody_c}
\end{subfigure}
\begin{subfigure}[b]{0.32\textwidth}
    \includegraphics[width=\textwidth,angle=0]{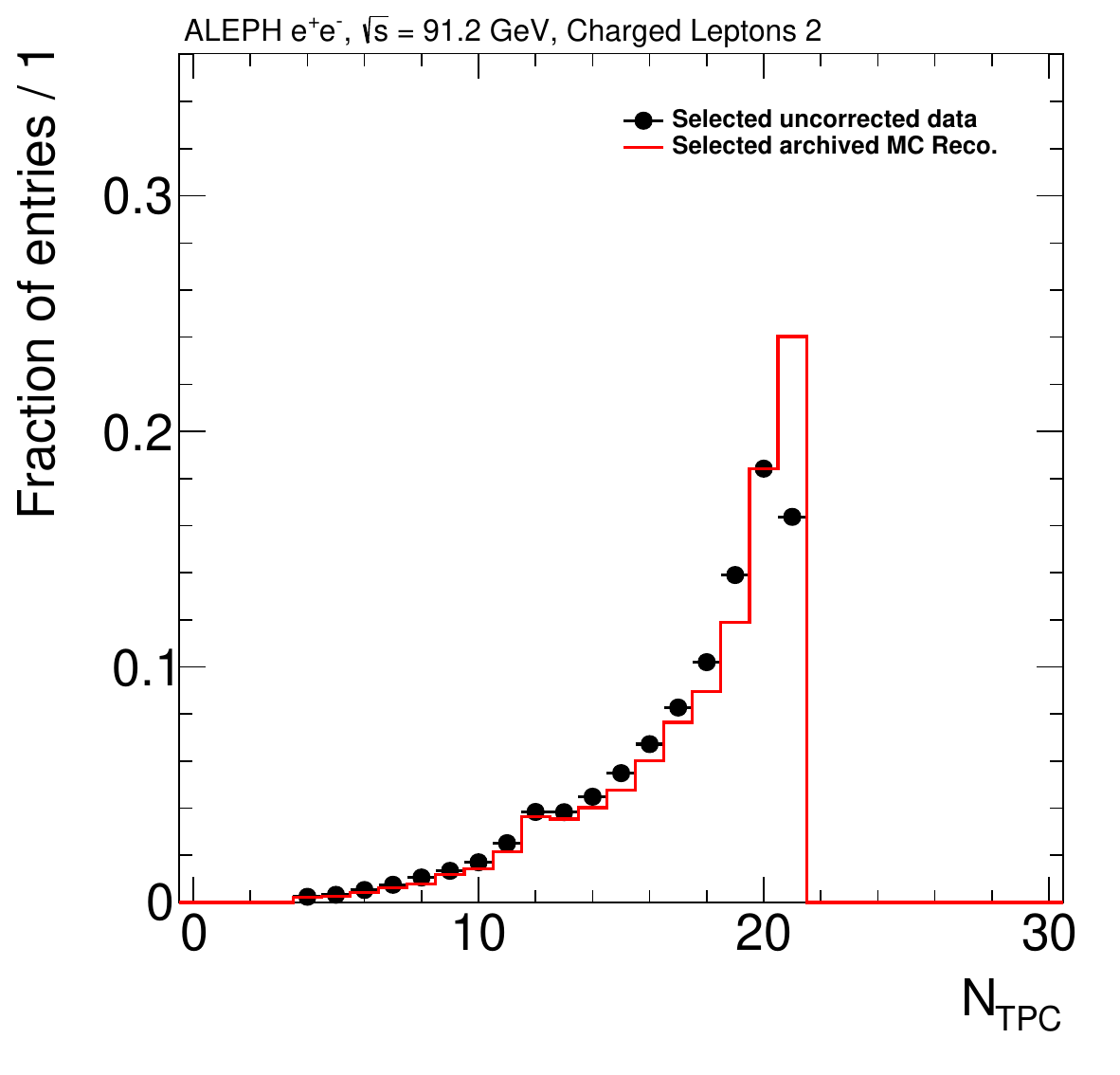}
    \caption{}
    \label{fig:kinem_pwflag2_mainbody_d}
\end{subfigure}
\begin{subfigure}[b]{0.32\textwidth}
    \includegraphics[width=\textwidth,angle=0]{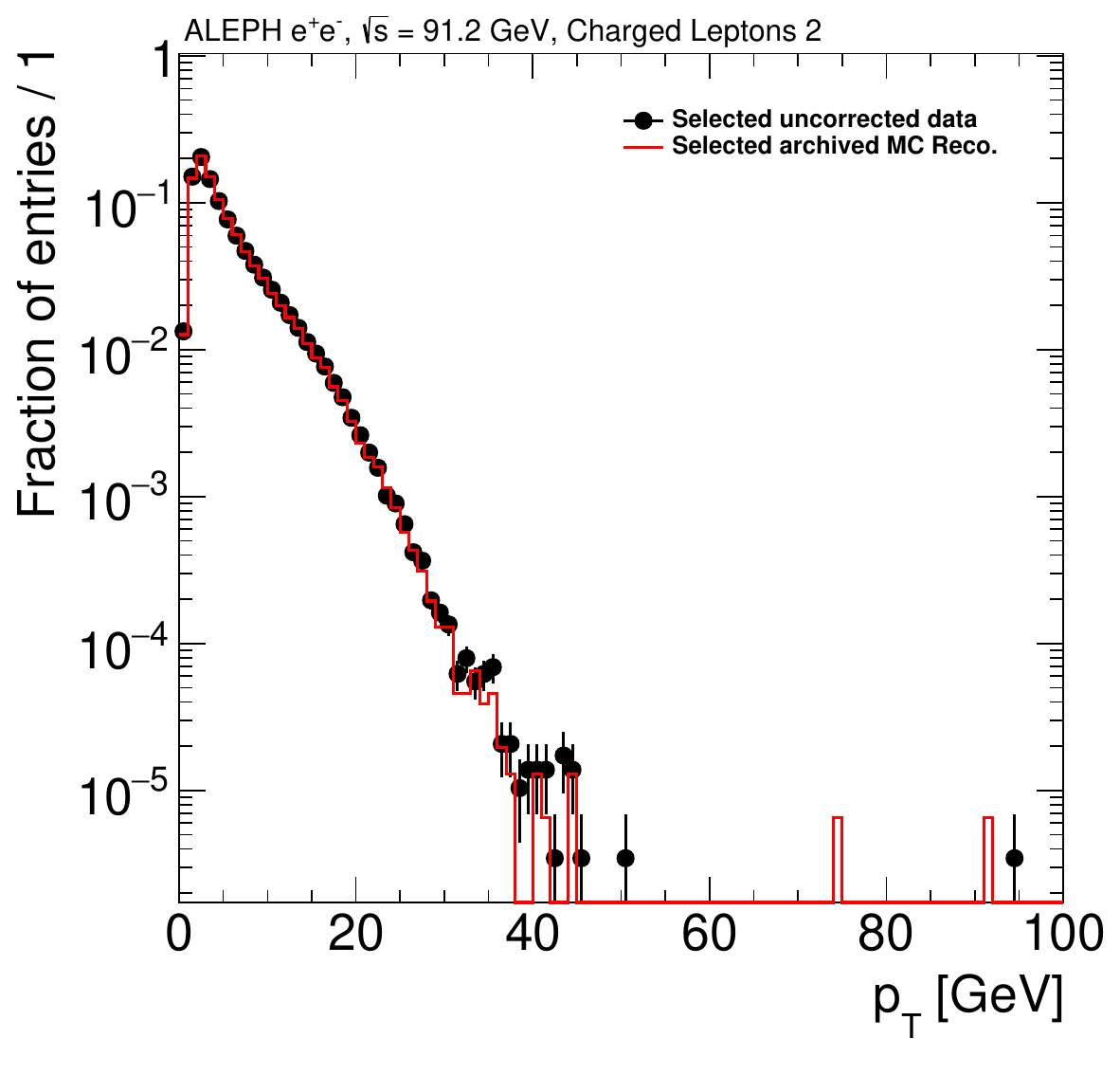}
    \caption{}
    \label{fig:kinem_pwflag2_mainbody_e}
\end{subfigure}
\begin{subfigure}[b]{0.32\textwidth}
    \includegraphics[width=\textwidth,angle=0]{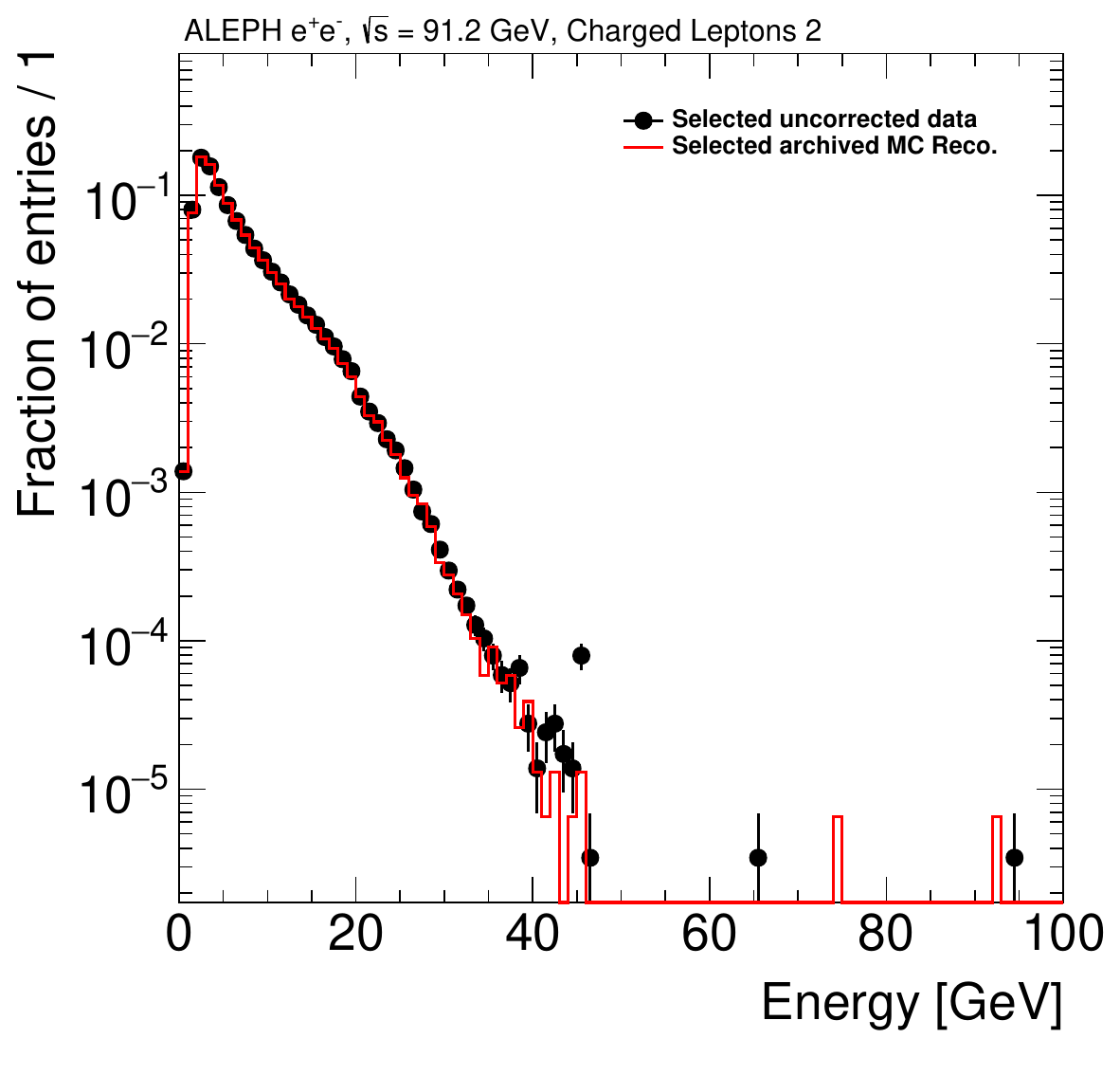}
    \caption{}
    \label{fig:kinem_pwflag2_mainbody_f}
\end{subfigure}
\caption{Observables used for selecting high-quality charged leptons with tracker hits and hits in the muon sub-detectors from penetration through the calorimeter: (a) the cosine of the polar angle $\theta$, (b) the transverse impact parameter $d_{0}$, (c) the longitudinal impact parameter $z_{0}$, (d) the number of hits in the time projection chamber ($N_{\mathrm{TPC}}$), (e) the transverse momentum~\pT, and (f) the energy (E).}
\label{fig:kinem_pwflag2_mainbody}
\end{figure}

The first category of charged leptons corresponds to those charged particle candidates with hits in the tracker and in the ECAL. This primarily represents electrons, since those particles produce characteristic showers in the electromagnetic calorimeters. The $\cos\theta$ spectrum in Figure~\ref{fig:kinem_pwflag1_mainbody_a} shows the acceptance of those particles with barrel and endcap calorimetry coverage. The other observables are also shown, and in general good agreement is seen between the archived data and the MC simulated events. The second category of charged leptons corresponds to those charged particle candidates with hits in the tracker and that penetrate the calorimeters to hit the muon chambers. This primarily represents muons. The $\cos\theta$ spectrum in Figure~\ref{fig:kinem_pwflag2_mainbody_a} reflects the acceptance of these particles via the barrel and endcap muon chambers. The other kinematic distributions are also shown and exhibit generally good agreement between the archived data and the MC simulated events. For both categories of charged leptons, data shows small excesses over MC at high \pT~and energy. The impact on the thrust distribution with and without this contribution was found to be negligible, and therefore no explicit cut was applied for this contribution.

\begin{figure}[t!]
\centering
\begin{subfigure}[b]{0.32\textwidth}
    \includegraphics[width=\textwidth,angle=0]{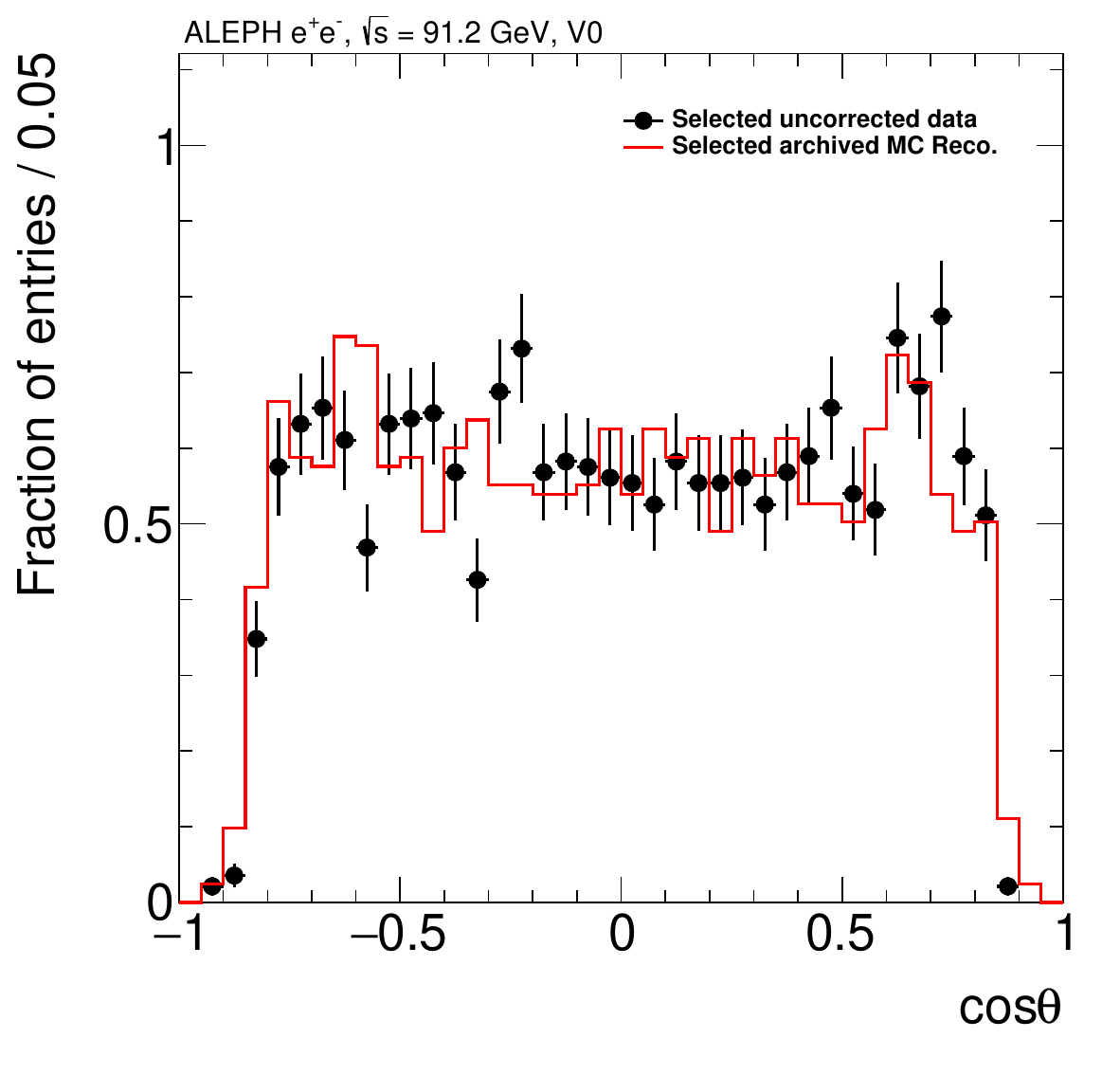}
    \caption{}
    \label{fig:kinem_pwflag3_mainbody_a}
\end{subfigure}
\begin{subfigure}[b]{0.32\textwidth}
    \includegraphics[width=\textwidth,angle=0]{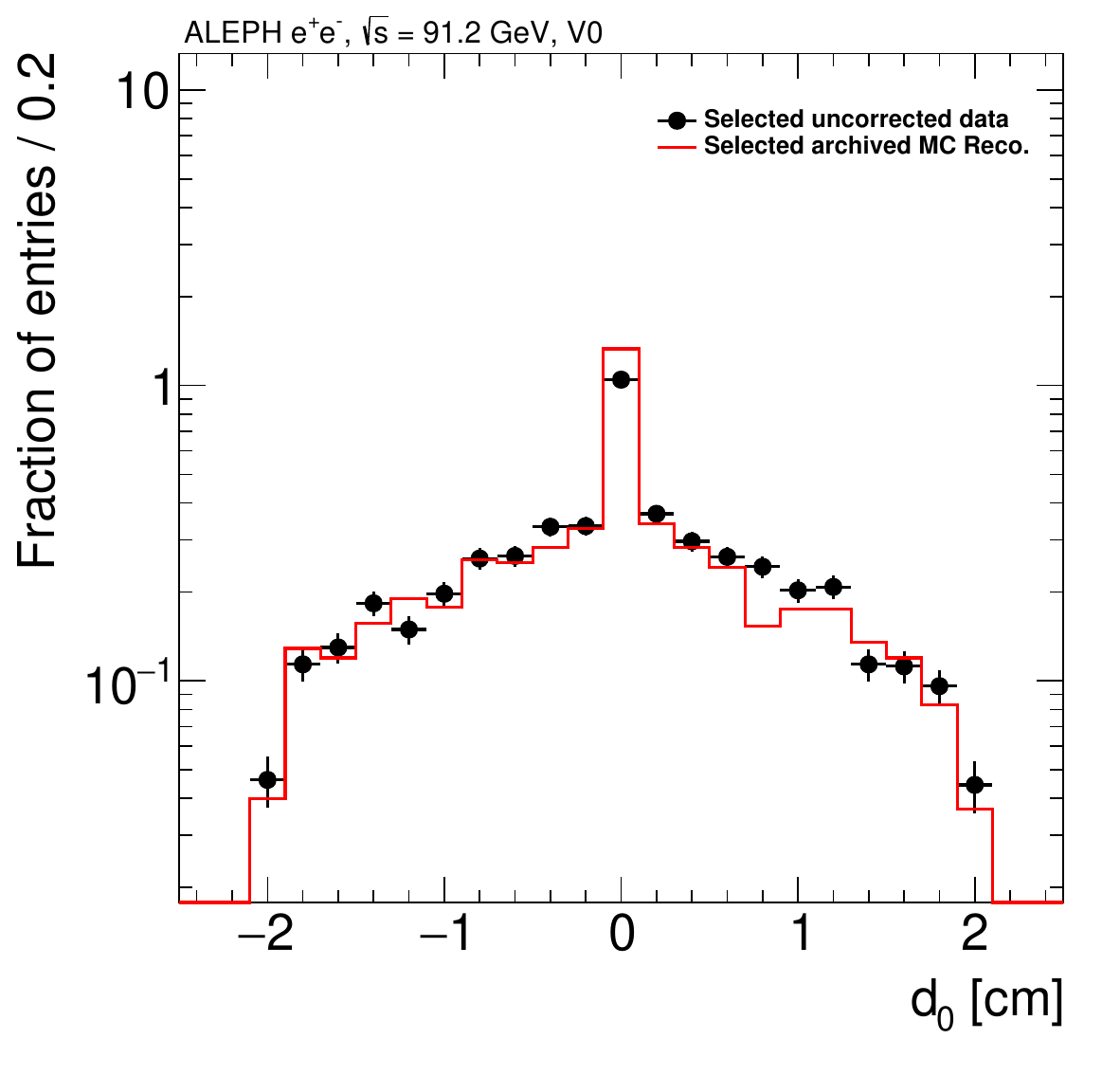}
    \caption{}
    \label{fig:kinem_pwflag3_mainbody_b}
\end{subfigure}
\begin{subfigure}[b]{0.32\textwidth}
    \includegraphics[width=\textwidth,angle=0]{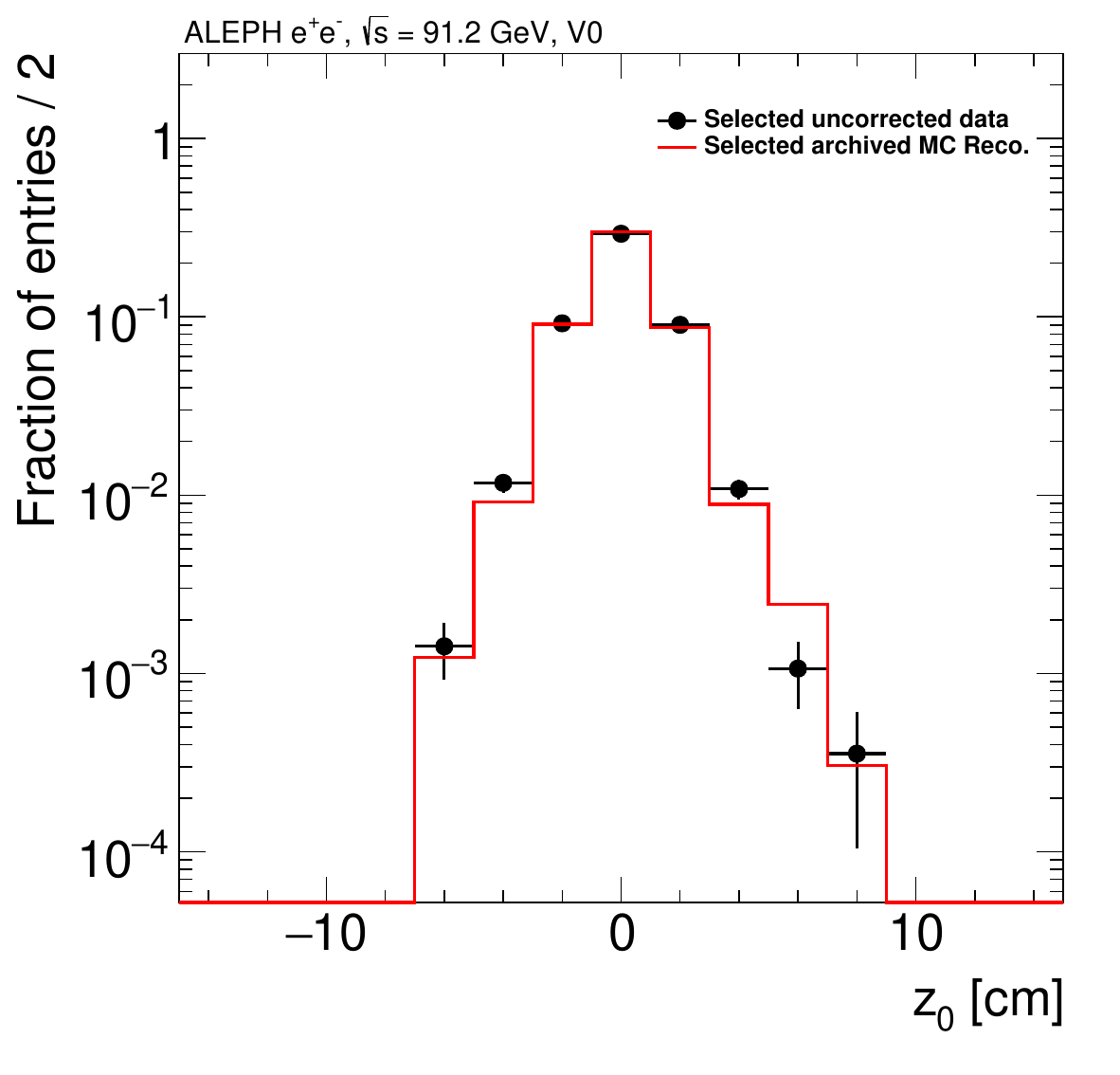}
    \caption{}
    \label{fig:kinem_pwflag3_mainbody_c}
\end{subfigure}
\begin{subfigure}[b]{0.32\textwidth}
    \includegraphics[width=\textwidth,angle=0]{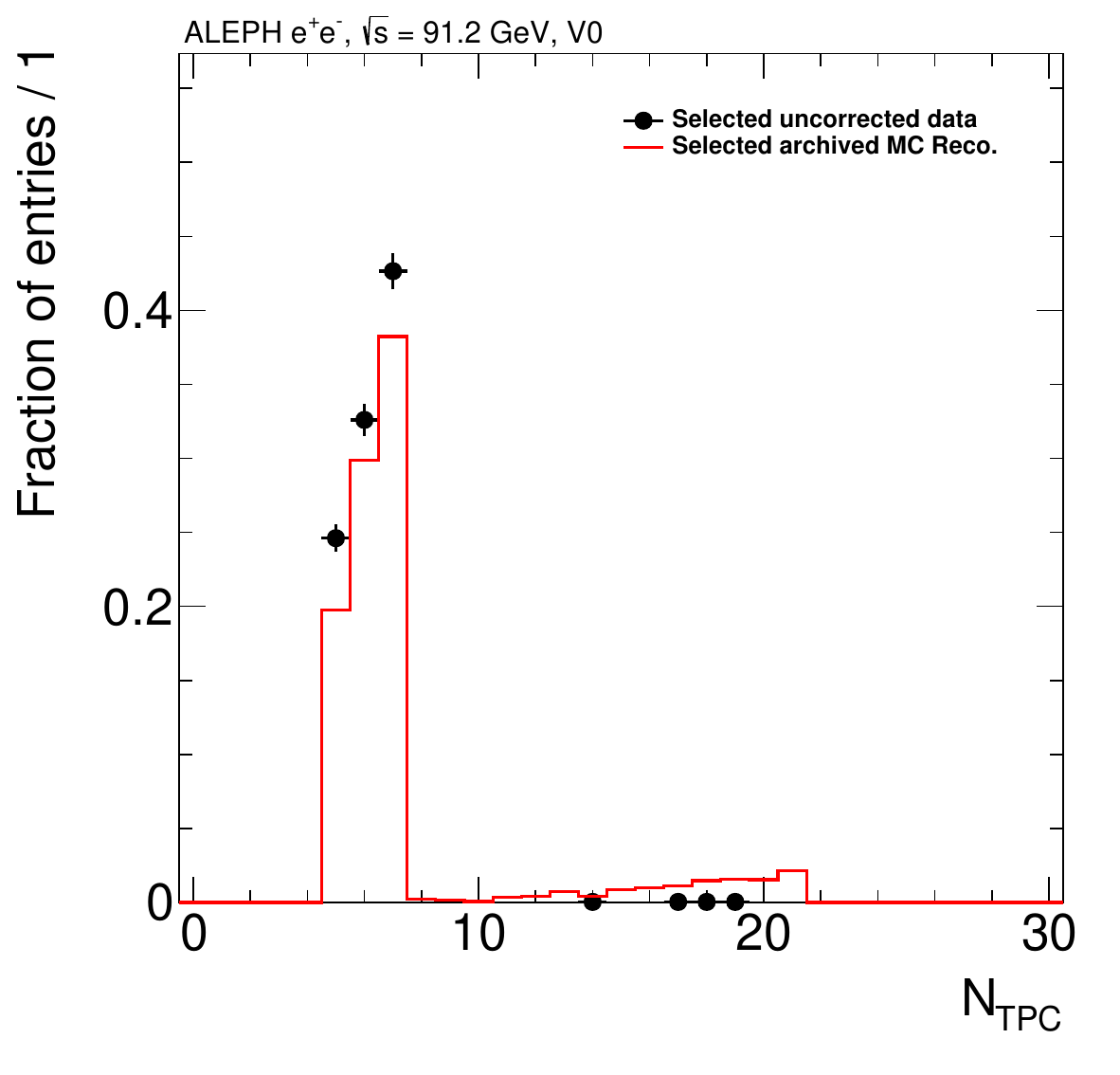}
    \caption{}
    \label{fig:kinem_pwflag3_mainbody_d}
\end{subfigure}
\begin{subfigure}[b]{0.32\textwidth}
    \includegraphics[width=\textwidth,angle=0]{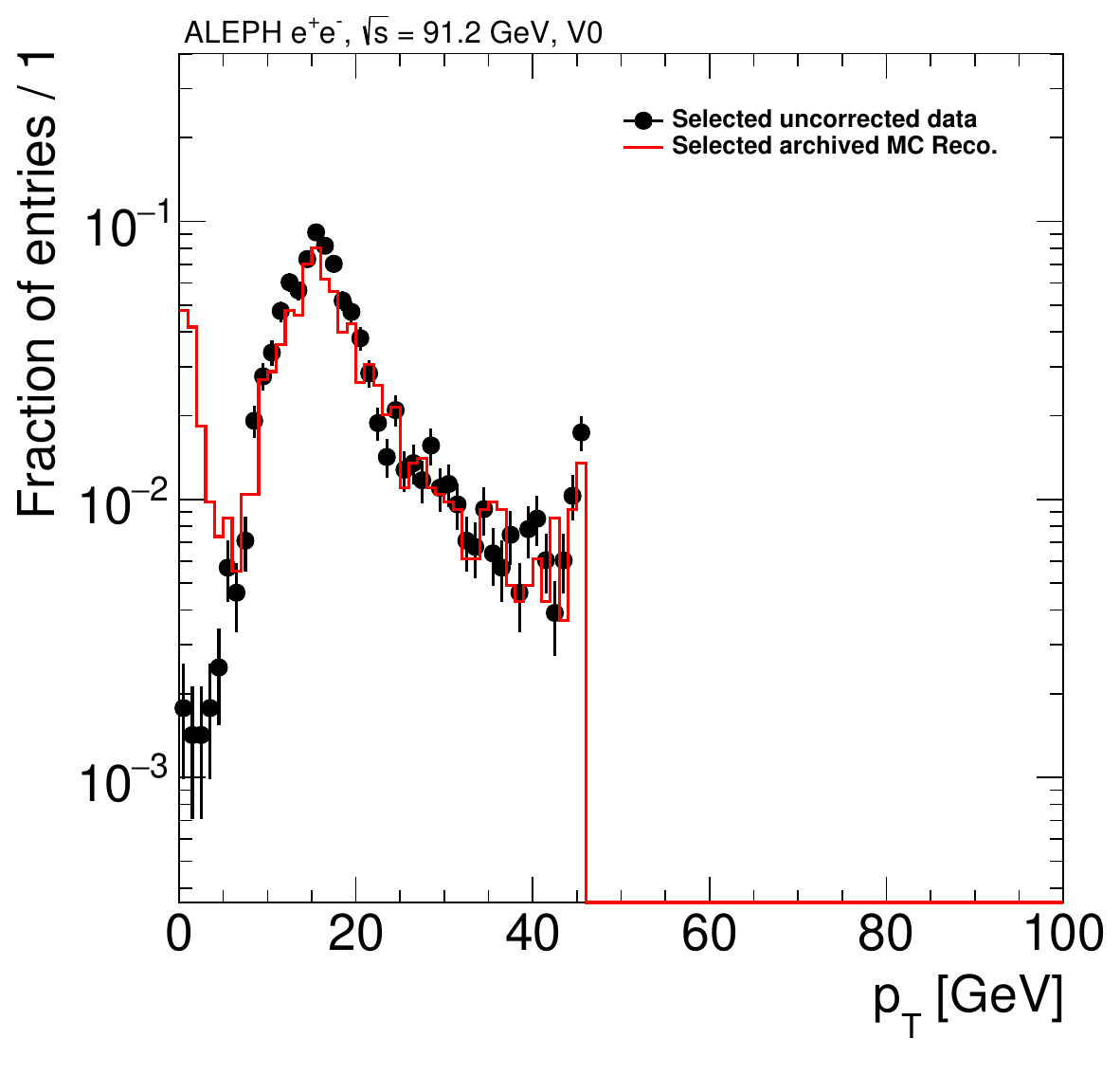}
    \caption{}
    \label{fig:kinem_pwflag3_mainbody_e}
\end{subfigure}
\begin{subfigure}[b]{0.32\textwidth}
    \includegraphics[width=\textwidth,angle=0]{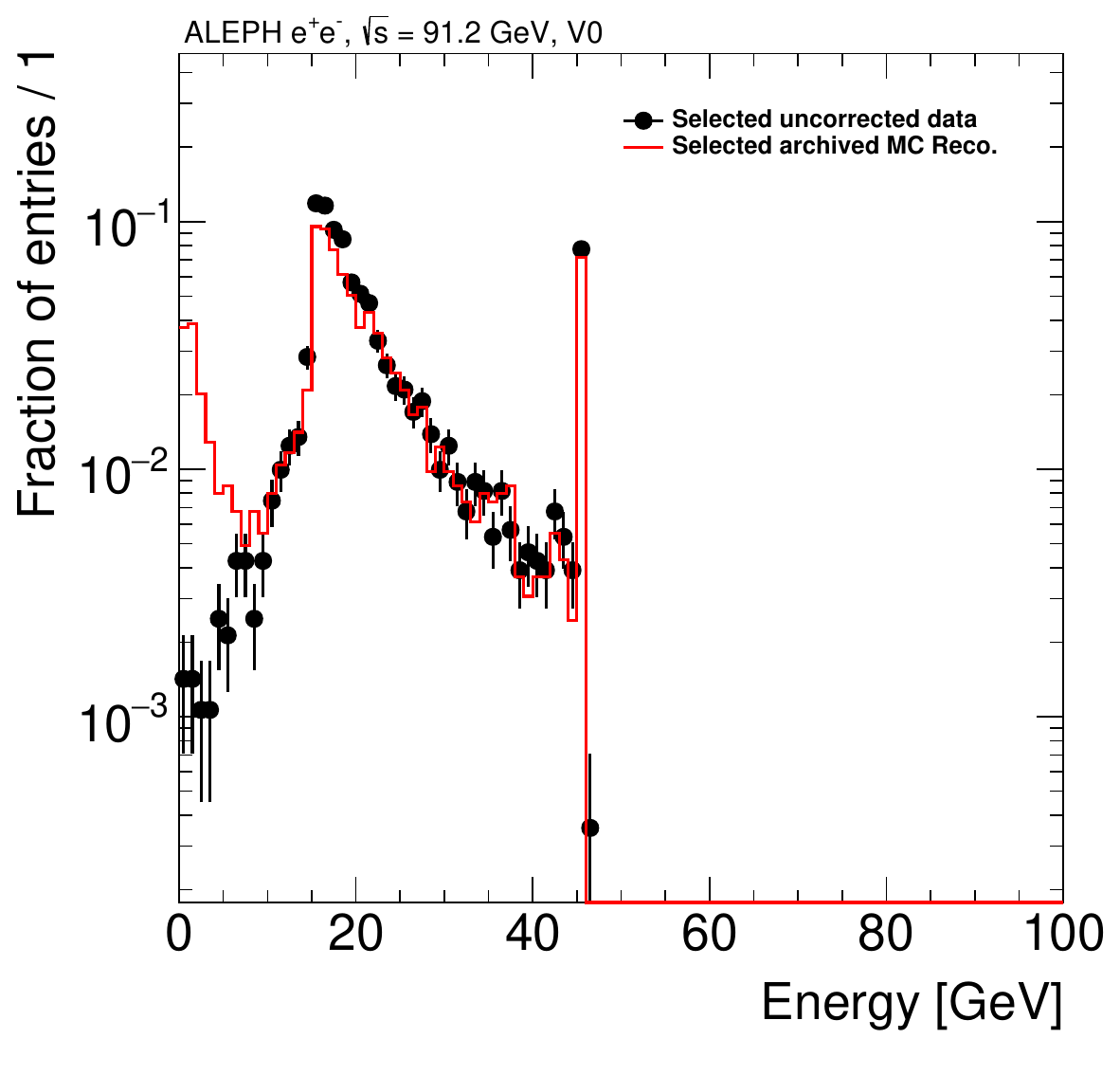}
    \caption{}
    \label{fig:kinem_pwflag3_mainbody_f}
\end{subfigure}
\caption{Observables used for selecting high-quality V0 particles: (a) the cosine of the polar angle $\theta$, (b) the transverse impact parameter $d_0$, (c) the longitudinal impact parameter $z_0$, (d) the number of hits in the time projection chamber ($N_{\mathrm{TPC}}$), (e) the transverse momentum~\pT, and (f) the energy (E).}
\label{fig:kinem_pwflag3_mainbody}
\end{figure}

The V$^{0}$ candidates correspond to neutral, long-lived particles that decay into pairs of charged tracks. These particles are reconstructed using displaced vertices, which are characterized by large impact parameters relative to those of prompt charged particles discussed earlier. Figure~\ref{fig:kinem_pwflag3_mainbody} shows the relevant kinematic distributions for these candidates. In general, good agreement is observed between data and MC simulation in the high \pT~region. However, at low \pT~$<5$ GeV, a significant discrepancy between data and simulation is evident, particularly in the distribution shown in Figure~\ref{fig:kinem_pwflag3_mainbody_e}. This difference may be attributed to limitations in the tracking efficiency for displaced objects or the modeling of low-momentum V$^{0}$ physics. The V$^{0}$ particles are found to have a small impact on the thrust distribution. Therefore, no additional cuts are applied for their contribution and further studies on the mismodeling are left for future works.

\begin{figure}[t!]
\centering
\begin{subfigure}[b]{0.4\textwidth}
    \includegraphics[width=\textwidth,angle=0]{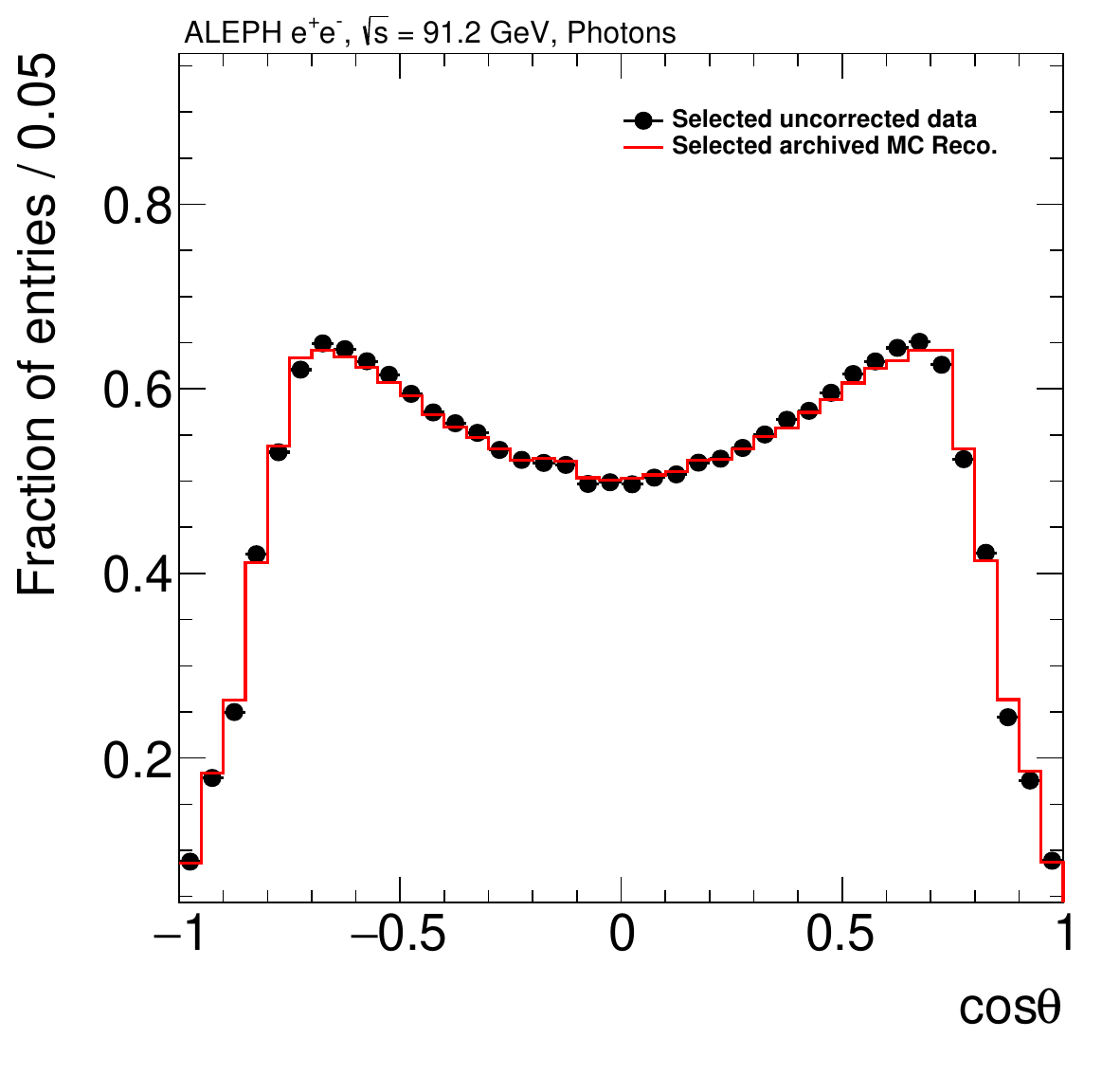}
    \caption{}
    \label{fig:kinem_pwflag4and5_mainbody_a}
\end{subfigure}
\begin{subfigure}[b]{0.4\textwidth}
    \includegraphics[width=\textwidth,angle=0]{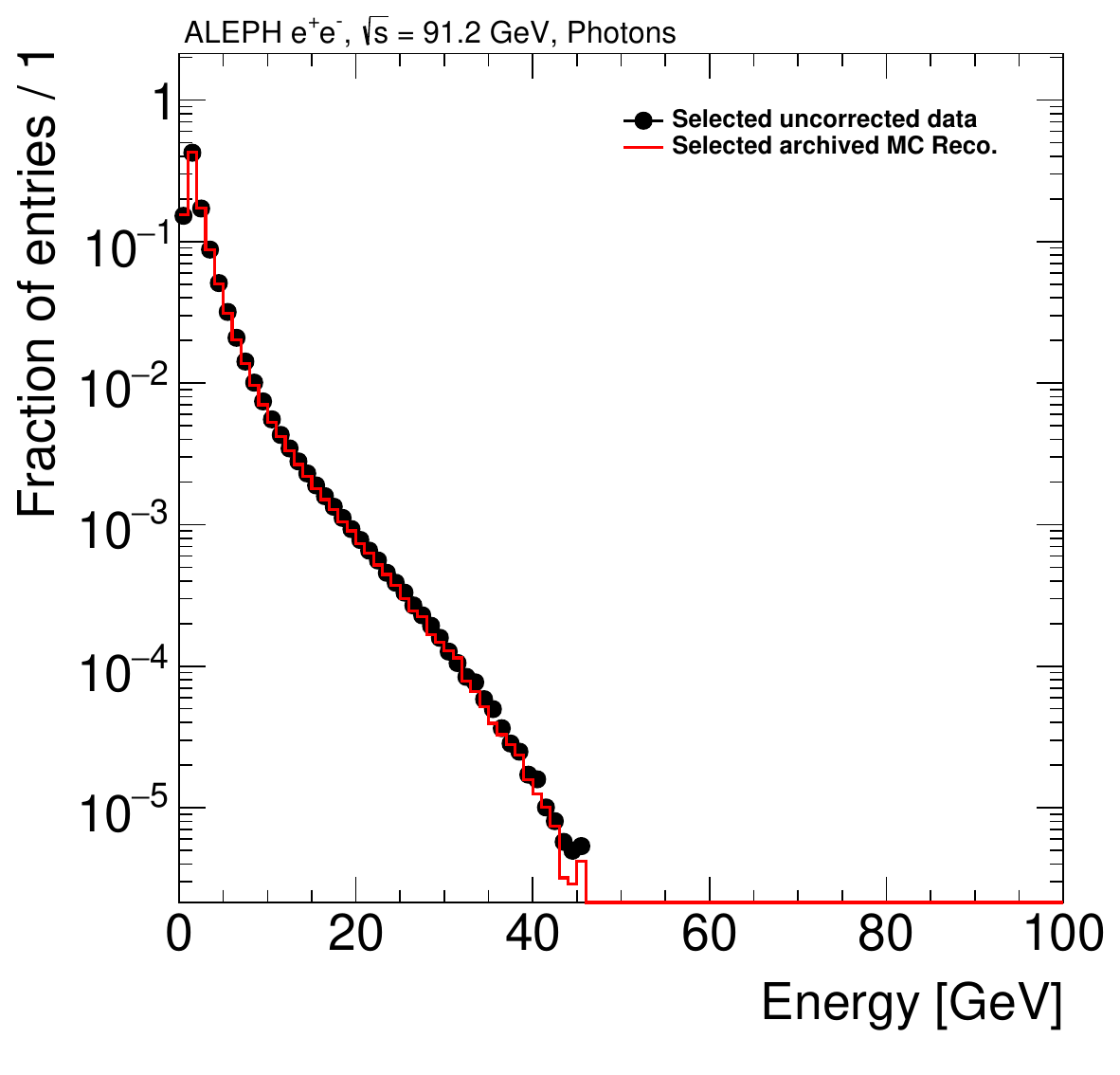}
    \caption{}
    \label{fig:kinem_pwflag4and5_mainbody_b}
\end{subfigure}
\begin{subfigure}[b]{0.4\textwidth}
    \includegraphics[width=\textwidth,angle=0]{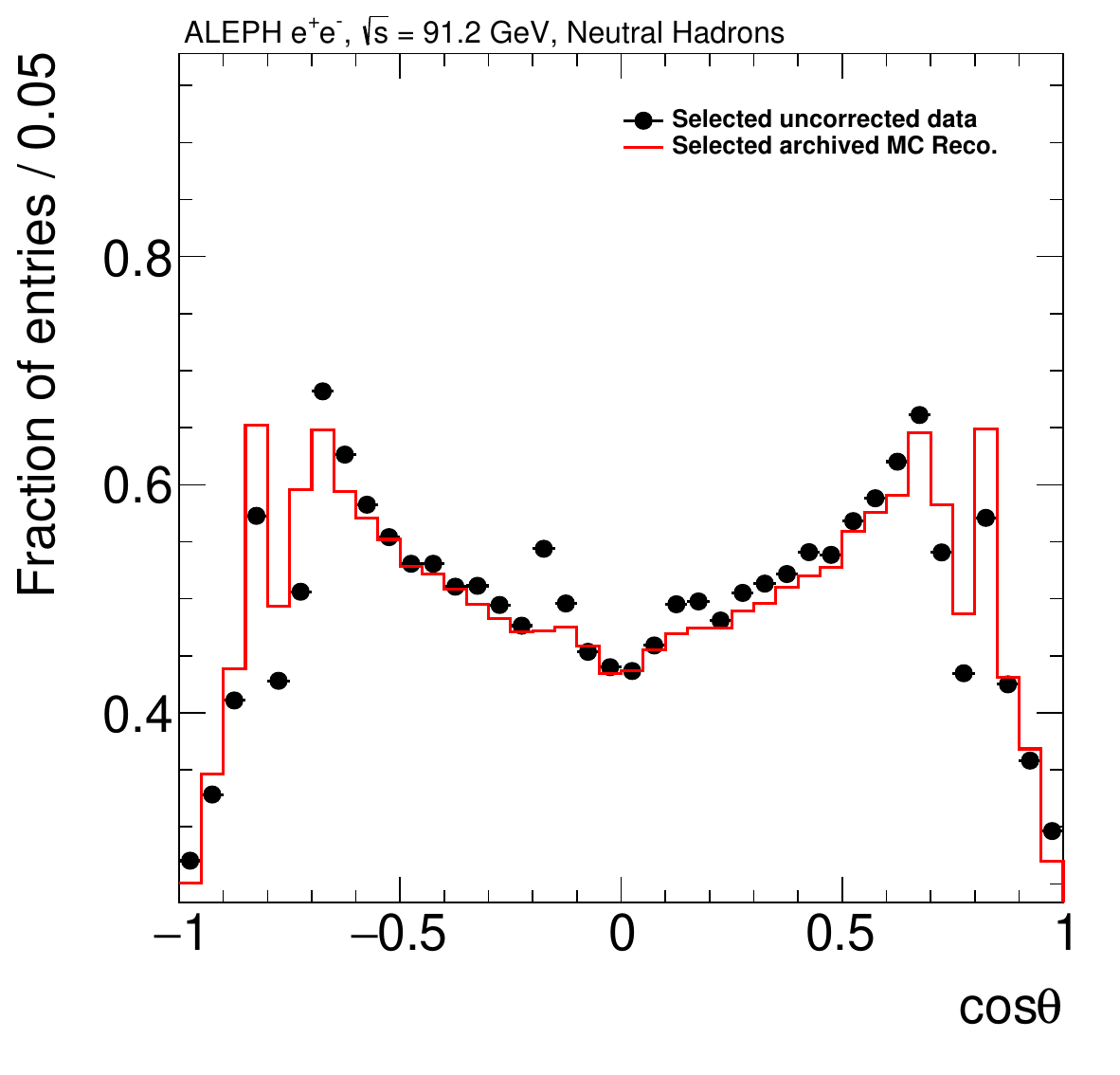}
    \caption{}
    \label{fig:kinem_pwflag4and5_mainbody_c}
\end{subfigure}
\begin{subfigure}[b]{0.4\textwidth}
    \includegraphics[width=\textwidth,angle=0]{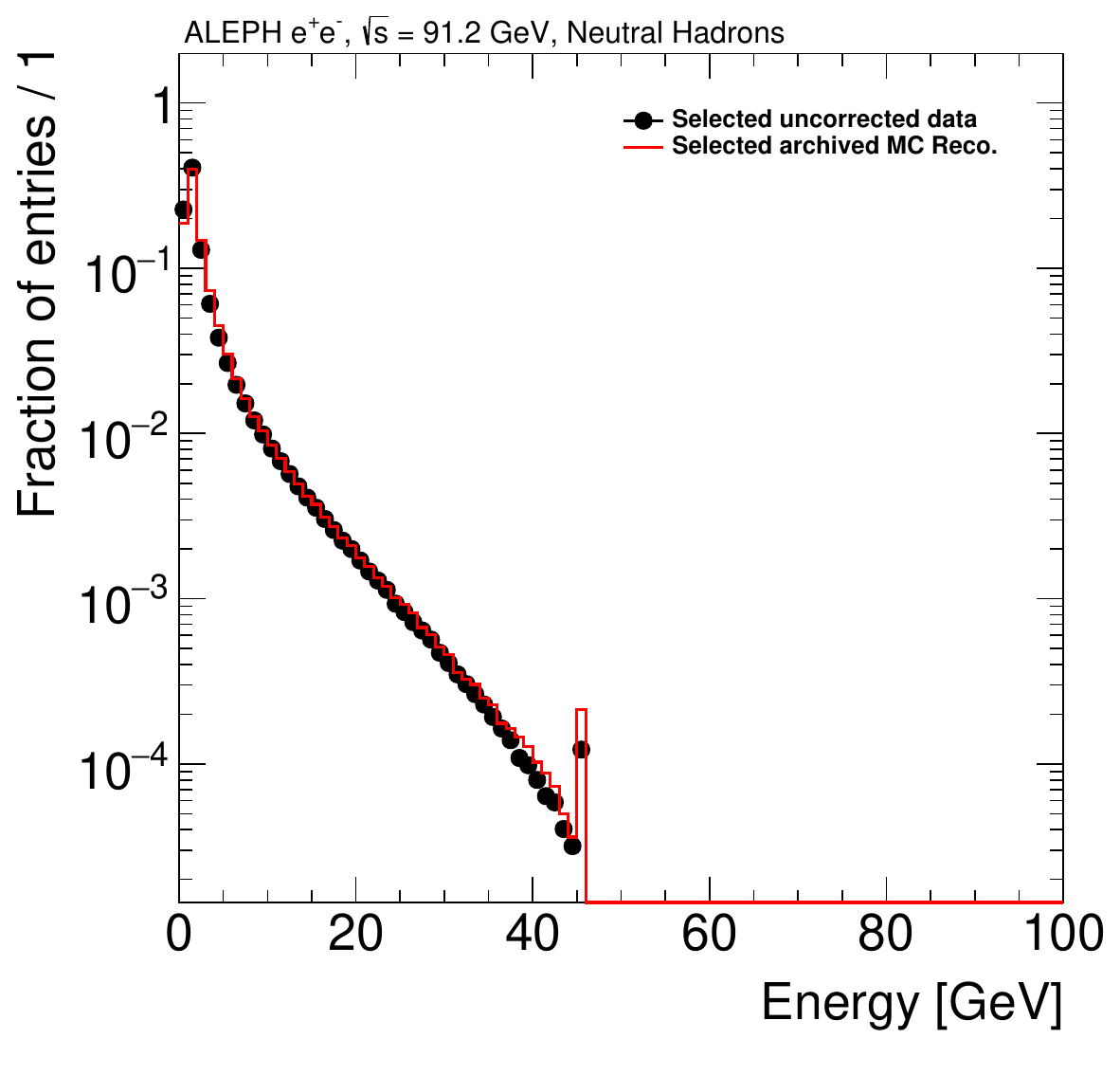}
    \caption{}
    \label{fig:kinem_pwflag4and5_mainbody_d}
\end{subfigure}
\caption{Observables used for the selection of high-quality photons and neutral hadrons. Panels (a) and (b) show, respectively, the cosine of the polar angle $\theta$ and the energy spectrum for photons. Panels (c) and (d) show the distributions for neutral hadrons.}
\label{fig:kinem_pwflag4and5_mainbody}
\end{figure}

The category of neutral particles corresponding to photon candidates includes those with no associated track in the tracker but with significant energy deposits in the electromagnetic calorimeter. These are primarily photons arising from QED radiation processes, such as initial-state radiation from the incoming leptons or final-state radiation from charged particles produced in the decay of the Z boson. The photons initiate electromagnetic showers similar to those of electrons but are distinguished by the absence of a matching charged track. The $\cos\theta$ spectrum shown in Figure~\ref{fig:kinem_pwflag4and5_mainbody_a} exhibits a pronounced enhancement in the forward and backward regions, which reflects the detector coverage of the ECAL and QED radiation pattern where photons are preferentially emitted collinear to the initial or final charged particles. The photon radiation leads to a characteristic smoothly falling energy spectrum as shown in Figure~\ref{fig:kinem_pwflag4and5_mainbody_b}. As with the charged leptons, the other kinematic observables are also presented and show generally good agreement between the archived data and the MC simulated events.

The category of neutral particles corresponding to neutral hadron candidates includes those with no associated track in the tracker and with energy deposits predominantly in the HCAL. Those candidates are typically longer lived hadrons that do not interact significantly with the ECAL. Instead, they interact hadronically with the HCAL material, producing hadronic showers that are broader and less localized than electromagnetic ones. Importantly, these particles are distinct from $V^0$ candidates, which refer to neutral particles such as $K^0_S$ and $\Lambda$ that decay into two charged tracks and can be reconstructed from a visible secondary vertex in the tracker. In contrast, neutral hadrons like neutrons leave no track or reconstructible decay signature and are instead identified via their calorimeter energy deposits. The $\cos\theta$ spectrum in Figure~\ref{fig:kinem_pwflag4and5_mainbody_c} reflects both the production kinematics of these neutral hadrons and the acceptance of the hadronic calorimeter, with yield modulated by its coverage in the barrel and endcap regions. A discrepancy is seen around $\cos\theta\sim$ 0.2, where data shows an peak that is not explained by the MC. The peak was checked in finer bin sizes of 0.01 and observed to be in a single bin. Particles that fall into that single bin are removed from both the data and MC when the thrust distribution is calculated.
The neutral hadron energy spectrum is shown in Figure~\ref{fig:kinem_pwflag4and5_mainbody_d}, with characteristic falling spectrum seen in both data and MC. As with other particle categories, the remaining kinematic observables are shown in the appendix and display generally good agreement between the archived data and the MC simulated events.


\begin{figure}[t!]
\centering
\begin{subfigure}[b]{0.4\textwidth}
    \includegraphics[width=\textwidth,angle=0]{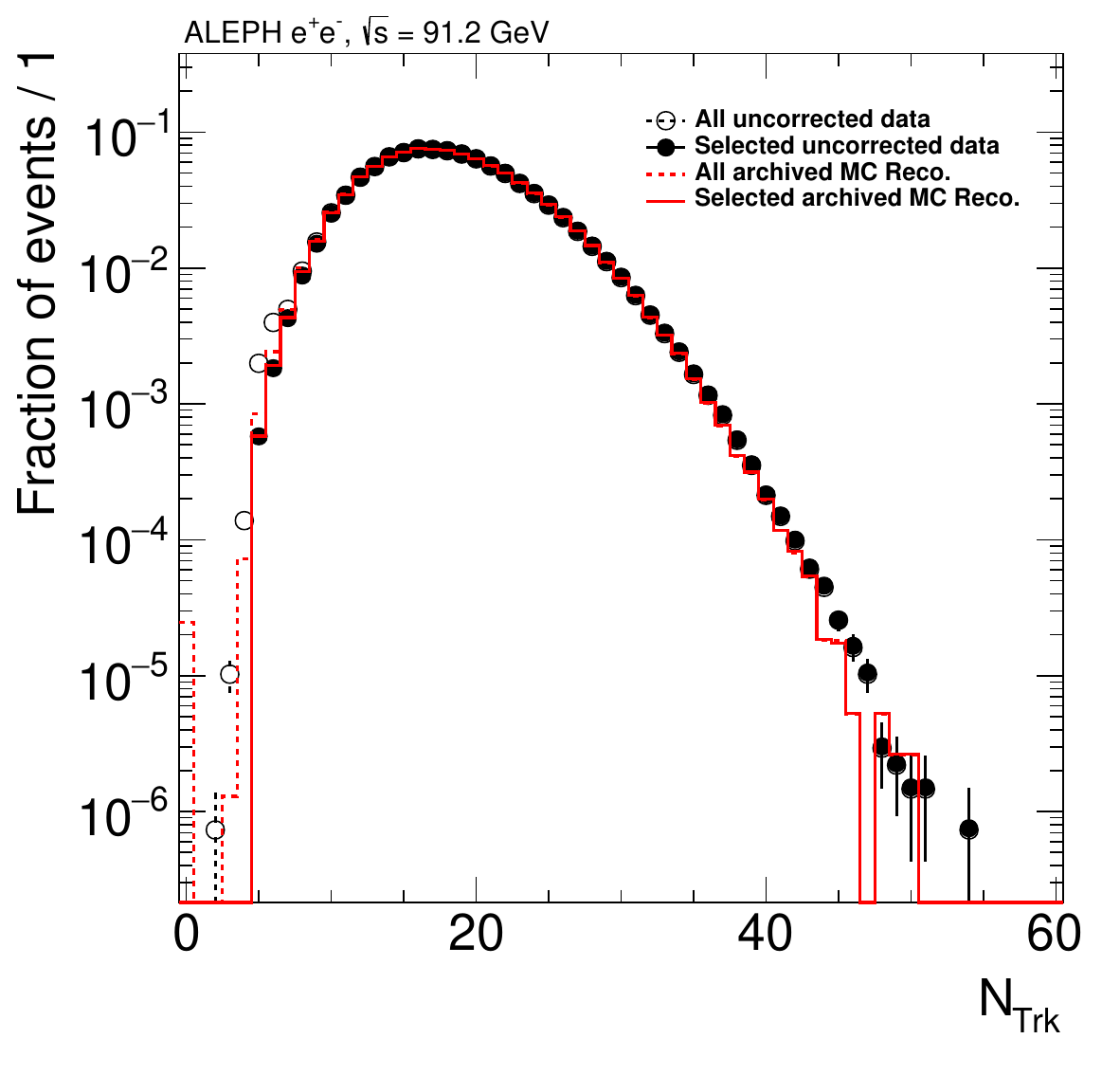}
    \caption{}
    \label{fig:EventObs_mainbody_a}
\end{subfigure}
\begin{subfigure}[b]{0.4\textwidth}
    \includegraphics[width=\textwidth,angle=0]{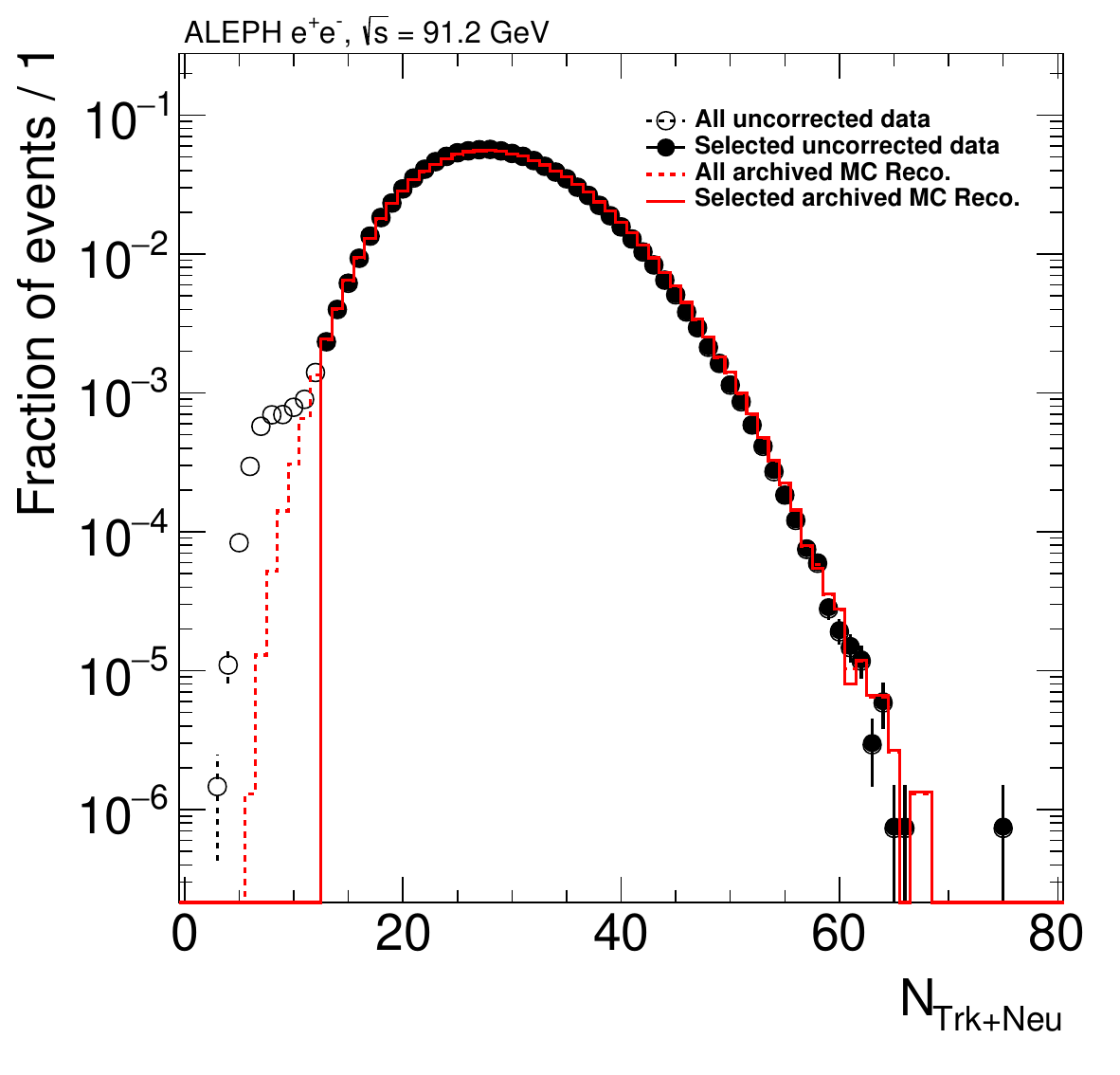}
    \caption{}
    \label{fig:EventObs_mainbody_b}
\end{subfigure}
\begin{subfigure}[b]{0.4\textwidth}
    \includegraphics[width=\textwidth,angle=0]{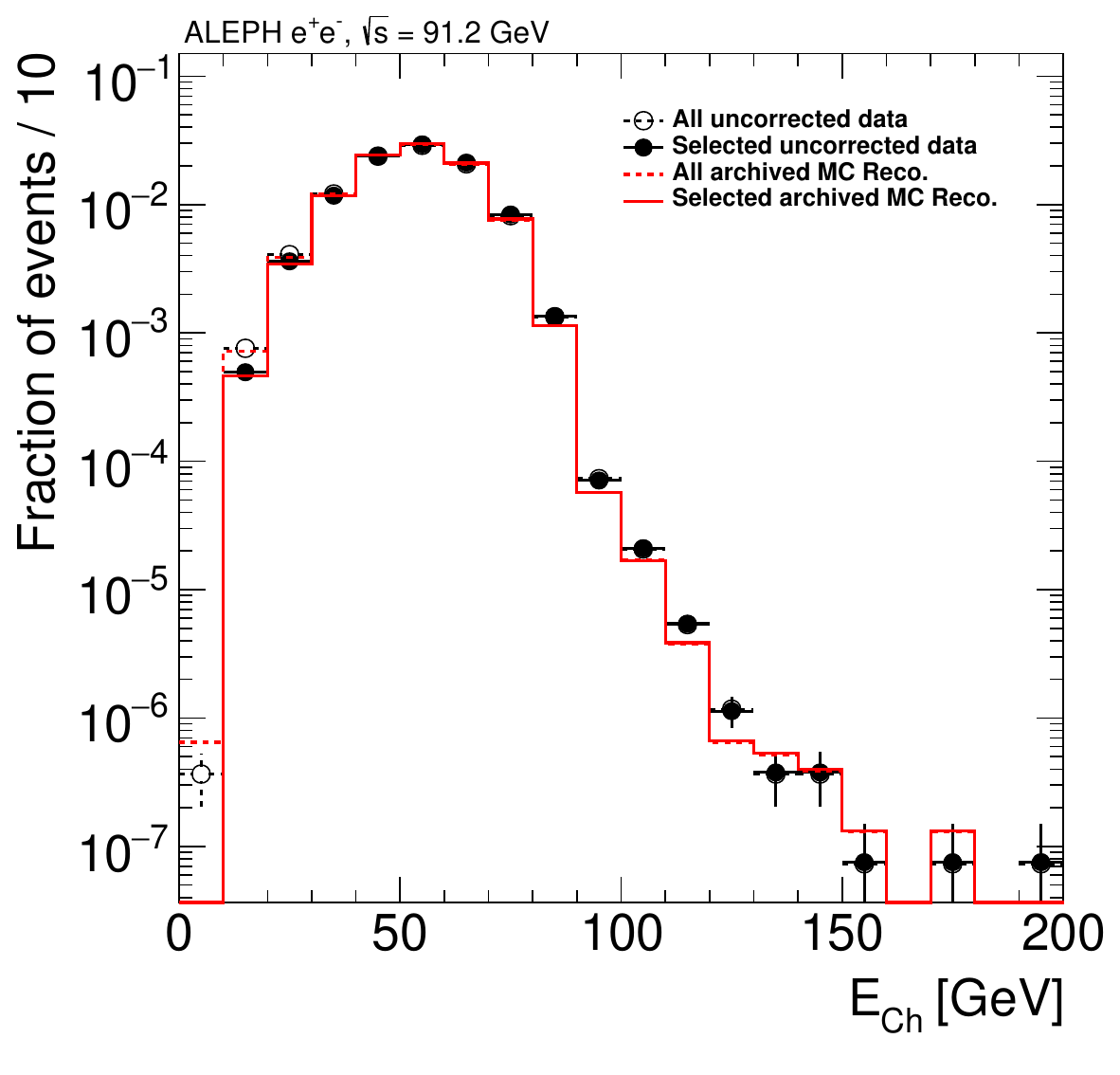}
    \caption{}
    \label{fig:EventObs_mainbody_c}
\end{subfigure}
\begin{subfigure}[b]{0.4\textwidth}
    \includegraphics[width=\textwidth,angle=0]{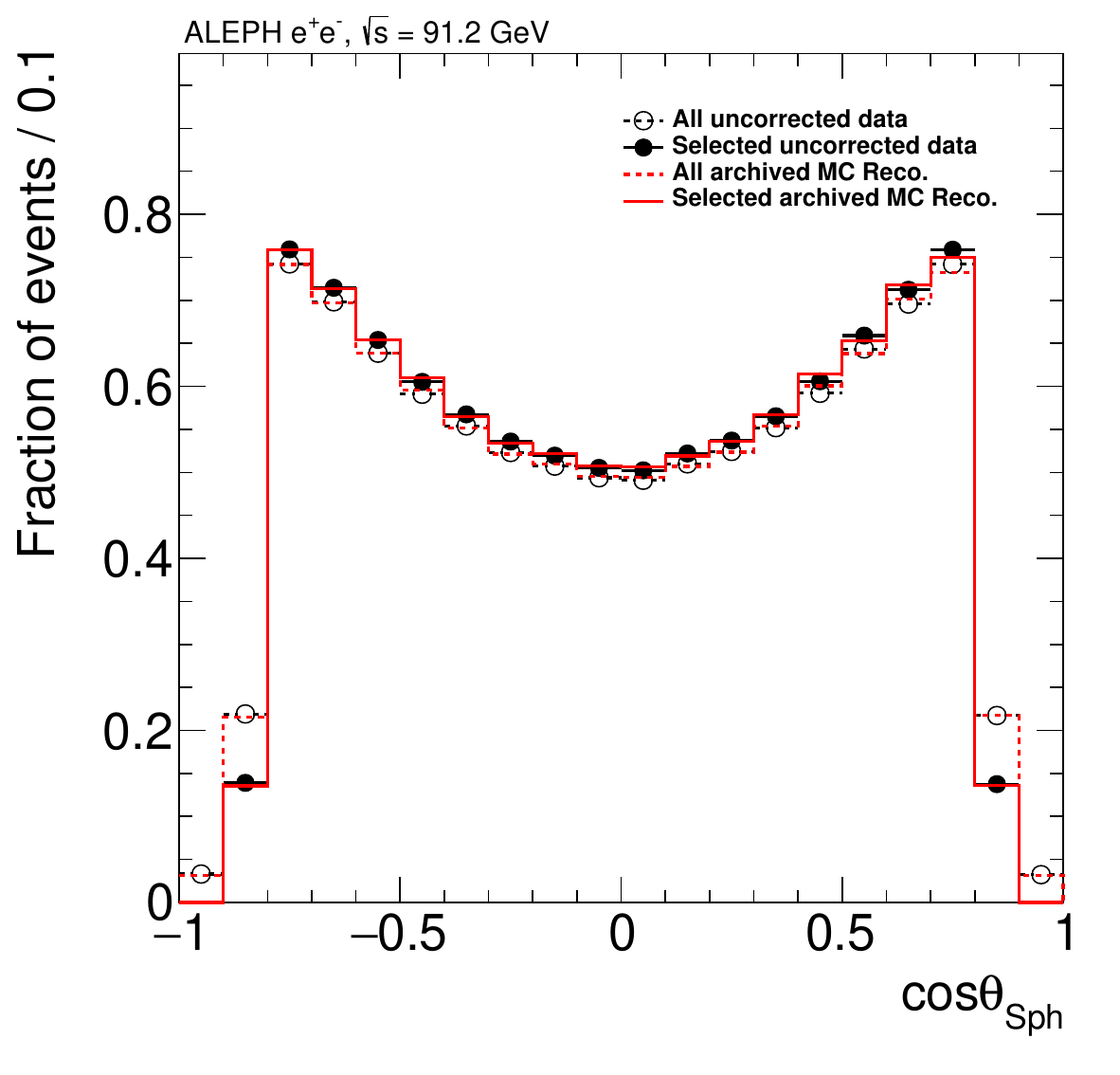}
    \caption{}
    \label{fig:EventObs_mainbody_d}
\end{subfigure}
\caption{Observables used for selecting high-quality hadronic $Z$ decay events: (a) the selected charged particle multiplicity, (b) the charged plus neutral particle multiplicity, (c) the total charged energy, and (d) the cosine of the sphericity axis polar angle.}
\label{fig:EventObs_mainbody}
\end{figure}

The selected objects are used to build event level observables that are used for identifying hadronic Z events as outlined in Table~\ref{tab:selections}. Those distributions are shown before and after event selection to assess the agreement with the archived MC. The selected charged and charged plus neutral particle multiplicities are shown in Figures~\ref{fig:EventObs_mainbody_a}-~\ref{fig:EventObs_mainbody_b}. The charged and total multiplicities peak at roughly 20 and 30 particles, respectively. The total charged particle energy is shown in Figure~\ref{fig:EventObs_mainbody_c}. The spectrum peaks around 50--60\,GeV, which can be understood by noting that roughly two-thirds of the pions produced are charged, so about $\frac{2}{3} \times 91\,\mathrm{GeV} \approx 60\,\mathrm{GeV}$. The remaining difference is due to tracking inefficiencies, which cause some energy loss. Lastly, the cosine of the polar angle of the sphericity axis is shown in Figure~\ref{fig:EventObs_mainbody_d}. Similar to the thrust distribution, the sphericity is an event shape observable measuring the overall event topology. The selection on the sphericity axis polar angle ensures that the event plane is contained within the detector acceptance. Those observables generally show good agreement with the archived MC.

\section{Analysis method}
\label{sec:analysismethod}

The analysis applies particle and event selections to obtain a pure sample of hadronic Z events. The thrust distribution is then built from the selected particles. An unbinned unfolding procedure corrects for the detector effects to extract the particle-level distribution.

\begin{figure}[t!]
\centering
\begin{subfigure}[b]{0.45\textwidth}
    \includegraphics[width=\textwidth,angle=0]{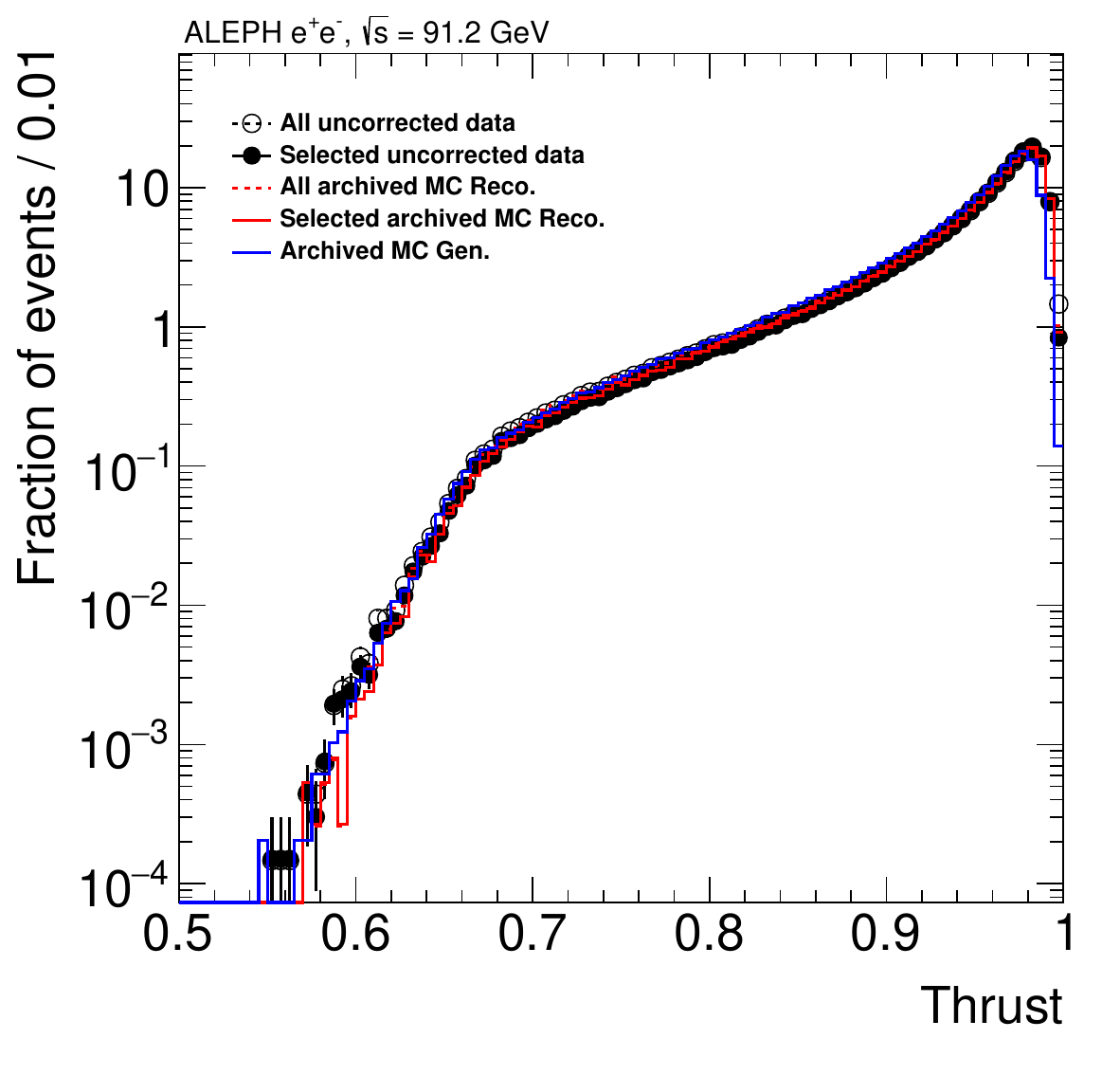}
    \caption{}
    \label{fig:detectorThrust_a}
\end{subfigure}
\begin{subfigure}[b]{0.45\textwidth}
    \includegraphics[width=\textwidth,angle=0]{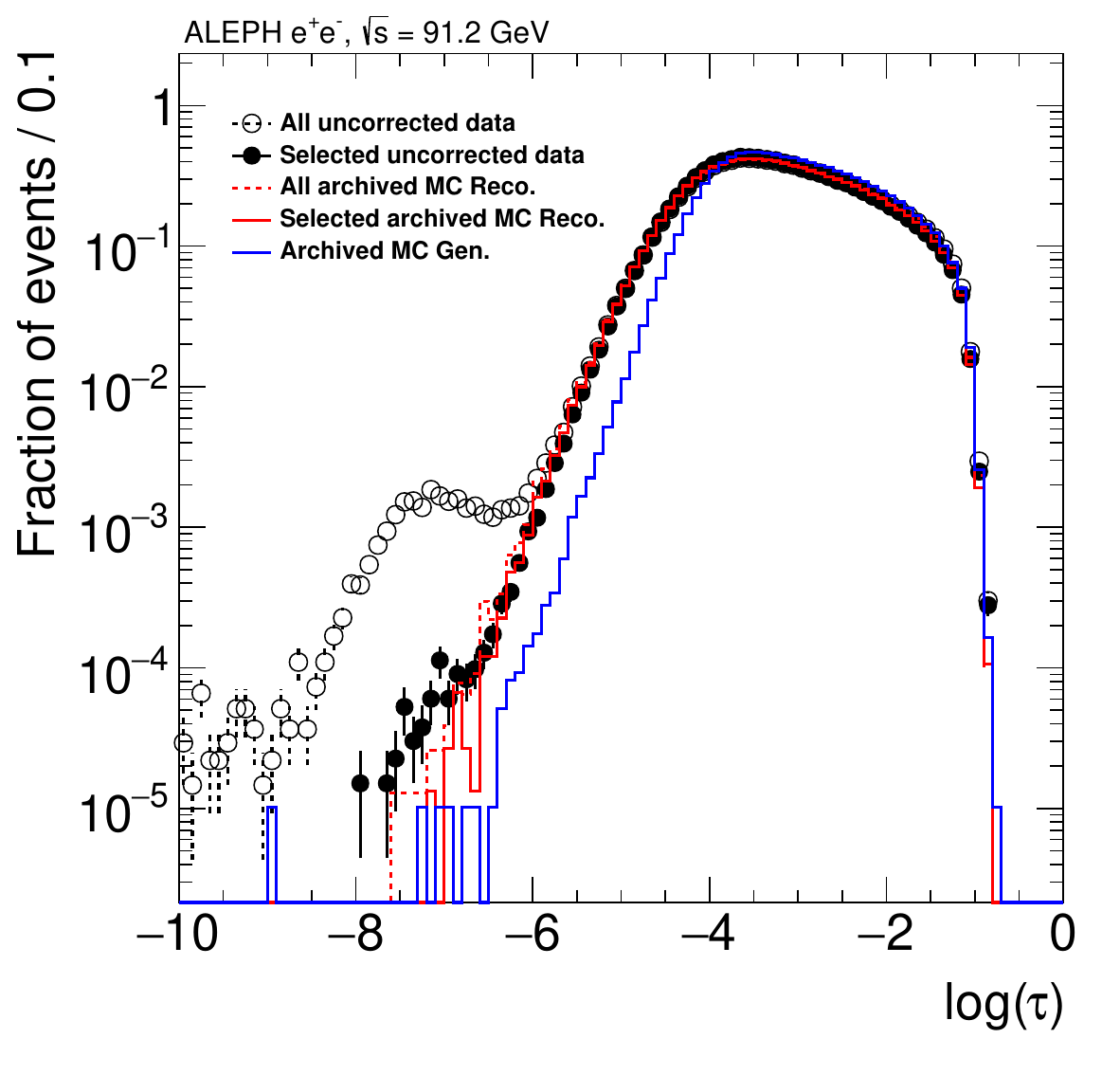}
    \caption{}
    \label{fig:detectorThrust_b}
\end{subfigure}
\caption{The thrust and $\log(\tau)$ distributions after nominal selections and without correction for detector effects, along with true particle-level thrust from MC for comparison.}
\label{fig:detectorThrust}
\end{figure}

\subsection{Thrust definition and construction}

Thrust is defined in the center-of-mass frame of an $e^+e^-$ collision as~\cite{PhysRevLett.39.1587}

\begin{equation}
    T = \max_{\hat{n}} \frac{\sum_i |\vec{p}_i \cdot \hat{n}|}{\sum_i |\vec{p}_i|}
\end{equation}

\noindent where the sum is over all particles in the event and the maximum is over 3-vectors $\vec{n}$ of unit norm. The vector $\hat{n}$ that maximizes thrust is known as the thrust axis and the plane which is defined as normal to $\hat{n}$ splits the event into 2 hemisphere. Since the thrust distribution roughly follows a log normal distribution, $\mathrm{P}(1- T) \sim \exp\left[- C \log^2 (1-T)\right])$, where $C$ is a constant, a second variable is defined as $\tau = 1-T$. The $\log\tau$ is used directly in the unfolding. A detailed description of the thrust distribution can be found in~\cite{Benitez:2024nav, Abbate:2010xh, Becher:2008cf}. The distribution is determined using all selected charged and neutral particles as outlined in Section~\ref{sec:eventReconAndSel}. Those objects are input to an optimization algorithm that iteratively determines the thrust axis. The algorithm implementation is based on the $\textsc{Herwig++}$ implementation~\cite{Brandt:1978zm, B_hr_2008, BUCKLEY20132803, rivet_thrust}, which is an exact algorithm to calculate thrust rather than one based on heuristics as is done in other implementations~\cite{Belle2Thrust}.

The detector level thrust and $\log\tau$ distributions are shown after event selections in Figure~\ref{fig:detectorThrust_a} and~\ref{fig:detectorThrust_b}, respectively. The distribution shows the characteristic shape with the dijet peak below the thrust maximum of 1 as a result of the Sudakov form factor suppression~\footnote{The Sudakov form factor is the no emission probability of radiating an additional soft particle which makes perfect dijet configurations less probable.} and broad tail from isotropic particle production. The event selections effectively remove events in the dijet region that arise from $Z\rightarrow \mu^{+}\mu^{-}\gamma$ and $Z \rightarrow  e^{+}e^{-}\gamma$ events.
The residual contamination from processes such as $e^{+}e^{-} \rightarrow \tau^{+}\tau^{-}$ is expected to be less than 0.26\% for these event selections~\cite{Barate:1996fi}.

\begin{figure}[t!]
\centering
\includegraphics[width=\textwidth,angle=0]{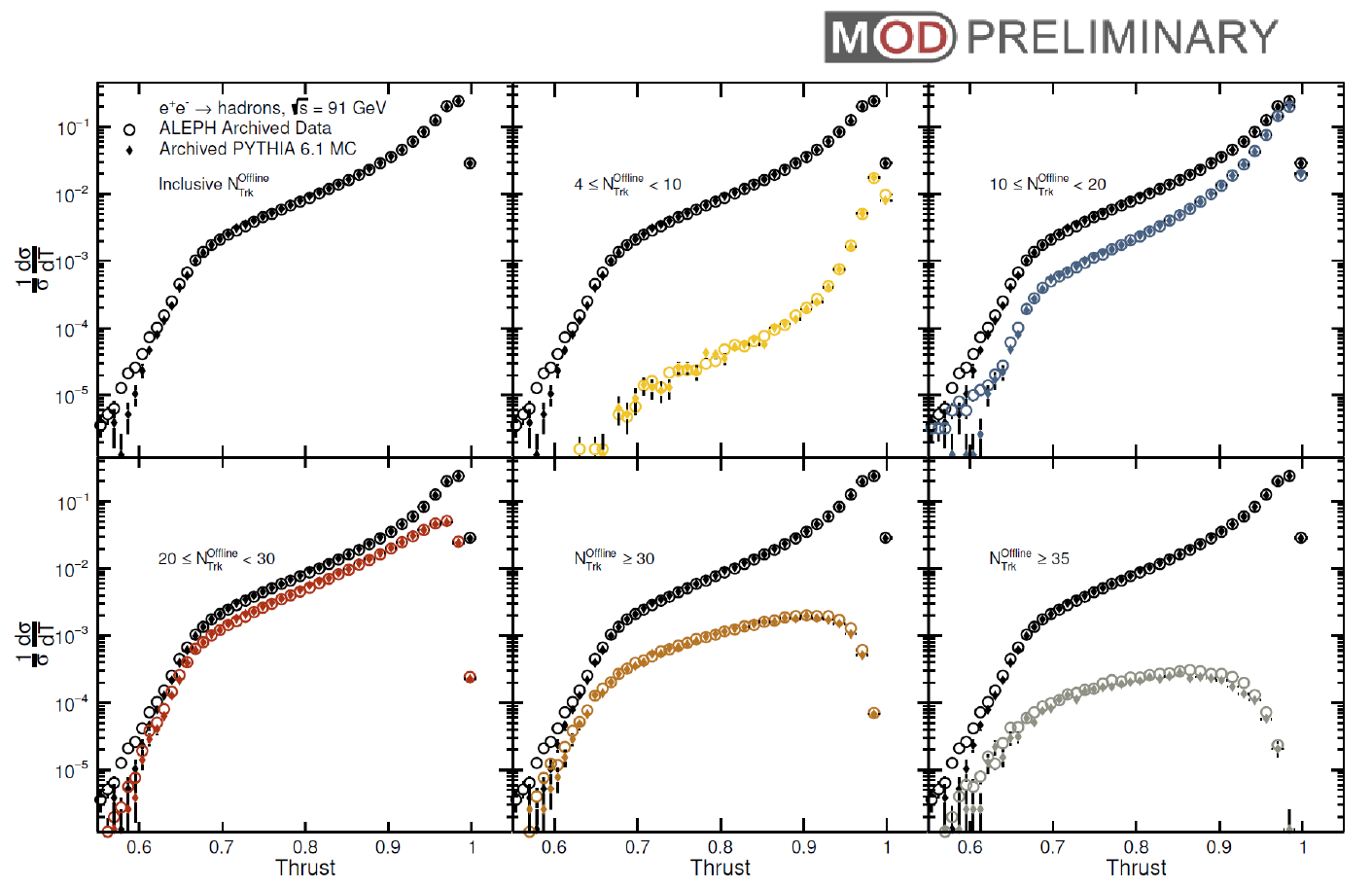}
\caption{Thrust distribution in various multiplicity categories before correcting for detector effects. The inclusive distribution is held constant for comparison across all panels.}
\label{fig:thrustPreUnfolding}
\end{figure}

The distribution is studied in multiplicity bins as well to understand the dependence on the underlying particle yield. The detector level thrust distributions are shown in the upper left corner of Figure~\ref{fig:thrustPreUnfolding}. The rest of the figure shows the overall distribution in comparison to the distribution for events in various multiplicity bins. In general, the higher multiplicity bins have broader thrust distributions, reflecting the more isotropic particle production. 





\subsection{Unfolding}
\label{sec:unfolding}

Detector effects are corrected using the unbinned unfolding algorithm $\textsc{UniFold}$~\cite{Andreassen:2019cjw}, a two-step iterative method based on classifier reweighting. First, a neural network (NN) is trained to distinguish real from simulated events at the detector level. Its output is converted into a likelihood ratio using the likelihood ratio trick~\cite{Neyman:1933wgr, Hastie:2009itz, Miller:2022haf, Sugiyama_Suzuki_Kanamori_2012, Cranmer:2015bka, PhysRevD.103.116013, Rizvi:2023mws}, which is then used to reweight the simulated events. Those weights are propagated to the corresponding generator-level events. Second, another NN is trained to distinguish between the reweighted generator-level distribution and the nominal one. The second NN is used to compute final weights that unfold the generator-level distribution. The $\textsc{PYTHIA}$ 6.1 archived MC is used for the nominal unfolding. 


\begin{figure}[t!]
\centering
\includegraphics[width=0.6\textwidth,angle=0]{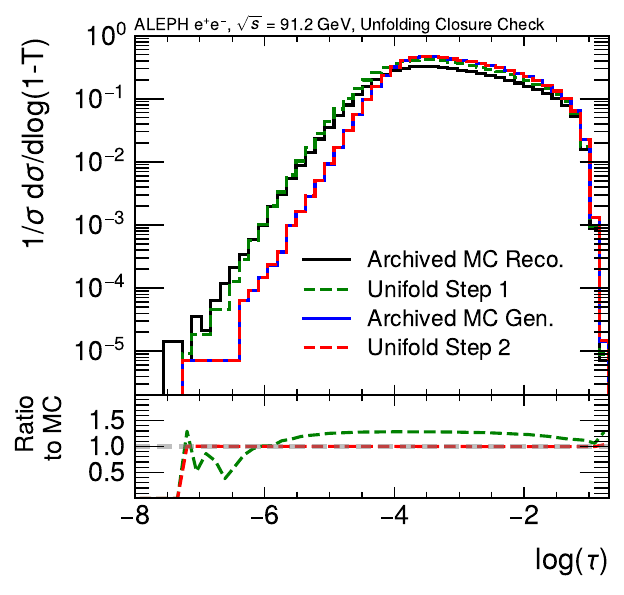}
\caption{Closure check of the unfolding procedure using reconstructed-level MC as pseudo-data. The ratio of the $\textsc{Unifold}$ Step 1 (green) and 2 (red) are shown to the respective reference distributions of the MC reco. (black) and gen. (blue) levels, respectively. Good agreement is seen after Step 1 and full closure after Step 2.}
\label{fig:unfolding_closure}
\end{figure}

As mentioned, $\textsc{UniFold}$ operates with a separate NN for each unfolding step. We use identical configurations for both networks. Hyperparameters are optimized by fixing a subset of parameters and measuring the validation loss after the first unfolding step during the first iteration. The following settings are fixed to the algorithm’s default values based on prior usage: NNs with ReLU activations, training up to 100 epochs with early stopping triggered after 10 epochs if the validation loss plateaued. The width of the networks are scanned from 50 to 200 nodes with a fixed depth of 3 layers, while the depth is scanned from 2 to 4 layers with a fixed width of 100 nodes. Additionally, the batch size is scanned from $2^8$ to $2^{11}$, and the learning rate from $10^{-4}$ to $5\times10^{-3}$. The resulting validation losses are all comparable, around 0.602. Based on these studies, we adopt the following standard configurations: networks with three dense layers with 100 nodes each, a batch size of 2048, and a learning rate of $5\times10^{-4}$. The unfolding is regularized by the number of \textsc{UniFold} iterations, analogous to the number of iterations in the D’Agostini method for IBU~\cite{dagostini2010improvediterativebayesianunfolding}, and the use of early stopping in the training. It has been shown that the number of iterations leads to increasing statistical uncertainty for IBU while decreasing statistical uncertainty from bootstrapping for \textsc{UniFold}~\cite{falcão2025highdimensionalunfoldinglargebackgrounds}. We checked the change in the unfolded thrust distribution with 3, 4, and 5 iterations of \textsc{UniFold}. No major difference is observed, so we use a nominal choice of 5 iterations.
The fraction of the datasets that are used for training is 80\%, with the remaining data used for validation. After training, the NN's are evaluated on the full datasets to evaulate the final result.


A closure check is performed by replacing the data sample with the full reco-level MC sample. This ensures that, in the absence of mismodeling, the unfolding procedure accurately recovers the known truth-level MC distribution. The reconstructed MC is treated as pseudo-data, unfolded using the same configuration as for data, and the result is compared to the corresponding truth-level MC. The result is shown in Figure~\ref{fig:unfolding_closure}. Good agreement is seen for both steps of the $\textsc{Unifold}$ procedure to the expected behavior. Some differences are observed in the step 1 closure but this does not lead to a major change in the step 2 closure.

\begin{figure}[t!]
\centering
\begin{subfigure}[b]{0.48\textwidth}
    \includegraphics[width=\textwidth,angle=0]{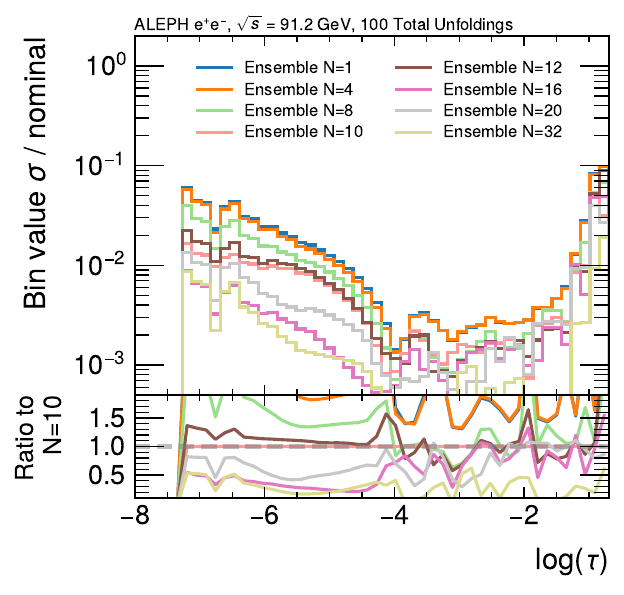}
    \caption{}
    \label{fig:unfolding_ensemble_a}
\end{subfigure}
\begin{subfigure}[b]{0.48\textwidth}
    \includegraphics[width=\textwidth,angle=0]{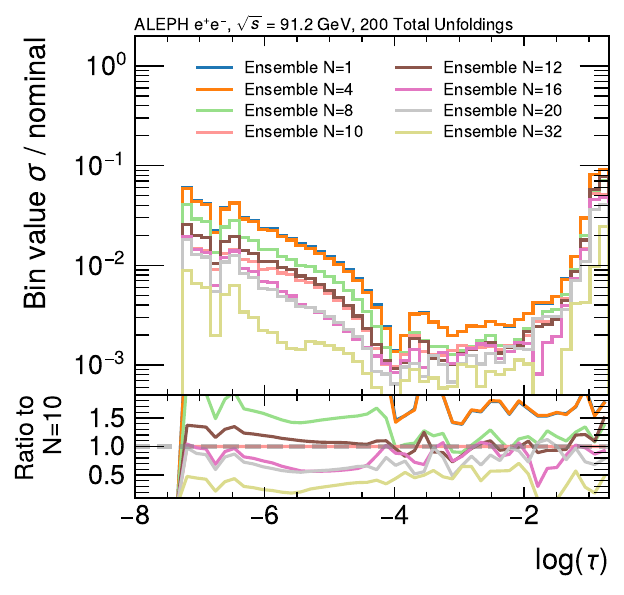}
    \caption{}
    \label{fig:unfolding_ensemble_b}
\end{subfigure}
\caption{Impact of neural network weight initialization on the unfolding. Ensembles of $N$ trainings with different random seeds are used to quantify the variation from random initialization. For all unfoldings throughout the analysis, the median event weight across the ensemble is used as the unfolded result. The total number of trainings is varied from (a) 100 to (b) 200 trainings to test for different number of ensembles. It is found that with 100 total trainings and $N=10$ the variation is subdominant compared to other uncertainties.}
\label{fig:unfolding_ensemble}
\end{figure}

The random initialization of neural network weights introduces variation in the final measurement. To quantify this effect, we perform $N$ independent $\textsc{UniFold}$ trainings with different random seeds and define this set of trainings as a single ensemble. We then generate $M$ such ensembles. For each ensemble, the median event weight across the $N$ trainings is taken as the unfolded result. The uncertainty is evaluated as the standard deviation of the binned distributions across the $M$ ensembles. The results are shown in Figure~\ref{fig:unfolding_ensemble} for different values of $N$. The total number of trainings is $X=100$ for all curves in Figure~\ref{fig:unfolding_ensemble_a} and 200 for those in Figure~\ref{fig:unfolding_ensemble_b}, meaning that $M = X/N$. We found that 100 total trainings and $N = 10$ trainings per ensemble were sufficient to constrain the statistical uncertainty from random initialization to a subdominant level. Consequently, all unfoldings are performed using ensembles of $N = 10$ trainings.

For all systematic variations, the unfolding is performed using the same ensembling procedure with $N = 10$ trainings per variation. This ensures that the impact of random network initialization is consistently accounted for in both the nominal result and all systematic shifts. The median event weight across each ensemble is used for the final unfolded result in every case.

\subsection{Implicit corrections}
\label{sec:analysismethod_corrections}
In previous binned analyses of archived ALEPH data, corrections were applied to account for a hadronic event selection on the reco-level archived Monte Carlo (MC), the inclusion of ISR/FSR in the generator-level (gen-level) MC, and tracking inefficiencies~\cite{Badea:2019vey, Chen:2021uws, Bossi:2025xsi}. In this analysis, those corrections are applied implicitly through the unfolding procedure as detailed below.

The reco-level MC is derived from a subset of the gen-level MC after a hadronic event selection is applied. The first step of the unfolding process utilizes the reco-level MC. A unique matching exists between events at the reco-level and their corresponding gen-level counterparts prior to the hadronic event selection. The weights obtained from step 1 are then propagated back and applied to those corresponding gen-level events. Subsequently, step 2 of the unfolding is performed using the full gen-level MC (with events that passed the hadronic event selection carrying the weights from step 1). This effectively unfolds the distribution to the gen-level phase space before selections.

The archived MC samples were generated with ISR/FSR, primarily for radiative return studies. However, the theoretical predictions we aim to compare with do not include ISR and optionally include FSR. Therefore, a cleaning procedure was applied to remove these contributions. This involved tracing the MC particle history to identify and remove conversion electrons originating from radiated photons. We apply this cleaning to both the gen-level and reco-level archived MC before the unfolding procedure. The new MC samples used for theoretical variations are produced without ISR, ensuring that the phase space is appropriately reweighted to account for this difference. The final unfolding uses the MC samples with ISR/FSR cleaning applied, which implicitly corrects the result for those effects.

The reconstruction efficiency of single tracks, influenced by non-uniform detector efficiency and mis-reconstruction biases, can systematically shift the measurement. Typically, a tracking efficiency correction is applied to mitigate this effect:

\begin{align}
    \varepsilon(p_{\mathrm{T}}, \theta, \phi, N_{\mathrm{Trk}}^{\mathrm{Offline}}) = \left[ \frac{d^3 N^{\mathrm{reco}}}{dp_{\mathrm{T}} d\theta d\phi} \middle/ \frac{d^3 N^{\mathrm{gen}}}{dp_{\mathrm{T}} d\theta d\phi} \right]_{N_{\mathrm{Trk}}^{\mathrm{Offline}}},
\end{align}

\noindent where $N^{\mathrm{reco}}$ represents the number of charged particles at the reconstruction level, and $N^{\mathrm{gen}}$ is the corresponding number at the gen-level. This efficiency correction establishes a correspondence between the reco and gen-level across the $p_{\mathrm{T}}$, $\theta$, and $\phi$ spectra. In this analysis, this correction is implicitly applied by the unfolding procedure through the iterative pulling and pushing of weights between steps 1 and 2.
\section{Systematic uncertainties}
\label{sec:systematics}

Uncertainties arise from a combination of statistical, experimental, unfolding, and theoretical components. Those components are outlined below and added in quadrature.






\subsection{Dataset statistical}


The uncertainty from limited dataset size is addressed in two ways. First, Poisson uncertainty on the final unfolding weights is estimated using the standard $\sqrt{\sum w^2}$ formula and added in quadrature with other uncertainties. Second, bootstrapping is used: the unfolding is repeated with data and MC event weights resampled using independent $\textsc{Poisson}(1)$ random factors, simulating statistical fluctuations. This is done separately for data and MC across multiple ensembles. We found that 4 ensembles for both data and MC was sufficient to yield a subdominant bootstrap uncertainty.

\subsection{Unfolding ensembling}

The uncertainty arising from statistical variation due to random neural network initialization is estimated by computing the spread across M = 10 ensembles of N = 10 trainings as described in Section~\ref{sec:unfolding}. The uncertainty is quoted as the standard deviation of the binned distributions obtained from each ensemble’s median prediction.

\begin{figure}[t!]
\centering
\begin{subfigure}[b]{0.32\textwidth}
    \includegraphics[width=\textwidth,angle=0]{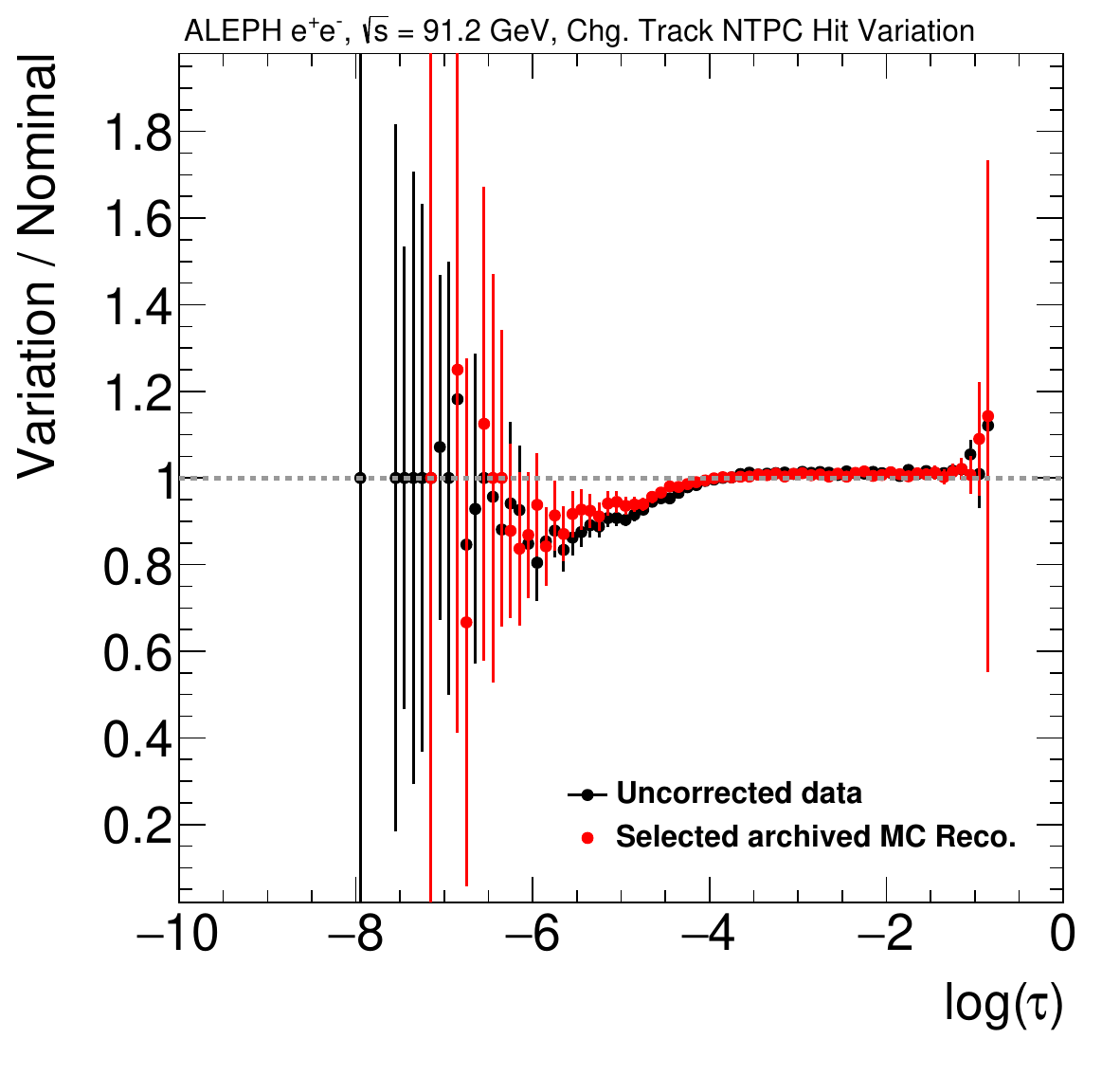}
    \caption{N$_{\rm TPC} \ge 4 \rightarrow 7$}
    \label{fig:ExpVariations_mainbody_a}
\end{subfigure}
\begin{subfigure}[b]{0.32\textwidth}
    \includegraphics[width=\textwidth,angle=0]{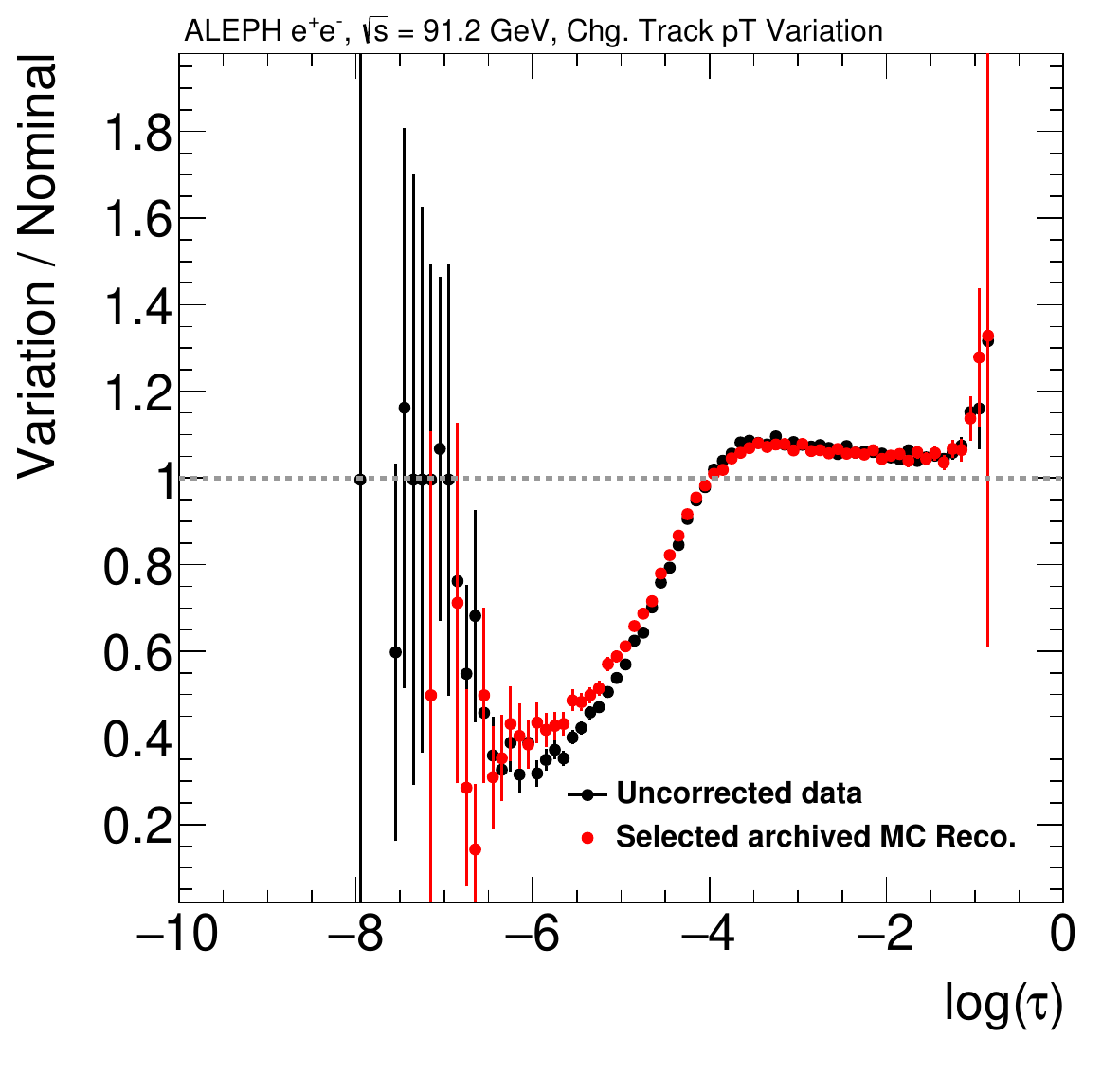}
    \caption{Track $p_{\rm T} \ge 0.2 \rightarrow 0.4$ GeV}
    \label{fig:ExpVariations_mainbody_b}
\end{subfigure}
\begin{subfigure}[b]{0.32\textwidth}
    \includegraphics[width=\textwidth,angle=0]{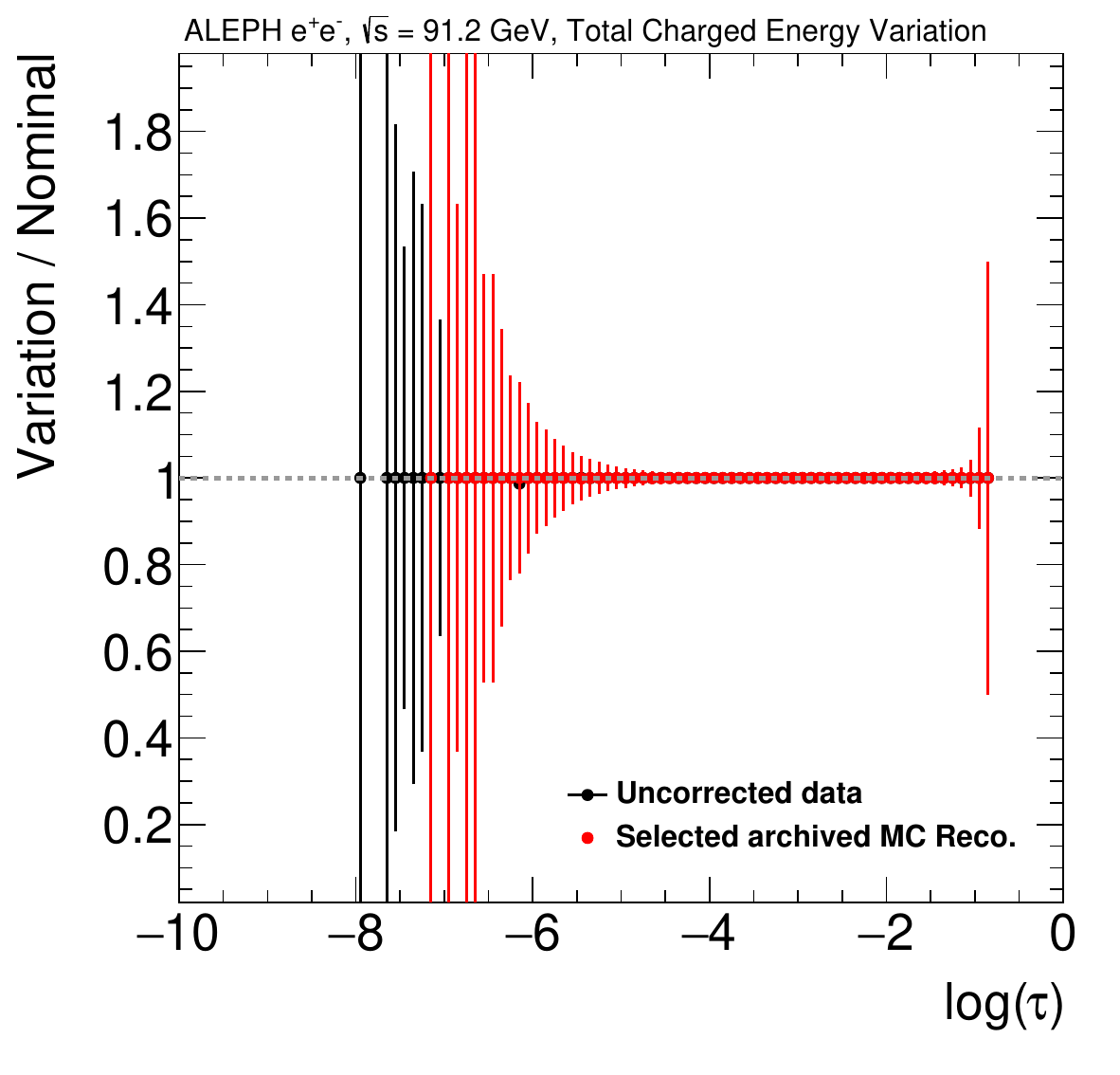}
    \caption{E$_{\rm ch} \ge 15 \rightarrow 10$ GeV}
    \label{fig:ExpVariations_mainbody_c}
\end{subfigure}
\begin{subfigure}[b]{0.32\textwidth}
    \includegraphics[width=\textwidth,angle=0]{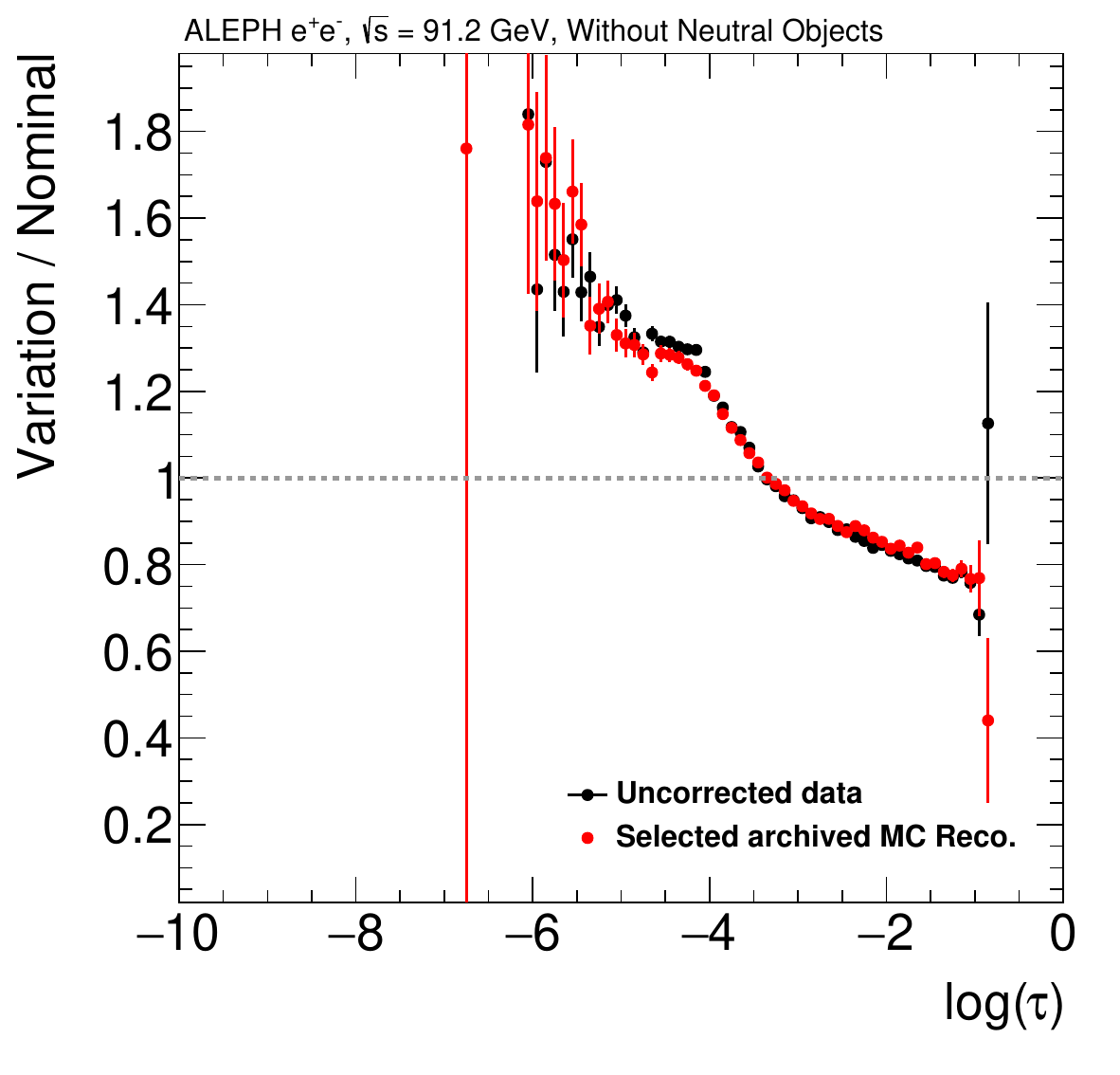}
    \caption{Thrust w/o neutral objects}
    \label{fig:ExpVariations_mainbody_d}
\end{subfigure}
\begin{subfigure}[b]{0.32\textwidth}
    \includegraphics[width=\textwidth,angle=0]{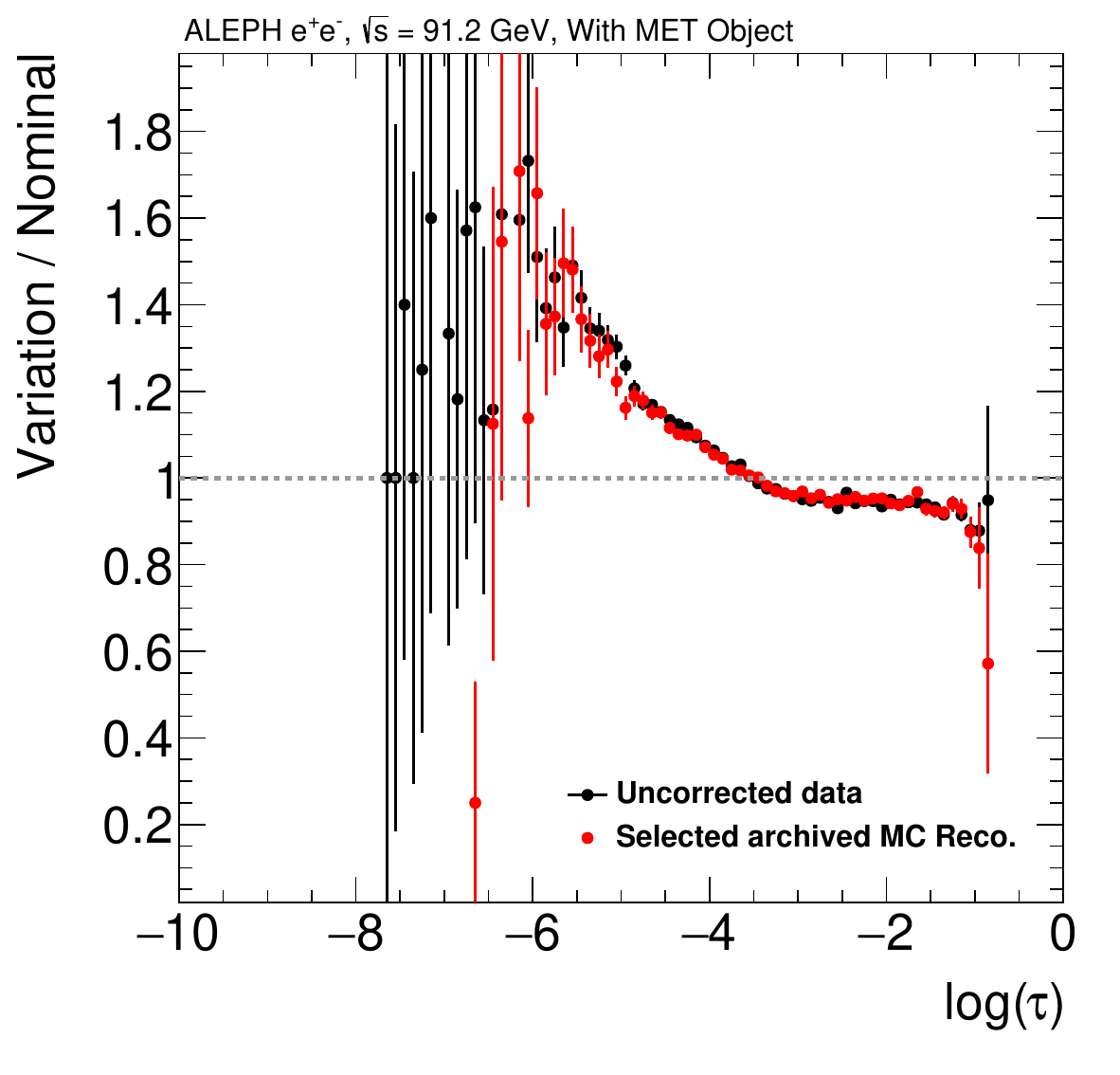}
    \caption{Thrust w/ $\vec{p}_{\mathrm{MET}}$}
    \label{fig:ExpVariations_mainbody_e}
\end{subfigure}
\caption{
Impact of variations in track and event selections on the reconstructed $\log\tau$ distribution~\cite{Barate:1996fi, ALEPH:2003obs}, shown as the ratio to the nominal distribution. Each panel corresponds to a different experimental variation: (a) increased minimum number of TPC hits, (b) increased minimum track $p_{\mathrm{T}}$, (c) reduced total charged energy threshold, (d) exclusion of neutral objects in thrust calculation, and (e) inclusion of the $\vec{p}_{\mathrm{MET}}$.}
\label{fig:ExpVariations_mainbody}
\end{figure}

\subsection{Experimental variations}

Experimental uncertainties are evaluated by varying the track and event selections following the procedure in~\cite{Barate:1996fi, ALEPH:2003obs}. The thrust is recalculated for any variation that modifies the particle level selections. The unfolding procedure is repeated for each variation and the resulting change is taken as a systematic uncertainty. The impact of each variation before unfolding is shown in Figure~\ref{fig:ExpVariations_mainbody} as the ratio of the variation to the nominal $\log\tau$ distribution. For charged tracks, the required number of hits in the TPC is varied from 4 to 7 and the minimum \pT~from 0.2 to 0.4 GeV. With the nominal particle selections, the total required charged energy is varied from 15 to 10 GeV. To account for potential mismodeling of neutral objects, the thrust is recalculated without neutral objects. To account for the impact of the missing momentum, the thrust is recalculated including the $\vec{p}_{\mathrm{MET}}$ vector.



\begin{figure}[t!]
\centering
\begin{subfigure}[b]{0.45\textwidth}
    \includegraphics[width=\textwidth,angle=0]{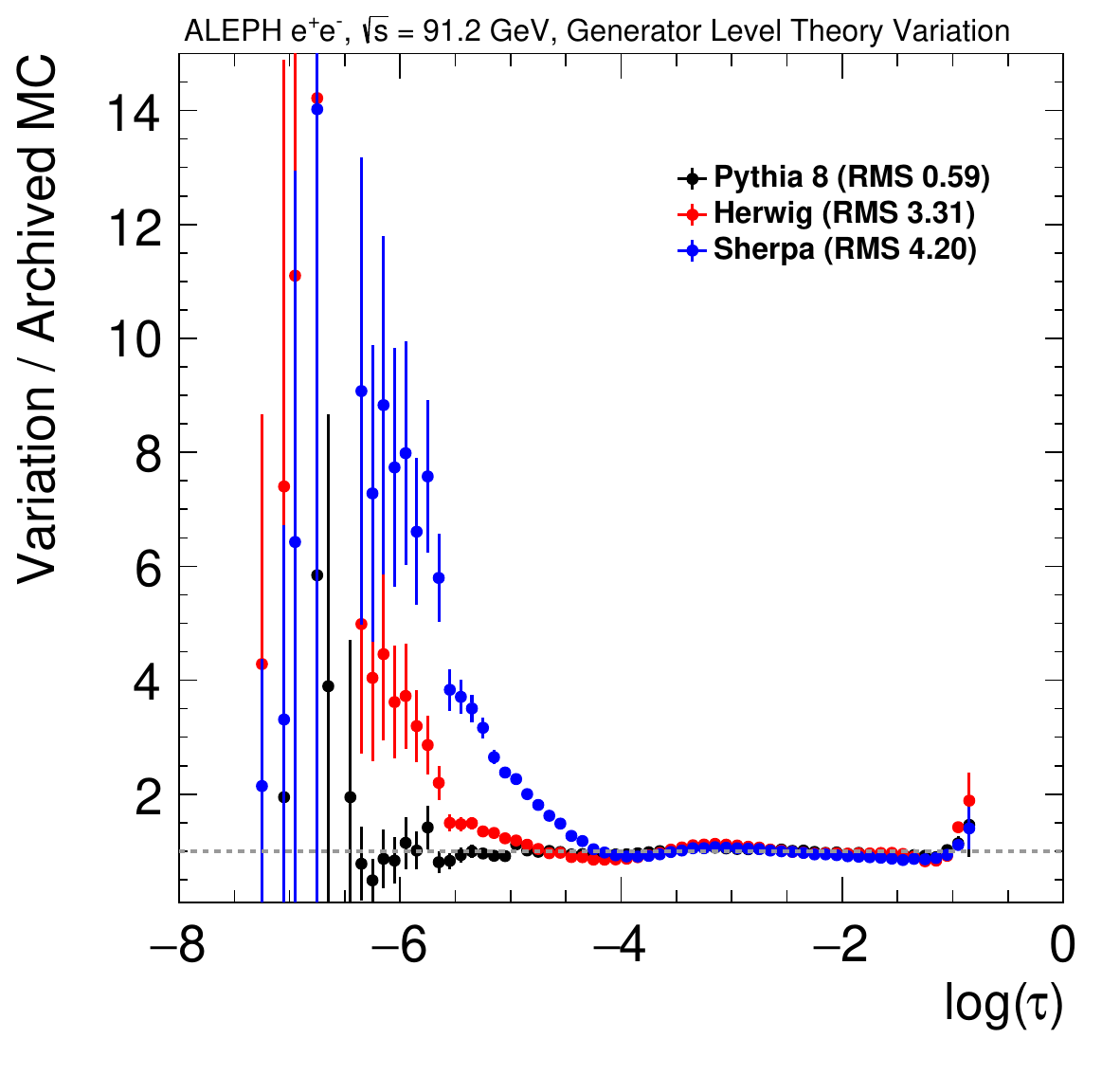}
    \caption{}
    \label{fig:logtau_check_gen_theory_variation_a}
\end{subfigure}
\begin{subfigure}[b]{0.45\textwidth}
    \includegraphics[width=\textwidth,angle=0]{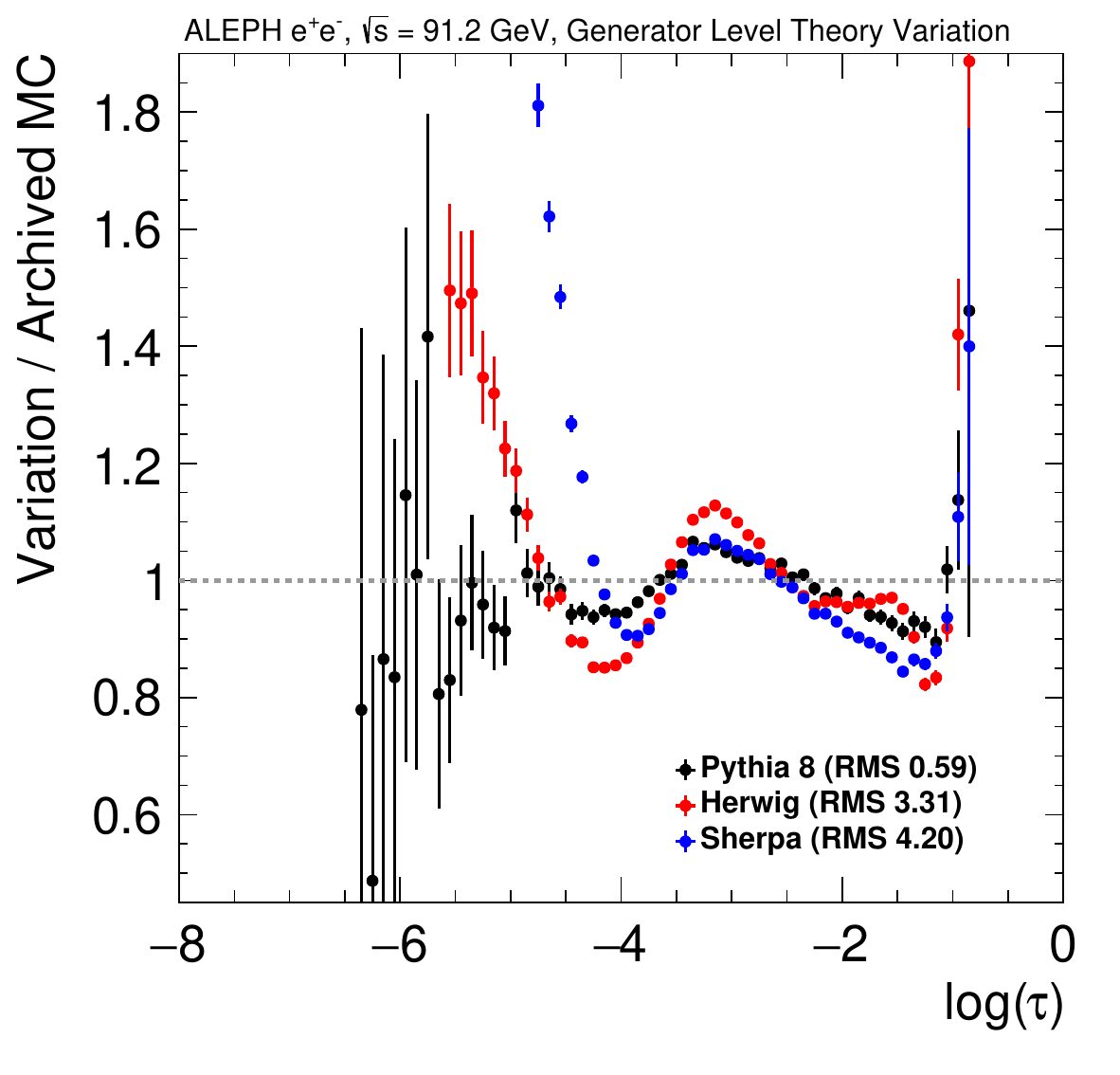}
    \caption{}
    \label{fig:logtau_check_gen_theory_variation_b}
\end{subfigure}
\caption{Generator level comparisons of the $\log(\tau)$ distributions from different MC simulations to the Archived \textsc{Pythia} 6.1. The ratios of $\textsc{Pythia}$ 8 (black), $\textsc{Herwig}$ (red), and $\textsc{Sherpa}$ (blue) relative to the archived MC are shown. The figures are shown zoomed out (left) and zoomed in (right) on the y-axis to highlight the structure in the tail and the central region of the distribution. The root mean squared (RMS) difference from unity is indicated in the legend as a measure of the overall agreement between the samples.}
\label{fig:logtau_check_gen_theory_variation}
\end{figure}

\subsection{Theoretical modeling}

The theoretical uncertainty is estimated by varying the prior used in the unfolding procedure. Ideally, the uncertainty would be assessed by varying the reconstruction and generator level MC sample used in the unfolding procedure. However, since the GALEPH detector simulation is not currently available, an alternative approach is developed by reweighting the archived $\textsc{Pythia}$ 6.1 MC. For this, an alternative MC is generated using the $\textsc{Pythia}$ 8, $\textsc{Herwig}$, and $\textsc{Sherpa}$ programs. Neither ISR or FSR are included in the simulations. The $\log(\tau)$ distribution is checked in Figure~\ref{fig:logtau_check_gen_theory_variation} for each MC in comparison to the archived MC. The root mean squared (RMS) difference with respect to a ratio of 1 is also shown in the legend as a measure of the overall level of agreement.

\begin{figure}[t!]
\centering
\begin{subfigure}[b]{0.30\textwidth}
    \includegraphics[width=\textwidth,angle=0]{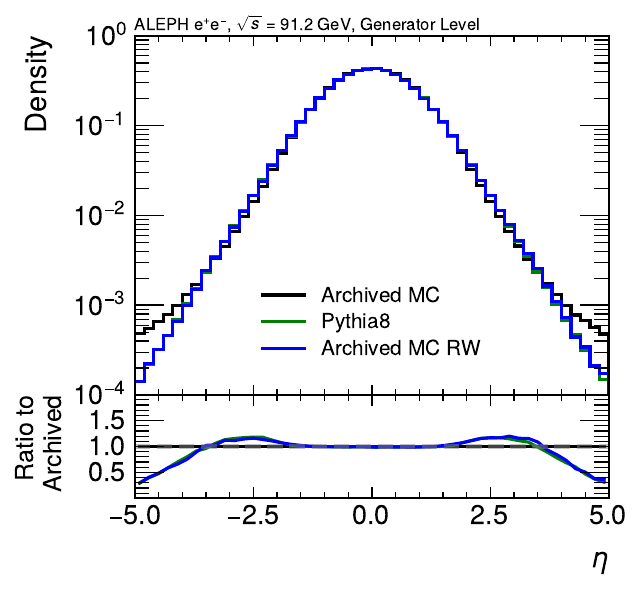}
    \caption{}
    \label{fig:theoryreweighting_pythia8_a}
\end{subfigure}
\begin{subfigure}[b]{0.30\textwidth}
    \includegraphics[width=\textwidth,angle=0]{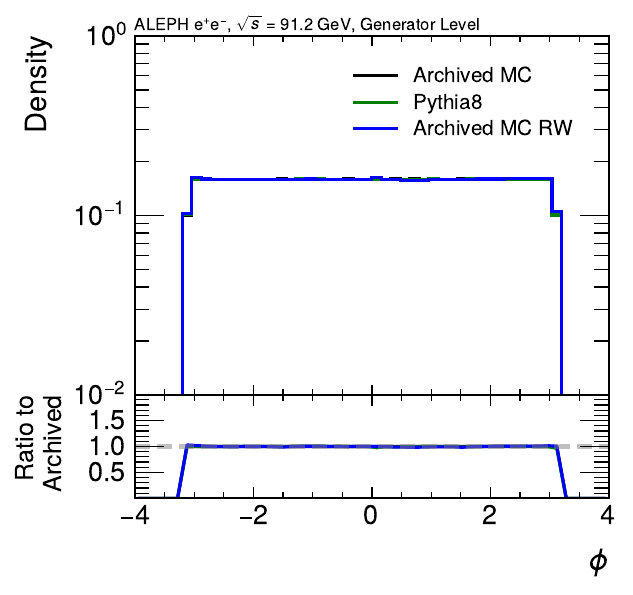}
    \caption{}
    \label{fig:theoryreweighting_pythia8_b}
\end{subfigure}
\begin{subfigure}[b]{0.30\textwidth}
    \includegraphics[width=\textwidth,angle=0]{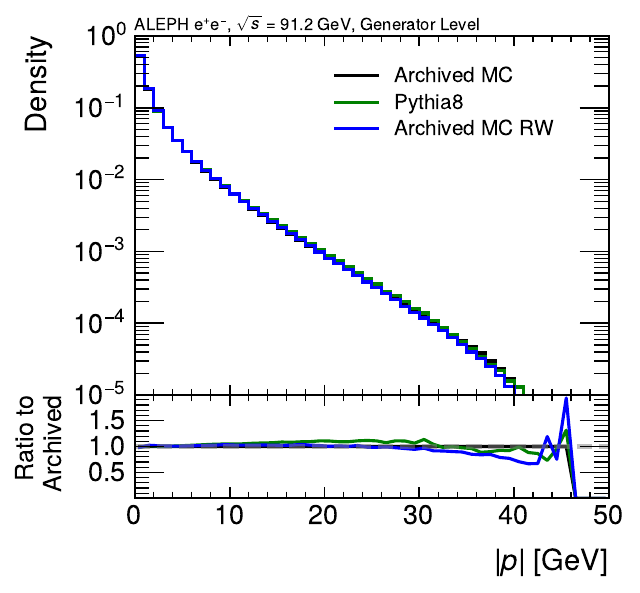}
    \caption{}
    \label{fig:theoryreweighting_pythia8_d}
\end{subfigure}
\begin{subfigure}[b]{0.30\textwidth}
    \includegraphics[width=\textwidth,angle=0]{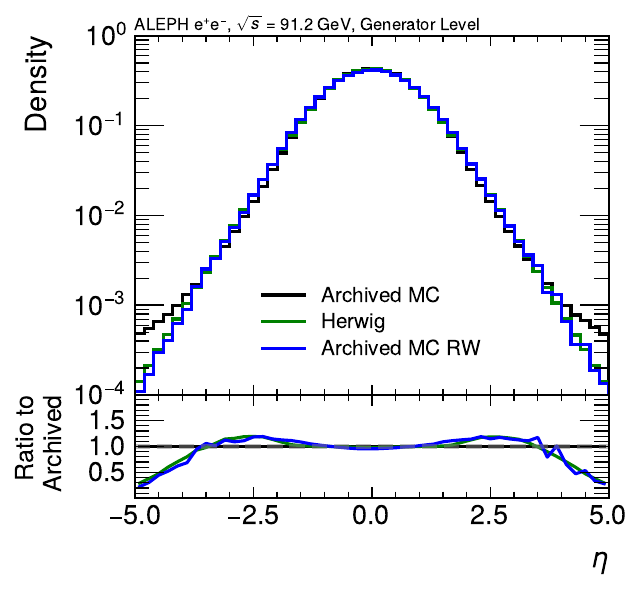}
    \caption{}
    \label{fig:theoryreweighting_herwig_a}
\end{subfigure}
\begin{subfigure}[b]{0.30\textwidth}
    \includegraphics[width=\textwidth,angle=0]{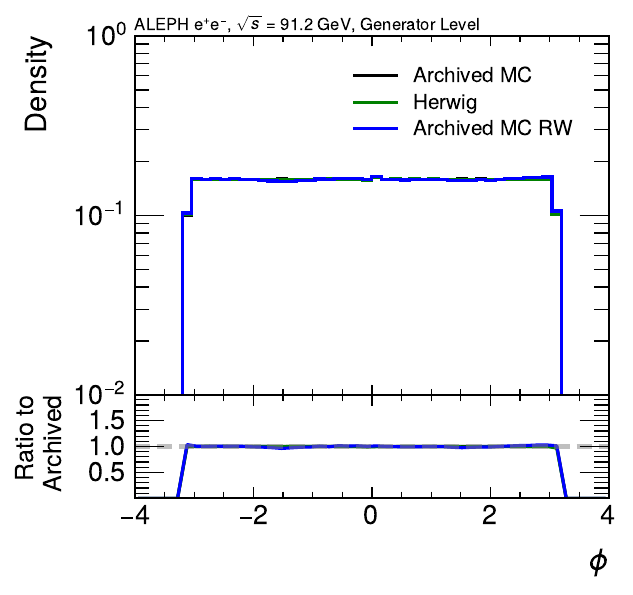}
    \caption{}
    \label{fig:theoryreweighting_herwig_b}
\end{subfigure}
\begin{subfigure}[b]{0.30\textwidth}
    \includegraphics[width=\textwidth,angle=0]{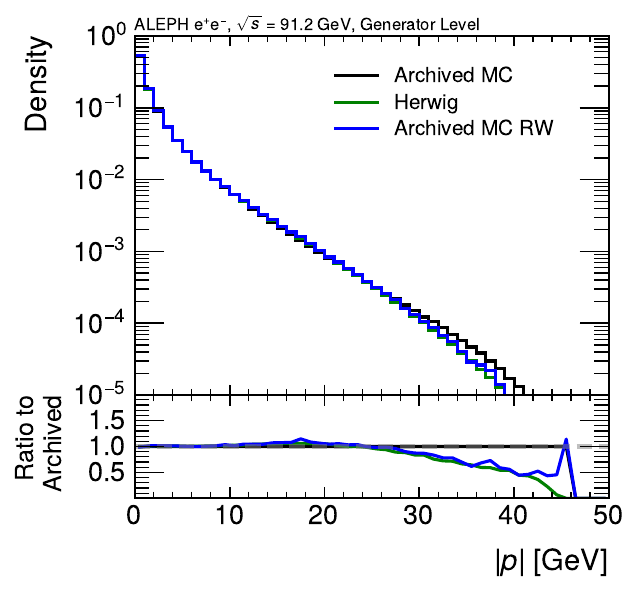}
    \caption{}
    \label{fig:theoryreweighting_herwig_c}
\end{subfigure}
\begin{subfigure}[b]{0.30\textwidth}
    \includegraphics[width=\textwidth,angle=0]{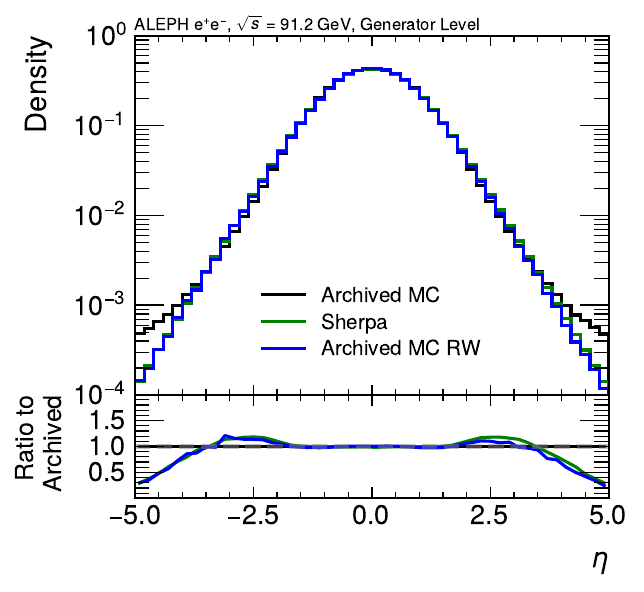}
    \caption{}
    \label{fig:theoryreweighting_sherpa_a}
\end{subfigure}
\begin{subfigure}[b]{0.30\textwidth}
    \includegraphics[width=\textwidth,angle=0]{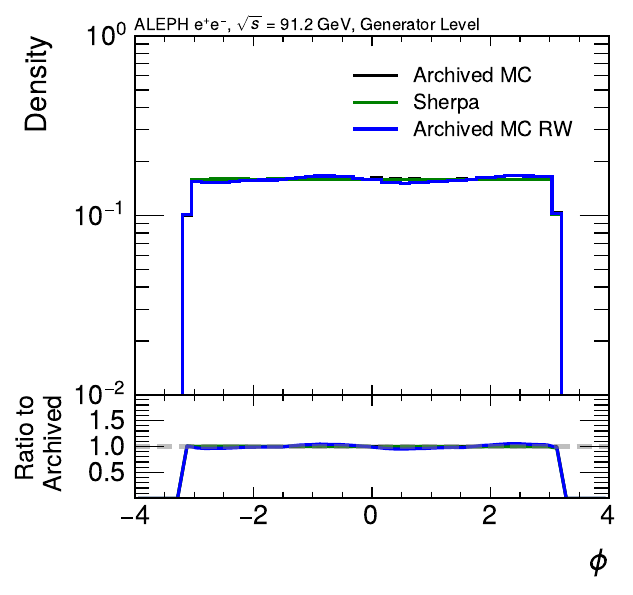}
    \caption{}
    \label{fig:theoryreweighting_sherpa_b}
\end{subfigure}
\begin{subfigure}[b]{0.30\textwidth}
    \includegraphics[width=\textwidth,angle=0]{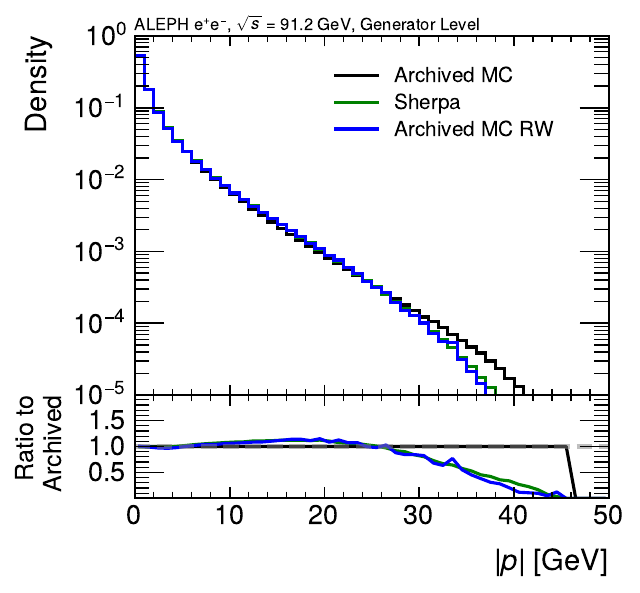}
    \caption{}
    \label{fig:theoryreweighting_sherpa_c}
\end{subfigure}
\caption{Kinematic distributions of $\log|p|$, $\eta$, and $\phi$ used as inputs to the reweighting procedure, comparing the archived $\textsc{Pythia}$ 6.1 MC, the alternative MC, and the reweighted archived MC after each step of the procedure. The reweighting successfully brings the archived MC into agreement with the alternative MC across the input observables. The application of the reweighting to the $\log(\tau)$ distribution is shown in (d).}
\label{fig:reweight_kinematics}
\end{figure}


\begin{figure}[t!]
\centering
\begin{subfigure}[b]{0.45\textwidth}
    \includegraphics[width=\textwidth,angle=0]{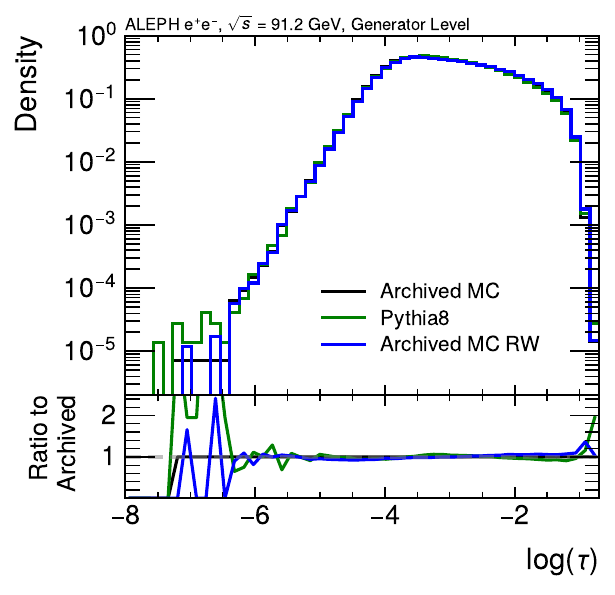}
    \caption{}
    \label{fig:reweight_logtau_pythia8}
\end{subfigure}
\begin{subfigure}[b]{0.45\textwidth}
    \includegraphics[width=\textwidth,angle=0]{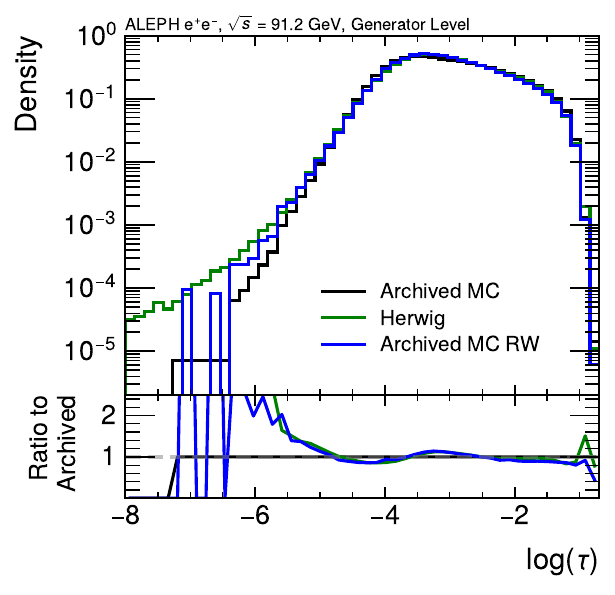}
    \caption{}
    \label{fig:reweight_logtau_herwig}
\end{subfigure}
\begin{subfigure}[b]{0.45\textwidth}
    \includegraphics[width=\textwidth,angle=0]{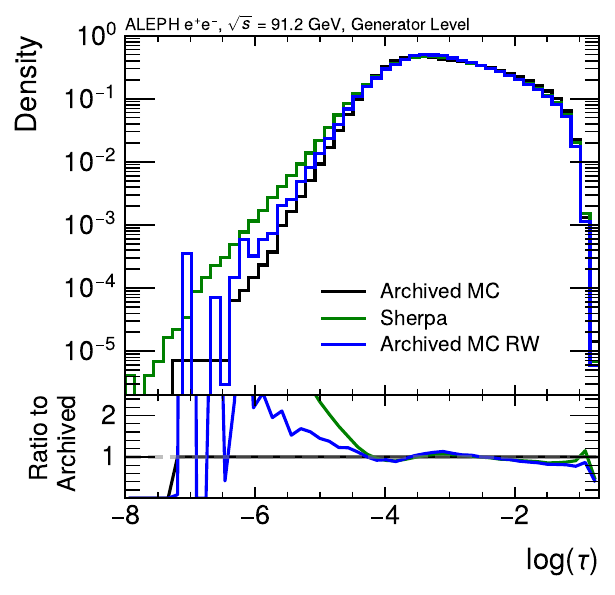}
    \caption{}
    \label{fig:reweight_logtau_sherpa}
\end{subfigure}
\caption{Kinematic distributions of $\log|p|$, $\eta$, and $\phi$ used as inputs to the reweighting procedure, comparing the archived $\textsc{Pythia}$ 6.1 MC, the alternative MC, and the reweighted archived MC after each step of the procedure. The reweighting successfully brings the archived MC into agreement with the alternative MC across the input observables. The application of the reweighting to the $\log(\tau)$ distribution is shown in (d).}
\label{fig:reweight_logtau}
\end{figure}

The reweighting procedure is performed at particle level for each alternative MC using a neural network\footnote{A particle edge transformer architecture is used.} trained to distinguish events from the archived and alternative MC samples. For each event, the $(\log|p|, \eta, \phi)$ of all particles are provided as input to the network. The training is done in two stages. First, a pretraining phase where the network is trained to reweight the archived MC to itself, effectively initializing the model close to an identity mapping within the relevant phase space. This step improves the efficiency and stability of the subsequent training. In the second step, the network is trained to reweight the archived MC to match the alternative MC. The network is trained for up to 200 epochs with early stopping based on validation loss. The loss curves are shown in the appendix Figure~\ref{fig:theoryReweighting_lossCurves}. The classifier’s output is used via the likelihood ratio trick to obtain event weights that reweight the archived MC into the alternative MC. Figure~\ref{fig:reweight_kinematics} shows the results of the reweighting on the variables used as inputs to the networks. Overall good agreement is found between the reweighted archived and the alternative MC. The event weights are then applied to the $\log(\tau)$ distribution in Figure~\ref{fig:reweight_logtau}. The reweighting achieves the best result for $\textsc{Pythia}$ 8, followed by the $\textsc{Herwig}$ and $\textsc{Sherpa}$. This is consistent with the expectation that the underlying physics modeling in $\textsc{Pythia}$ 8 is the closest to that in the archived $\textsc{Pythia}$ 6. For all three alternative MC, some discrepancies are seen after the reweighting in the tails of the distributions. It is left for future work to improve the reweighting procedure further. The reweighted response matrices are shown in the appendix Figure~\ref{fig:theoryUncert_responseMatrix}, but not are used directly in the result as this is an unbinned measurement.

\begin{figure}[t!]
\centering
\includegraphics[width=\textwidth,angle=0]{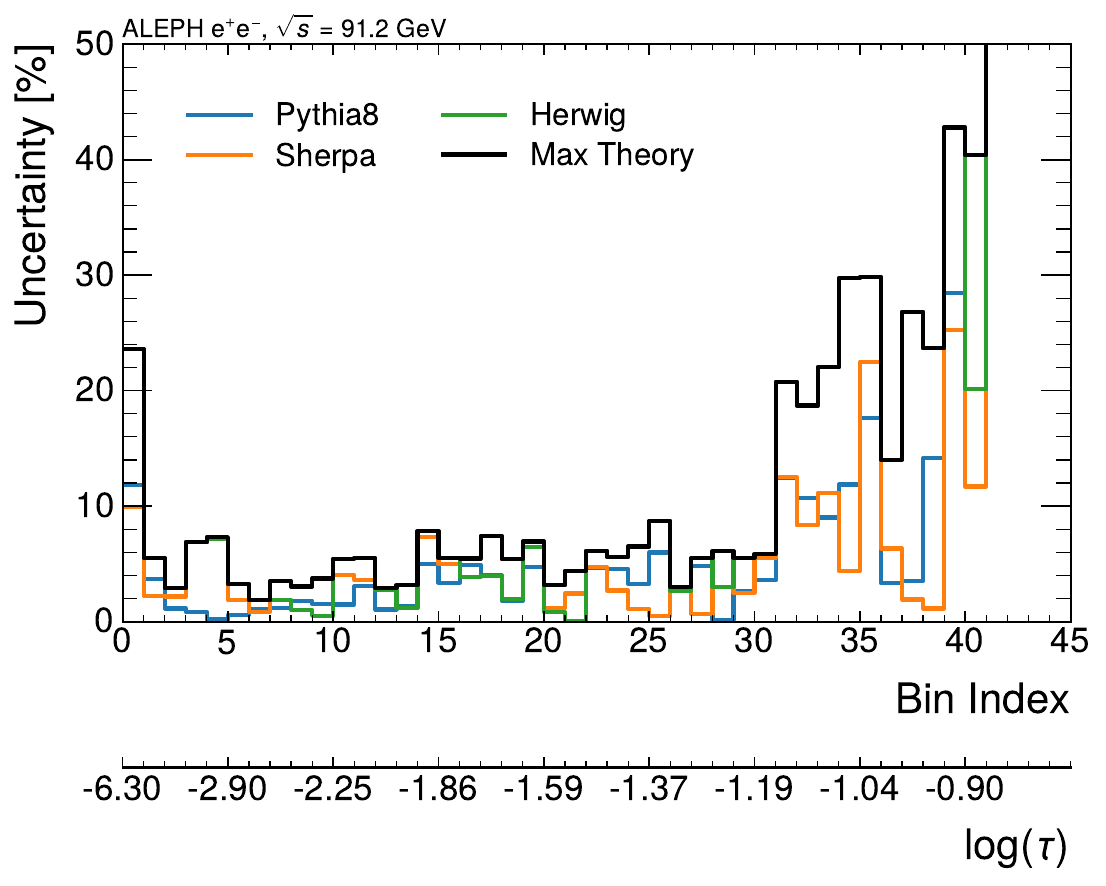}
\caption{Percentage uncertainty from theoretical variations, defined as the absolute bin-by-bin difference between the unfolded and nominal distributions divided by the nominal bin content. The unfolding and ensembling procedure is repeated using reweightings from three alternative MC generators. The largest deviations generally arise from the $\textsc{Herwig}$ and $\textsc{Sherpa}$ variations. A notable feature near $\log\tau \sim 3$ is a dip in the $\textsc{Pythia}$ 8 variation, absent in the others, which corresponds to the features seen in the generator-level behavior. The final theoretical uncertainty is conservatively defined as the maximum bin-by-bin deviation across all variations.}
\label{fig:theoryUncertComparison}
\end{figure}

The reweighting is then applied for all three alternative MC and the full unfolding plus ensembling procedure is run. For each variation, the uncertainty is defined as the difference between the unfolded and nominal binned distributions. The percentage uncertainty, the absolute value of the uncertainty divided by the bin content times 100, is shown Figure~\ref{fig:theoryUncertComparison}. The uncertainty is plotted versus bin index to improve visualization, as the bin widths in $\log\tau$ are much larger at low values than at high values, making the high-value region difficult to see when plotted on a physical axis. A second x-axis is shown that gives the corresponding $\log\tau$ values for the tick axes of the primary x-axis. The unfolded result using the $\textsc{Herwig}$ and $\textsc{Sherpa}$ reweightings in general show the largest difference with respect to the nominal result. An interesting feature at $\log\tau\sim$3 is that the $\textsc{Pythia}$ 8 reweighting has a dip in the error, while the other alternative reweightings do not. This feature aligns with the shape seen in Figure~\ref{fig:logtau_check_gen_theory_variation}, though a detailed understanding of the physics modeling that creates the feature is left for future work. The final theoretical uncertainty is conservatively estimated as the maximum bin-by-bin deviation across all variations and is recomputed for each choice of binning.

\subsection{Uncertainty breakdown}

\begin{figure}[t!]
\centering
\includegraphics[width=\textwidth,angle=0]{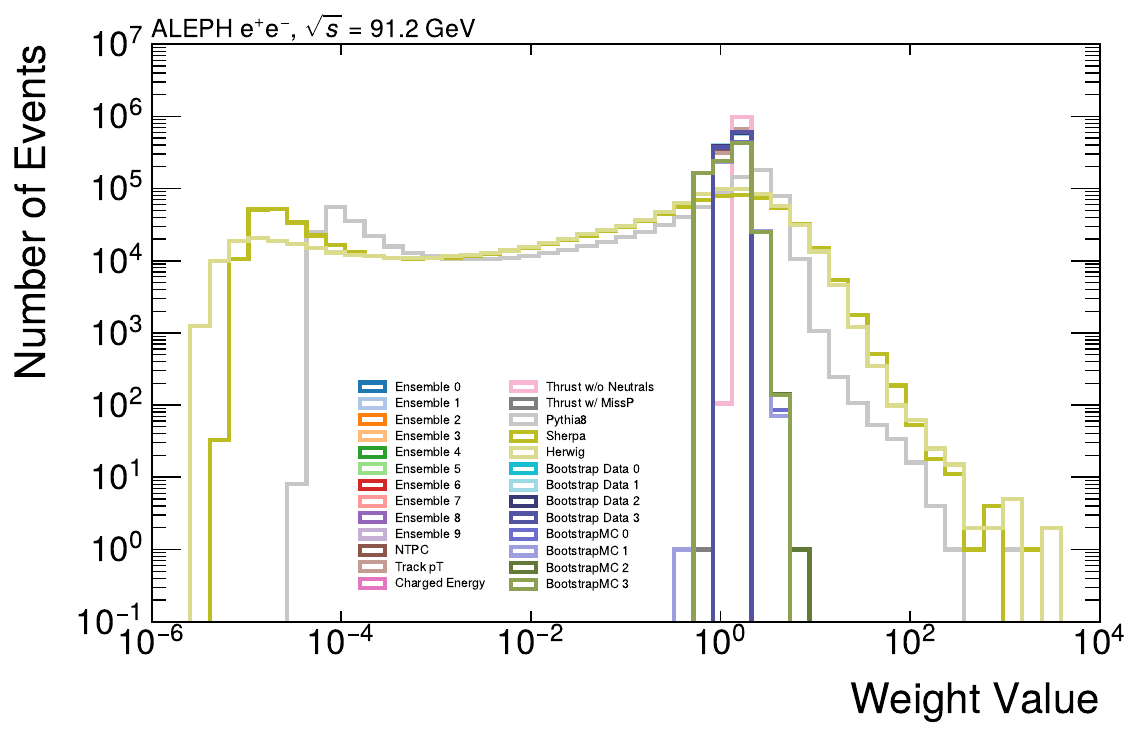}
\caption{Weights from the $\textsc{Unifold}$ unfolding procedure applied to the thrust distribution. The unfolded weights for the ensembling, systematic variation, bootstrapping, and theoretical variations. The unfolded weights are applied to the generator level MC to construct the final unfolded measurement of the particle-level thrust distribution.}
\label{fig:unfoldedWeights}
\end{figure}

A summary of the uncertainties is shown in Table~\ref{tab:uncertainty_summary} and the final weights from the unfolding procedures are shown in Figure~\ref{fig:unfoldedWeights}. These weights represent the correction factors applied to the generator-level MC events to transform them into the unfolded particle-level distribution. The weights exhibit the expected behavior, with values close to unity in most regions and larger corrections in the tail regions where detector effects are more pronounced. The theoretical variations show a broader spread in the weights, arising from the reweighting used to vary the theoretical modeling. The ensemble of weights from multiple random initializations shows good consistency, confirming the stability of the unfolding procedure. Those weights are applied to the generator level MC to yield the unfolded results and systematic variations. 

\begin{table}[h!]
\centering
\begin{tabular}{|l|l|p{0.56\linewidth}|}
\hline
\textbf{Category} & \textbf{Source} & \textbf{Description} \\
\hline
\multirow{4}{*}{Statistical} 
  & Dataset statistics & Poisson uncertainty on unfolded weights. \\ 
  & Bootstrapping & Sample data and MC using $\textsc{Poisson}(1)$ weights. \\
  & Ensembling & Account for variation from NN initializations. \\
\hline
\multirow{5}{*}{Experimental} 
  & Track TPC hits & Vary the required track TPC hits from $\geq$ 4 to $\geq$ 7. \\
  & Track $p_{\mathrm{T}}$ & Vary track $p_{\mathrm{T}}$ $\geq$ 0.2 to $\geq$ 0.4 GeV. \\
  & Charged energy & Vary $E_{\mathrm{ch}}$ $\geq$ 15 to $\geq$ 10 GeV. \\
  & Neutral particles & Remove neutral particles from thrust calculation. \\
  & MET vector & Include $\vec{p}_{\mathrm{MET}}$ in thrust calculation. \\
\hline
Theoretical 
  & Unfolding prior & Archived MC is reweighted to match $\textsc{Pythia}$ 8, $\textsc{Herwig}$, and $\textsc{Sherpa}$. The final uncertainty is defined as the maximum bin-by-bin deviation from the nominal unfolded result. \\
\hline
\end{tabular}
\caption{Summary of systematic uncertainties for the $\log(\tau)$ measurement.}
\label{tab:uncertainty_summary}
\end{table}

\begin{figure}[th!]
\centering
\includegraphics[width=\textwidth,angle=0]{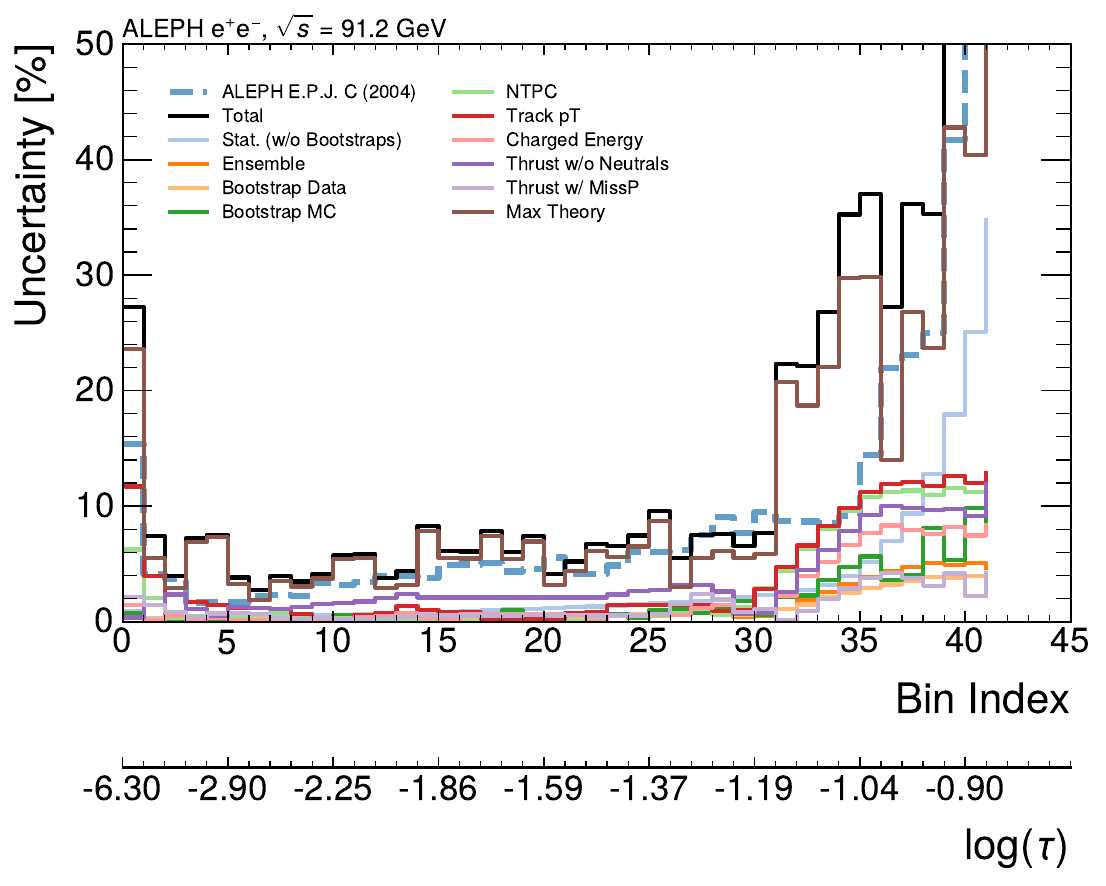}
\caption{Breakdown of the total uncertainty for the fully-corrected $\log\tau$ distribution, shown with binning matching the original ALEPH publication. Statistical components include the sum of weights squared (light blue), ensembling (dark orange), data bootstrapping (light orange), and MC bootstrapping (green). Detector systematics include variations in the number of TPC hits (light green), minimum track $p_{\mathrm{T}}$ (dark red), total charged energy (light pink), and thrust calculation (dark and light purple). The theoretical uncertainty from MC generator reweighting is shown in brown. The total uncertainty is shown in black and is dominated by the theoretical component. ALEPH’s published uncertainty is shown as a dashed blue line. Bin index is used on the x-axis for uniform spacing, with a secondary axis showing the corresponding $\log\tau$ values.}
\label{fig:uncertBreakdown}
\end{figure}

A breakdown of contributions to the overall $\log\tau$ uncertainty is shown in Figure~\ref{fig:uncertBreakdown}. While the unfolded result is unbinned, a binning identical to that used in the original ALEPH publication is applied here for visualization and comparison purposes. The total uncertainty from the ALEPH E.P.J. C (2004) measurement is shown as a dashed blue line~\cite{ALEPH:2003obs}. Statistical uncertainties are displayed as follows: light blue for the statistical Poisson error from the unfolded weights, dark orange for ensembling to capture the spread due to different neural network initializations, light orange for bootstrapping the data, and green for bootstrapping the MC. Detector-related systematics include variations in the number of TPC hits, shown in light green, and the minimum track $p_{\mathrm{T}}$, shown in dark red. The uncertainty due to the total charged energy variation is shown in light pink. Variations in the thrust calculation, excluding neutral particles and including the missing momentum vector, are shown in dark and light purple, respectively. The overall theoretical uncertainty, derived from generator reweighting, is shown in brown. As in previous figures, the x-axis represents the bin index and a secondary axis shows the $\log\tau$ values. Across most of the distribution, the total uncertainty remains below 10\%, increasing in the tails of the distribution. The total measured uncertainty, shown in black, is primarily driven by the theoretical component. Overall, the result is in good agreement with the ALEPH E.P.J. C (2004) measurement, with the largest differences attributable to theoretical uncertainty.

\section{Results}
\label{sec:results}

\begin{figure}[t!]
\centering
\includegraphics[width=\textwidth,angle=0]{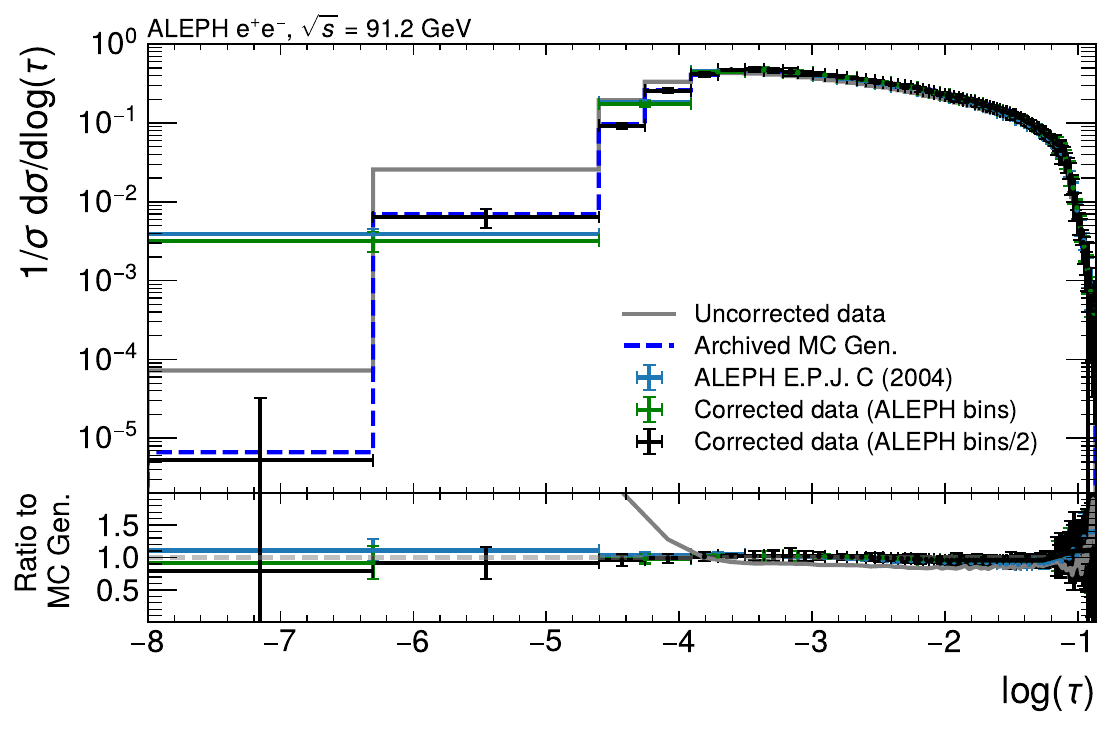}
\caption{Unfolded thrust distribution, obtained from an unbinned measurement and shown with two binnings: the nominal ALEPH binning (green) for direct comparison to the published result (light blue), and a factor two finer binning (black) to illustrate the flexibility of rebinning. The archived generator-level $\textsc{Pythia}$ 6.1 prediction is shown in dashed blue, and the uncorrected data in gray. The lower panel shows the ratio to the generator-level prediction using the same binning as the corresponding curve above. Bands around data points indicate the total uncertainty.}
\label{fig:unfolded_result}
\end{figure}

The unfolded thrust distribution is shown in Figure~\ref{fig:unfolded_result}. For visualization and comparison, the result is presented using two overlaid binning schemes: the nominal ALEPH binning, shown in green, and a binning with half the bin widths, shown in black. The coarser binning allows direct comparison with the published ALEPH measurement, shown in light blue~\cite{ALEPH:2003obs}, while the finer binning provides improved resolution, particularly in the two-jet region and in areas sensitive to multi-jet production. The uncertainty is recomputed for each binning. The uncorrected data and the archived generator-level prediction from $\textsc{Pythia}$ 6.1 are shown in gray and dashed blue, respectively, both using the finer binning to preserve the full granularity of the input. As the measurement is unbinned, the histogram representation is used solely for visualization and does not affect the unfolding itself. The lower panel displays the ratio of the unfolded result to the generator-level MC prediction, using the same binning as the corresponding curve in the upper panel. The total uncertainty from Figure~\ref{fig:uncertBreakdown} is shown as shaded bands around the unfolded data points.


The analysis code can be found on on GitHub~\cite{badea2025unfold_ee_logtau}. The event weights and archived generator level MC, corresponding to the unfolded result, will be released in the final publication along with usage guidelines. 
The weights can be used to reconstruct the observable with arbitrary binning, as done in the figure. While the weights are technically defined for each event, they are specific to the thrust observable and should not be applied to other observables without caution. This format will enable downstream flexibility, allowing users to rebin the distribution, compute logarithmic moments, or compare to alternative theoretical predictions without repeating the unfolding. 
\section{Conclusion}
\label{sec:conclusion}

In this analysis, we present the first fully corrected, unbinned measurement of the thrust distribution in $\ee$ collisions at 91 GeV with archived ALEPH data from LEP 1. This measurement provides new experimental input towards resolving the tension between thrust-based theoretical calculations of $\alpha_{s}(m_{Z})$ and the current experimental world average. It also supports ongoing efforts to constrain non-perturbative effects through logarithmic moments and to develop fully differential hadronization models using modern simulation techniques. 




Building on this work, there are many promising future directions to explore. A natural extension of this work is to perform similar measurements of other event-shape observables and at higher center-of-mass energies. In particular, measurements using archived LEP 2 data would provide an important cross-check and extend hadronization studies up to 209 GeV. Preparatory work already indicates the feasibility of such analyses~\cite{chen2024analysisnotetwoparticlecorrelation, Chen_2024}. In the longer term, this work strongly motivates measurements at current and future $e^{+}e^{-}$ collider programs~\cite{Benedikt:2651299, accardi2022opportunitiesprecisionqcdphysics}.

\FloatBarrier
\section*{Acknowledgments}

The authors would like to thank: Roberto Tenchini and Guenther Dissertori from the ALEPH Collaboration for their useful comments and suggestions on the use of ALEPH data, and Iain Stewart and Kyle Lee for their useful insight into the thrust theoretical calculations. This work has been supported by the Department of Energy, Office of Science, under Grant No. DE-SC0011088 (to Y.-C.C., Y.C., M.P., T.S., C.M., Y.-J.L.) and Schmidt Sciences Foundation (to A. Badea). MA, VM, and BN are supported by the U.S. Department of Energy, Office of Science under contract numbers DE-AC02-05CH11231 and DE-AC02-76SF00515.  This research used resources of the National Energy Research Scientific Computing Center, a DOE Office of Science User Facility supported by DOE contract number DE-AC02-05CH11231.  This work was also supported in part by the National Science Foundation under Grant No. 2311666.

\bibliographystyle{JHEP}
\bibliography{biblio}

\FloatBarrier\newpage

\appendix
\section{Supplemental material}
\label{sec:app_supplemental}

\begin{figure}[ht]
\centering
\begin{subfigure}[b]{0.32\textwidth}
    \includegraphics[width=\textwidth,angle=0]{figures/nominal/h_ntrk.pdf}
    \caption{}
\end{subfigure}
\begin{subfigure}[b]{0.32\textwidth}
    \includegraphics[width=\textwidth,angle=0]{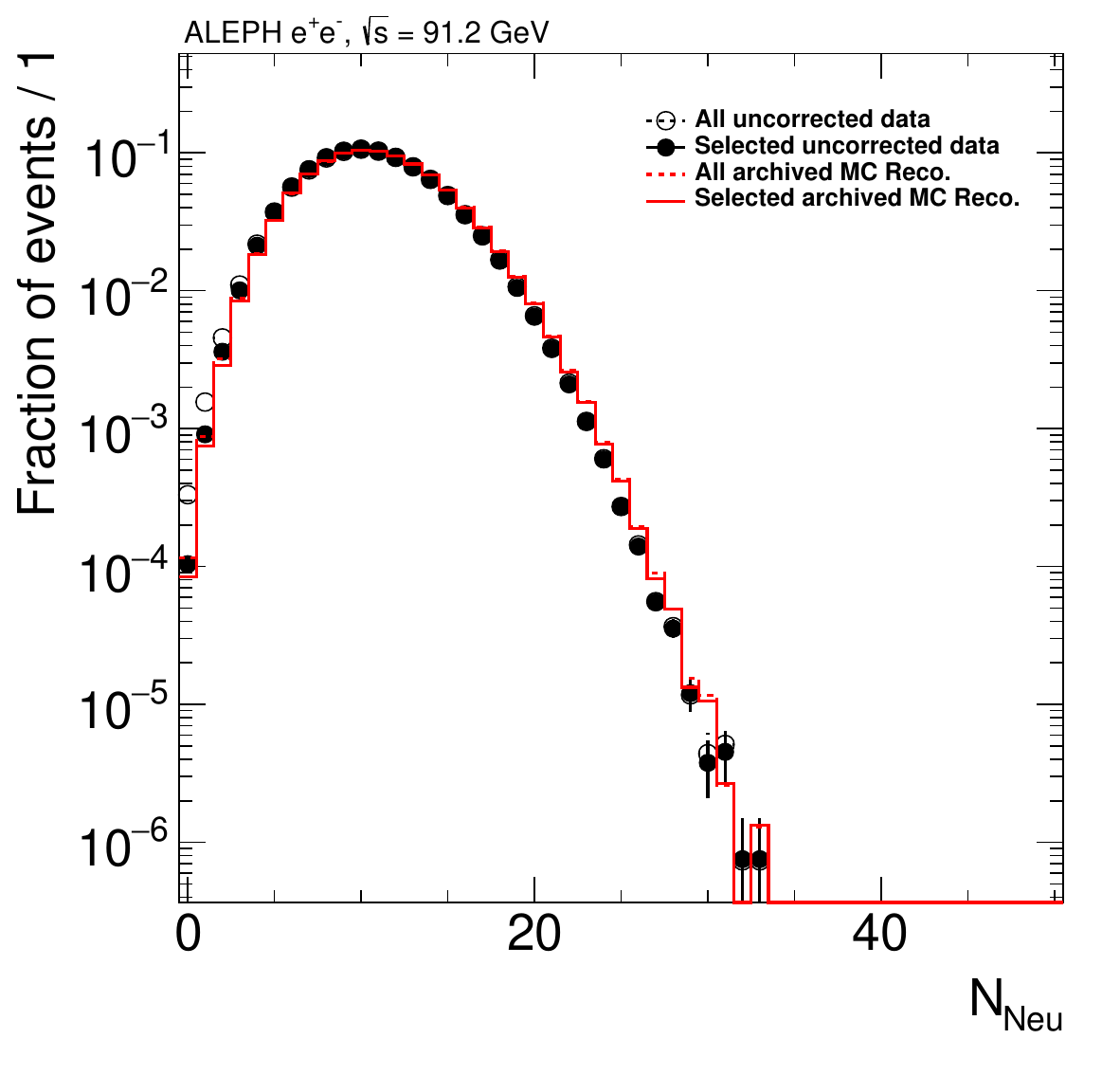}
    \caption{}
\end{subfigure}
\begin{subfigure}[b]{0.32\textwidth}
    \includegraphics[width=\textwidth,angle=0]{figures/nominal/h_ntrkPlusNeu.pdf}
    \caption{}
\end{subfigure}
\begin{subfigure}[b]{0.32\textwidth}
    \includegraphics[width=\textwidth,angle=0]{figures/nominal/h_eCh.pdf}
    \caption{}
\end{subfigure}
\begin{subfigure}[b]{0.32\textwidth}
    \includegraphics[width=\textwidth,angle=0]{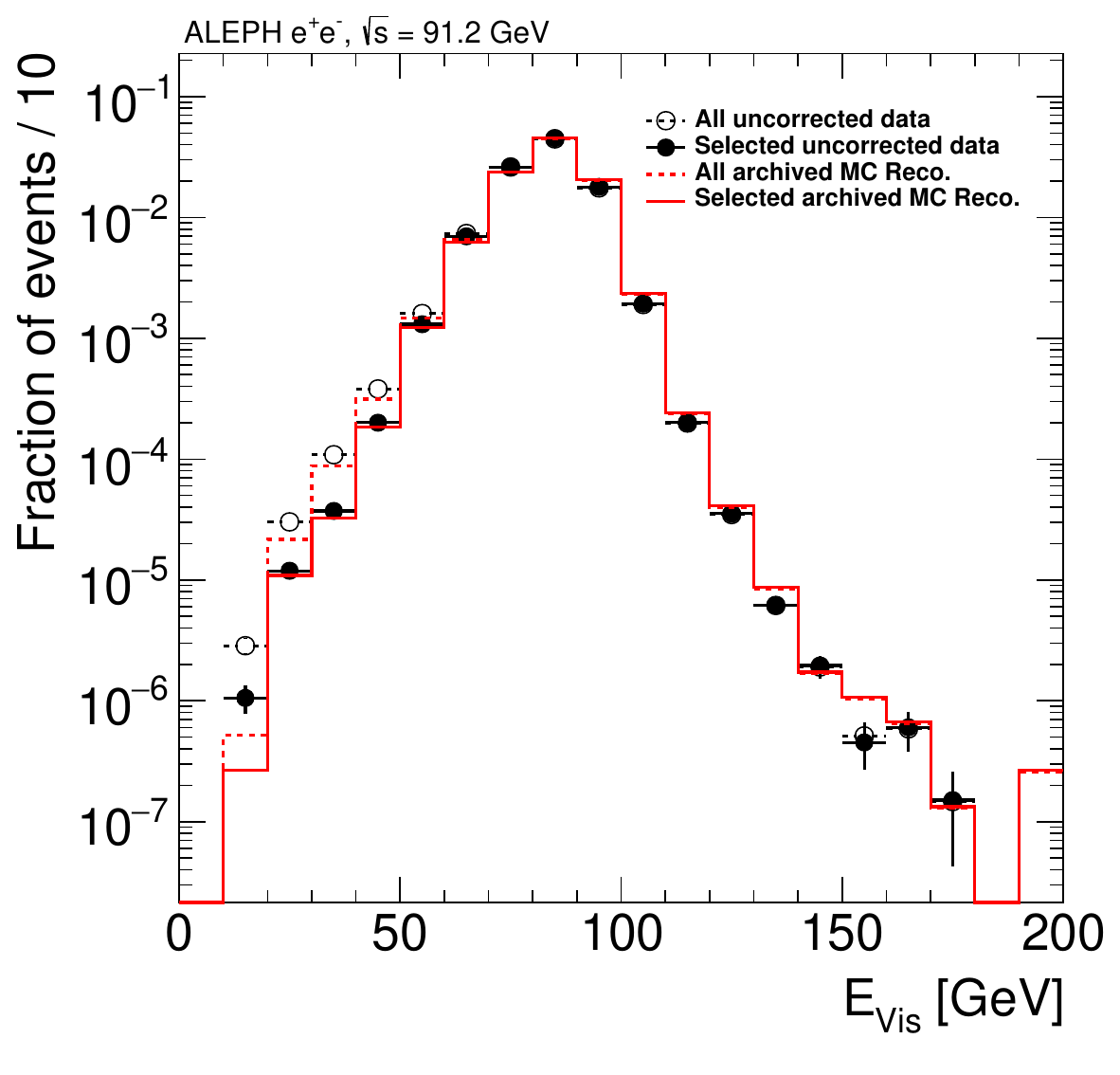}
    \caption{}
\end{subfigure}
\begin{subfigure}[b]{0.32\textwidth}
    \includegraphics[width=\textwidth,angle=0]{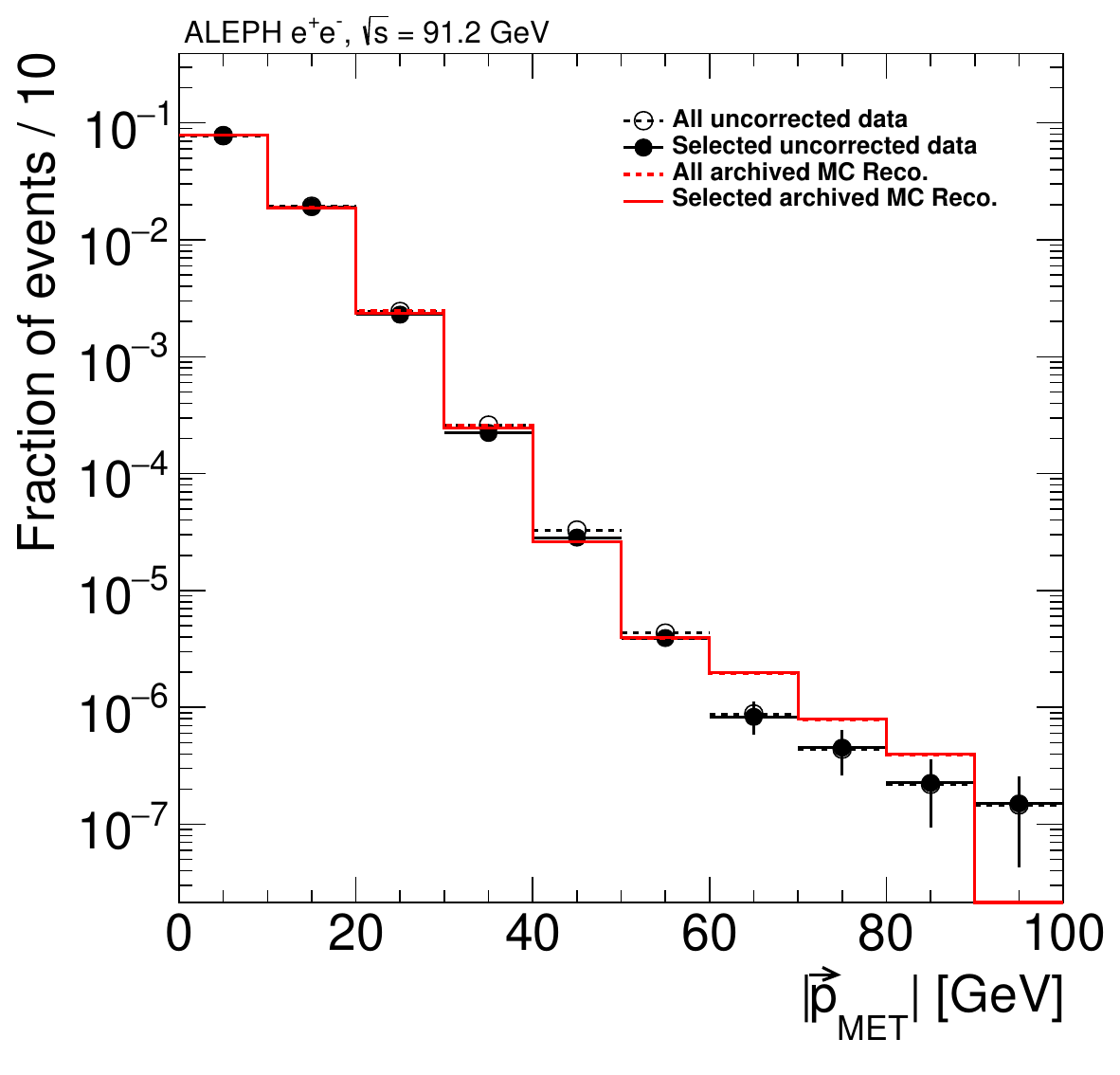}
    \caption{}
\end{subfigure}
\begin{subfigure}[b]{0.32\textwidth}
    \includegraphics[width=\textwidth,angle=0]{figures/nominal/h_cosThetaSph.pdf}
    \caption{}
\end{subfigure}
\begin{subfigure}[b]{0.32\textwidth}
    \includegraphics[width=\textwidth,angle=0]{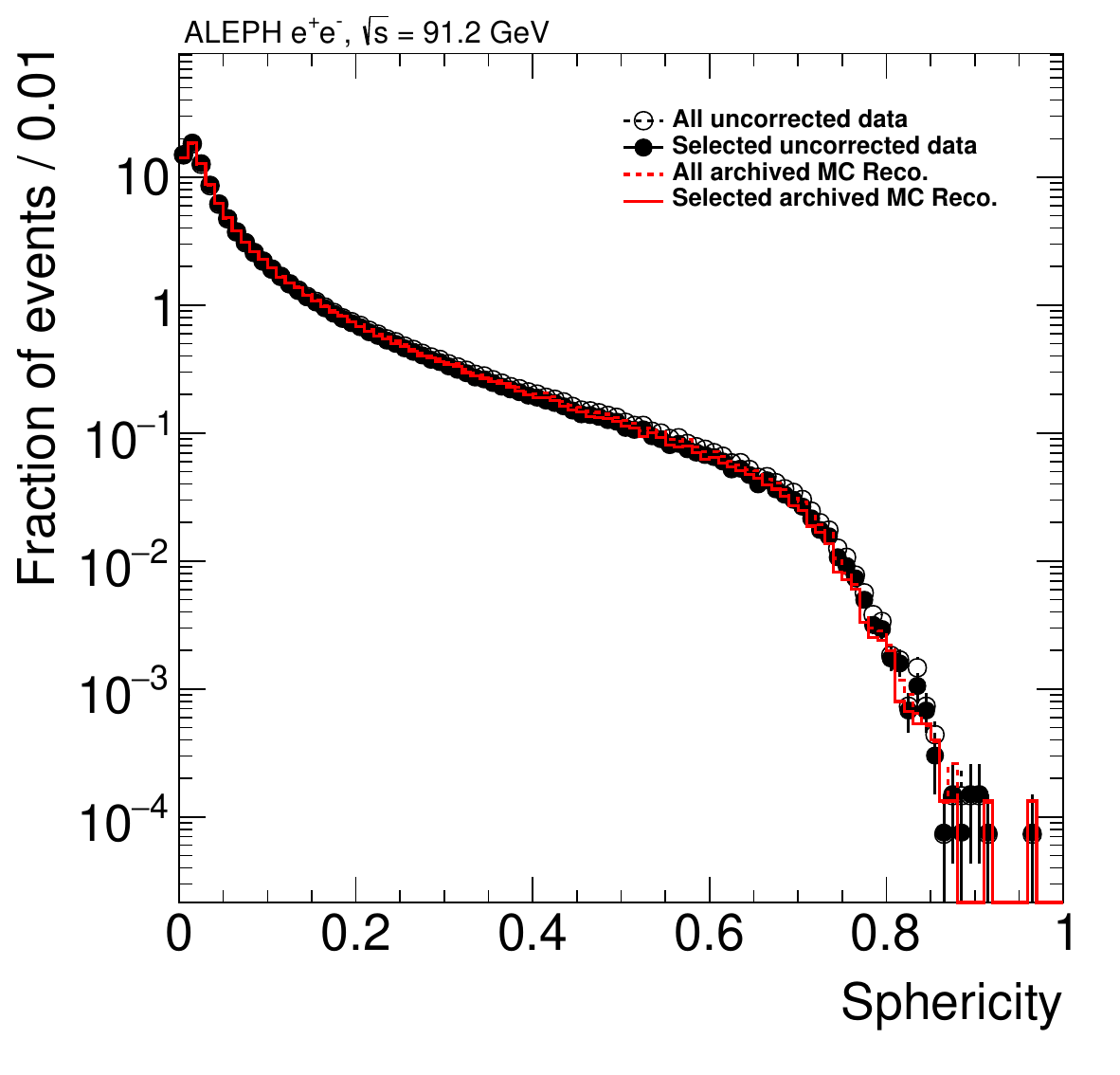}
    \caption{}
\end{subfigure}
\begin{subfigure}[b]{0.32\textwidth}
    \includegraphics[width=\textwidth,angle=0]{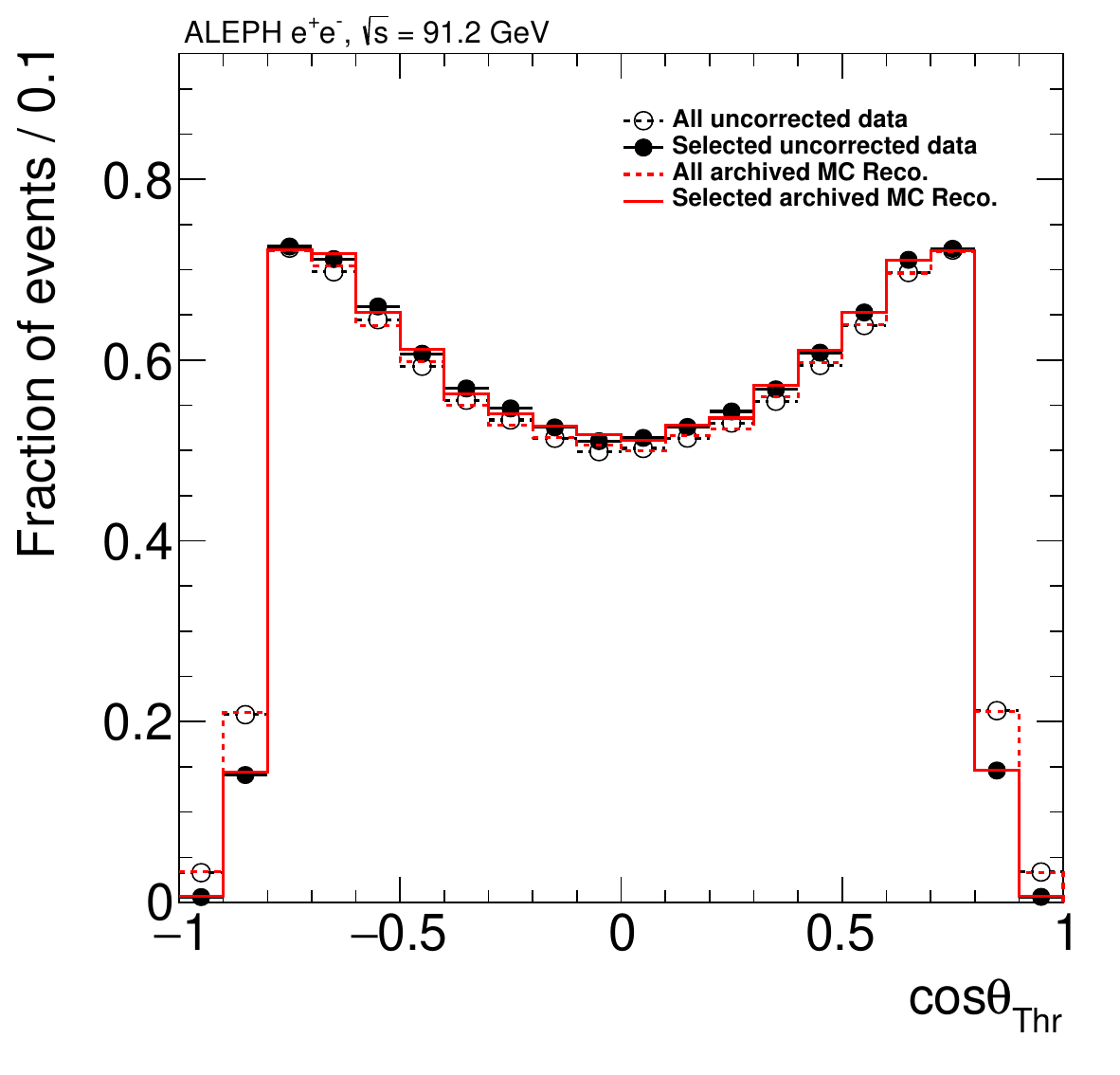}
    \caption{}
\end{subfigure}
\caption{Event level distributions with nominal object and event selections.}
\label{fig:app_eventObservables}
\end{figure}

The kinematic distributions of all charged tracks, leptons, photons, and neutral hadrons without selections or unfolding corrections are shown below in Figures~\ref{fig:kinem_pwflag0}-~\ref{fig:kinem_pwflag5}. The event level observables for selected particles are shown in Figure~\ref{fig:app_eventObservables} with nominal event selections and without unfolding corrections.

\begin{figure}[ht]
\centering
\begin{subfigure}[b]{0.32\textwidth}
    \includegraphics[width=\textwidth,angle=0]{figures/nominal/h_pwflag0_cosTheta.pdf}
    \caption{}
\end{subfigure}
\begin{subfigure}[b]{0.32\textwidth}
    \includegraphics[width=\textwidth,angle=0]{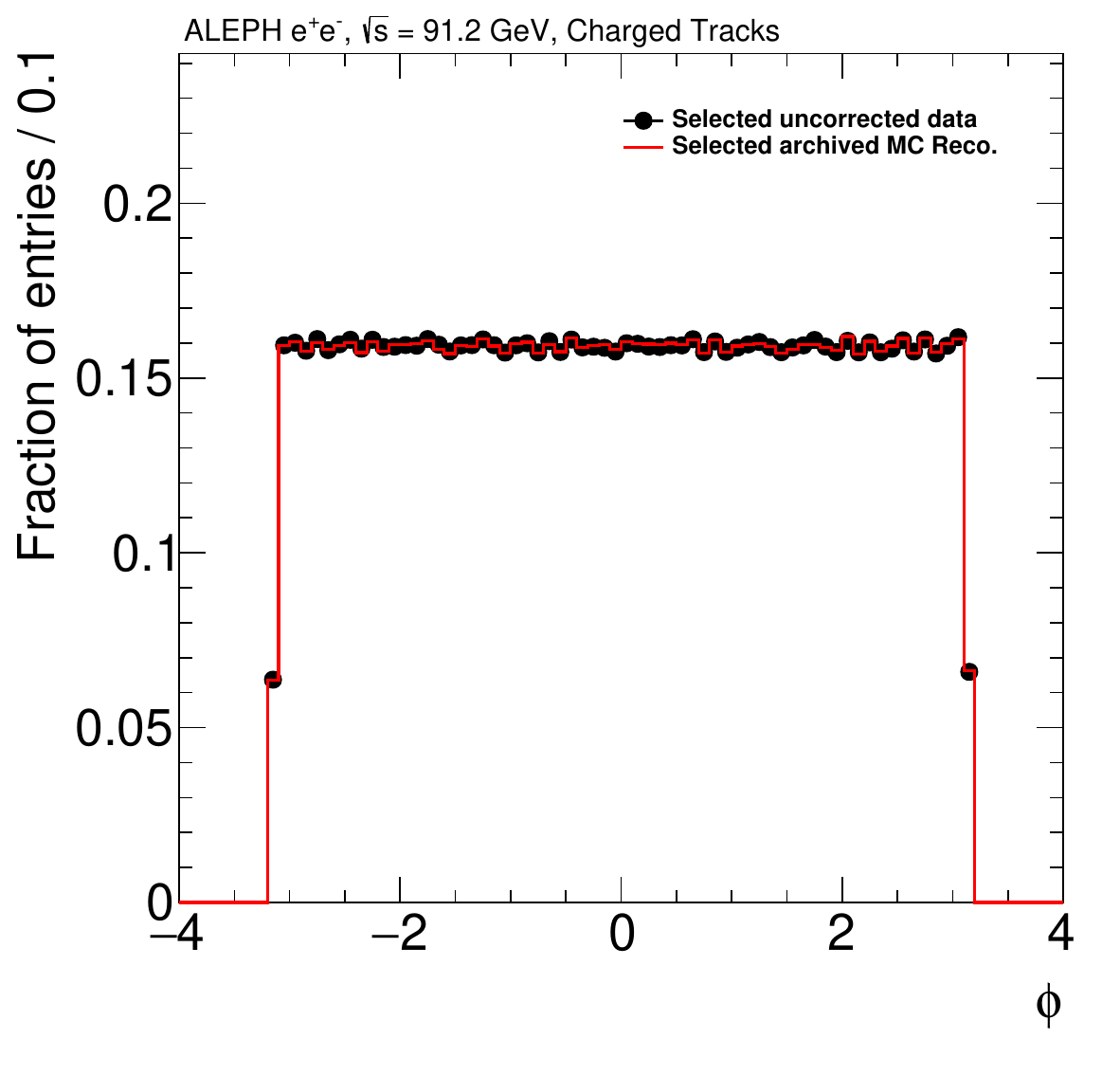}
    \caption{}
\end{subfigure}
\begin{subfigure}[b]{0.32\textwidth}
    \includegraphics[width=\textwidth,angle=0]{figures/nominal/h_pwflag0_pt.pdf}
    \caption{}
\end{subfigure}
\begin{subfigure}[b]{0.32\textwidth}
    \includegraphics[width=\textwidth,angle=0]{figures/nominal/h_pwflag0_ntpc.pdf}
    \caption{}
\end{subfigure}
\begin{subfigure}[b]{0.32\textwidth}
    \includegraphics[width=\textwidth,angle=0]{figures/nominal/h_pwflag0_d0.pdf}
    \caption{}
\end{subfigure}
\begin{subfigure}[b]{0.32\textwidth}
    \includegraphics[width=\textwidth,angle=0]{figures/nominal/h_pwflag0_z0.pdf}
    \caption{}
\end{subfigure}
\begin{subfigure}[b]{0.32\textwidth}
    \includegraphics[width=\textwidth,angle=0]{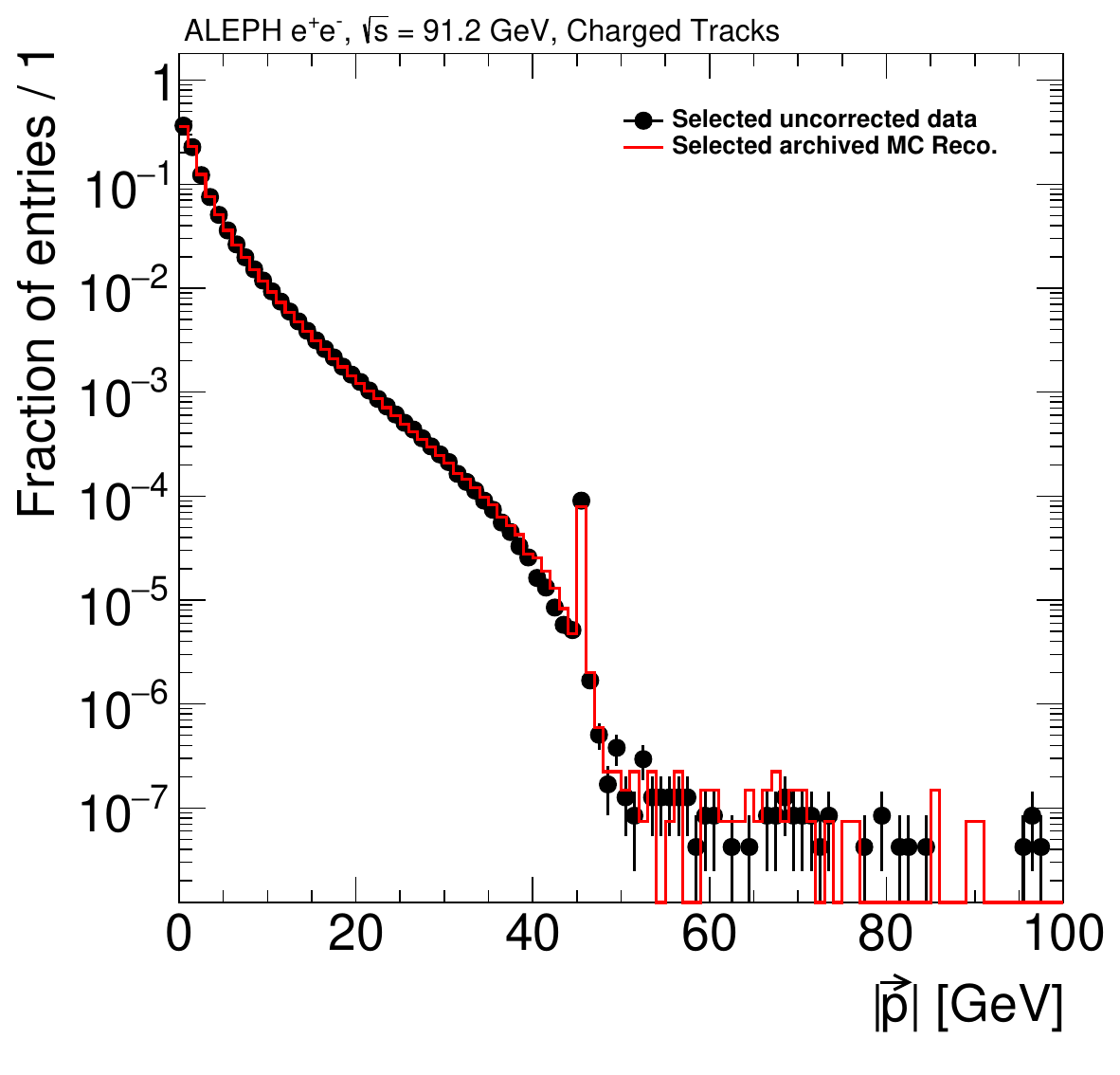}
    \caption{}
\end{subfigure}
\begin{subfigure}[b]{0.32\textwidth}
    \includegraphics[width=\textwidth,angle=0]{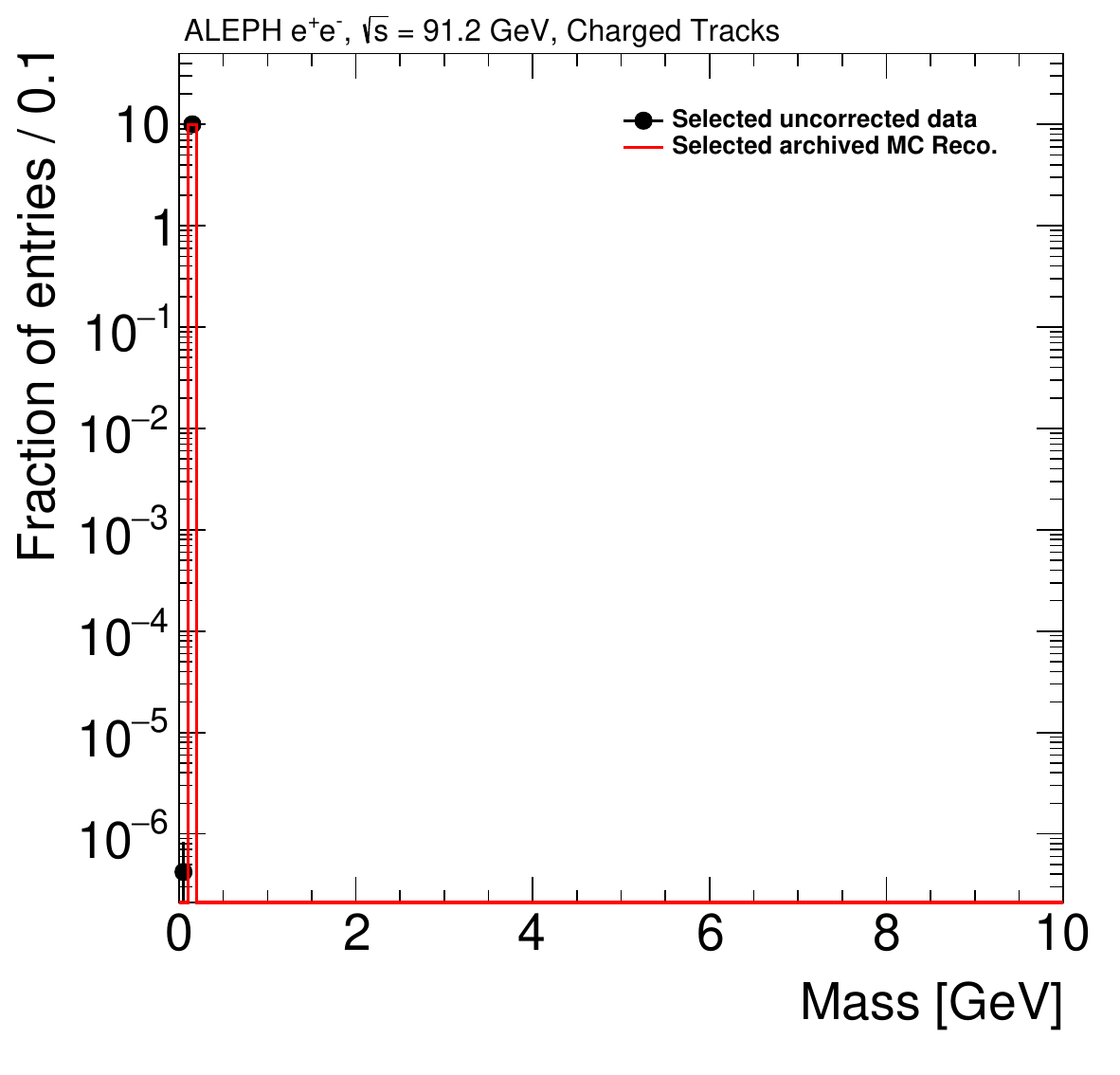}
    \caption{}
\end{subfigure}
\begin{subfigure}[b]{0.32\textwidth}
    \includegraphics[width=\textwidth,angle=0]{figures/nominal/h_pwflag0_energy.pdf}
    \caption{}
\end{subfigure}
\caption{Kinematic distributions for all charged tracks.}
\label{fig:kinem_pwflag0}
\end{figure}

\begin{figure}[ht]
\centering
\begin{subfigure}[b]{0.32\textwidth}
    \includegraphics[width=\textwidth,angle=0]{figures/nominal/h_pwflag1_cosTheta.pdf}
    \caption{}
\end{subfigure}
\begin{subfigure}[b]{0.32\textwidth}
    \includegraphics[width=\textwidth,angle=0]{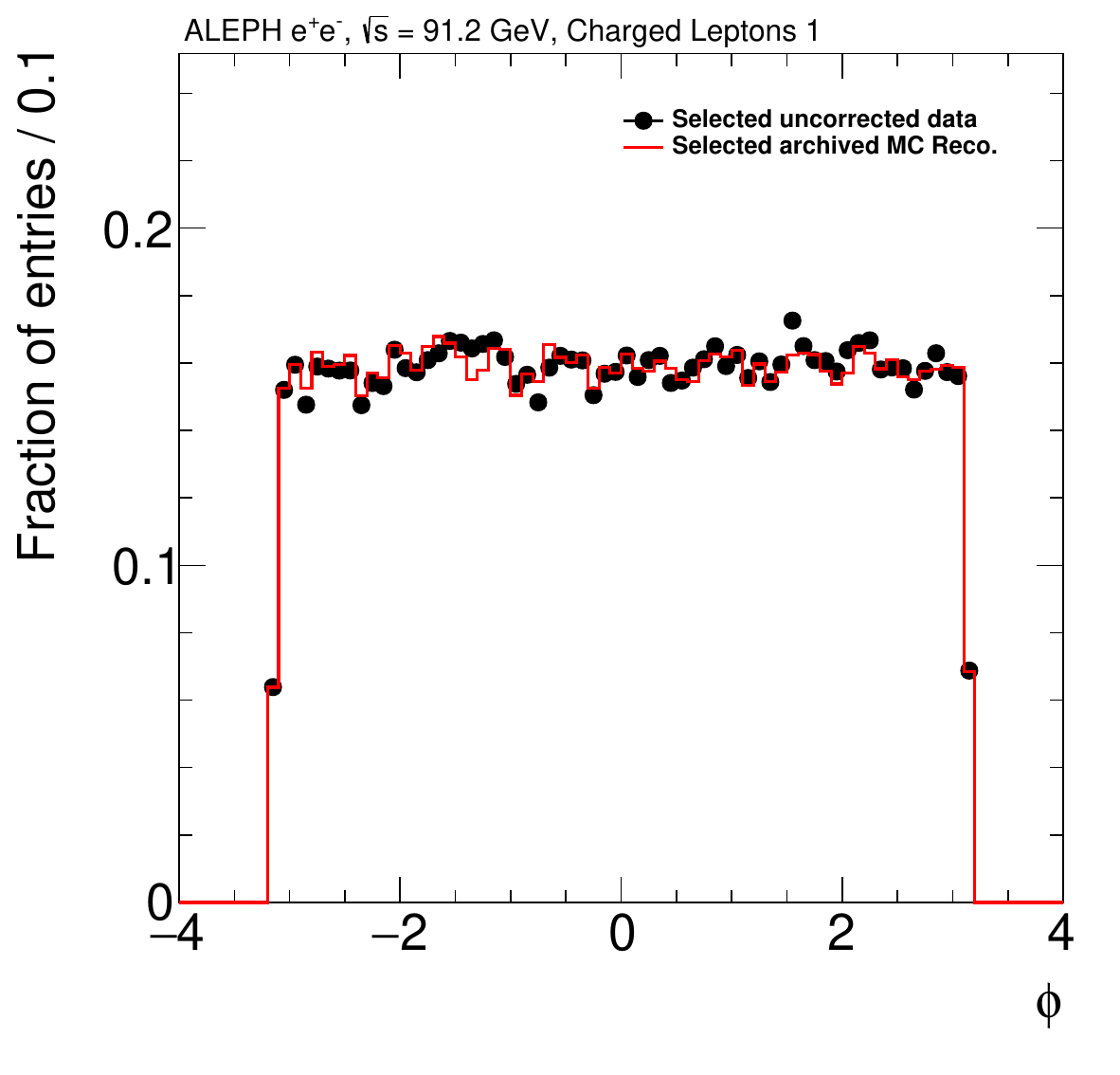}
    \caption{}
\end{subfigure}
\begin{subfigure}[b]{0.32\textwidth}
    \includegraphics[width=\textwidth,angle=0]{figures/nominal/h_pwflag1_pt.pdf}
    \caption{}
\end{subfigure}
\begin{subfigure}[b]{0.32\textwidth}
    \includegraphics[width=\textwidth,angle=0]{figures/nominal/h_pwflag1_ntpc.pdf}
    \caption{}
\end{subfigure}
\begin{subfigure}[b]{0.32\textwidth}
    \includegraphics[width=\textwidth,angle=0]{figures/nominal/h_pwflag1_d0.pdf}
    \caption{}
\end{subfigure}
\begin{subfigure}[b]{0.32\textwidth}
    \includegraphics[width=\textwidth,angle=0]{figures/nominal/h_pwflag1_z0.pdf}
    \caption{}
\end{subfigure}
\begin{subfigure}[b]{0.32\textwidth}
    \includegraphics[width=\textwidth,angle=0]{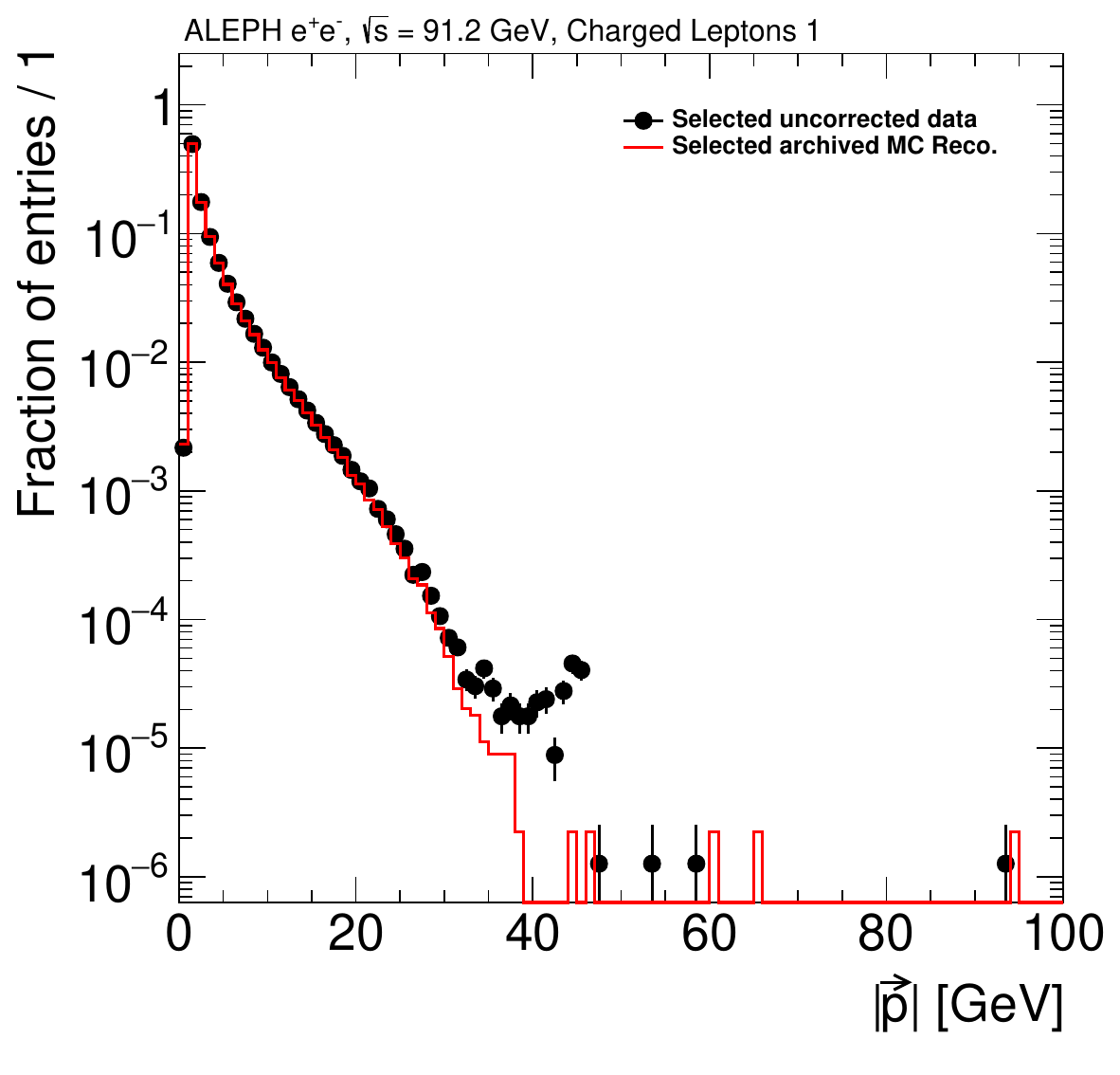}
    \caption{}
\end{subfigure}
\begin{subfigure}[b]{0.32\textwidth}
    \includegraphics[width=\textwidth,angle=0]{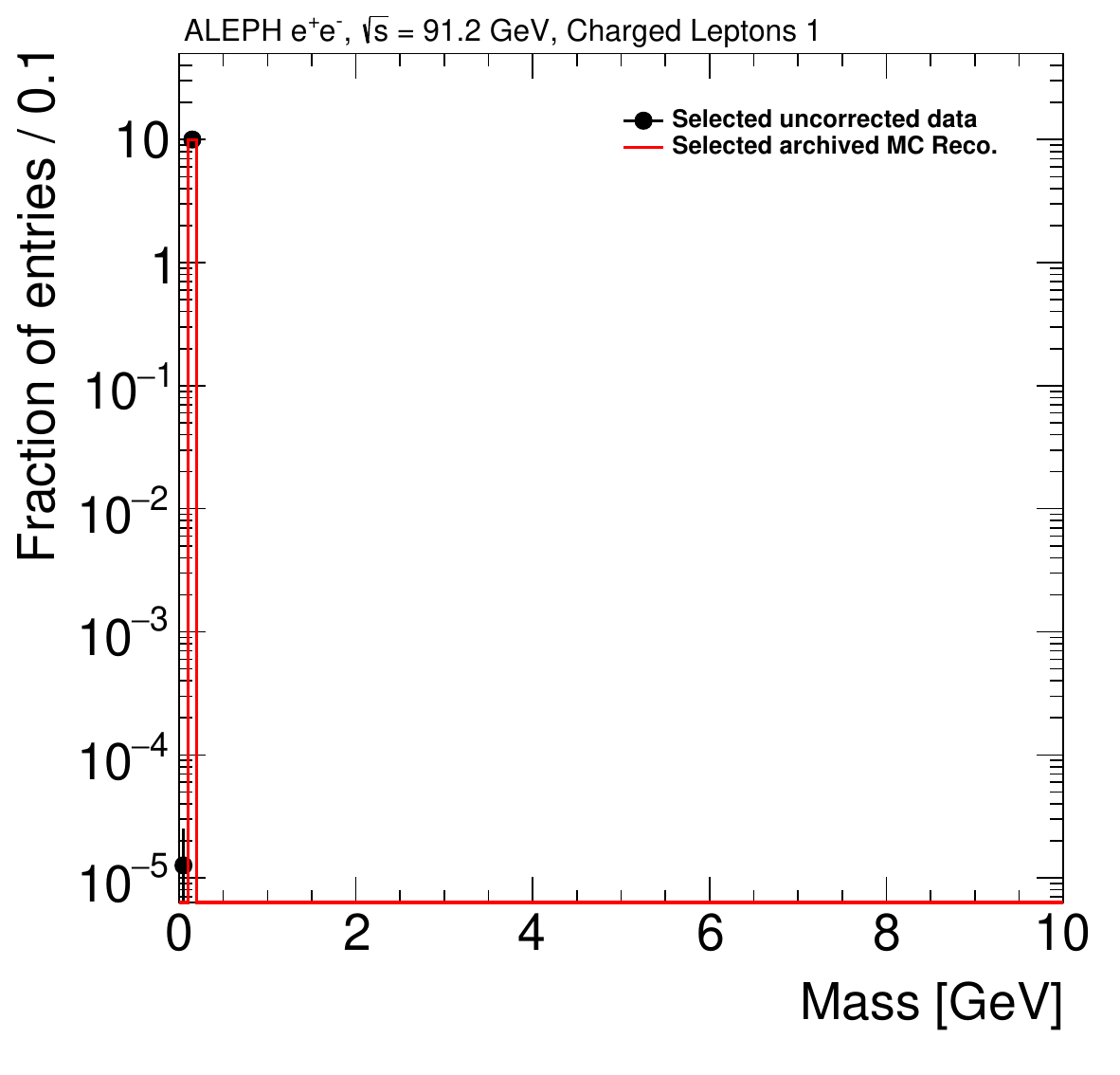}
    \caption{}
\end{subfigure}
\begin{subfigure}[b]{0.32\textwidth}
    \includegraphics[width=\textwidth,angle=0]{figures/nominal/h_pwflag1_energy.pdf}
    \caption{}
\end{subfigure}
\caption{Kinematic distributions for all charged leptons 1.}
\label{fig:kinem_pwflag1}
\end{figure}

\begin{figure}[ht]
\centering
\begin{subfigure}[b]{0.32\textwidth}
    \includegraphics[width=\textwidth,angle=0]{figures/nominal/h_pwflag2_cosTheta.pdf}
    \caption{}
\end{subfigure}
\begin{subfigure}[b]{0.32\textwidth}
    \includegraphics[width=\textwidth,angle=0]{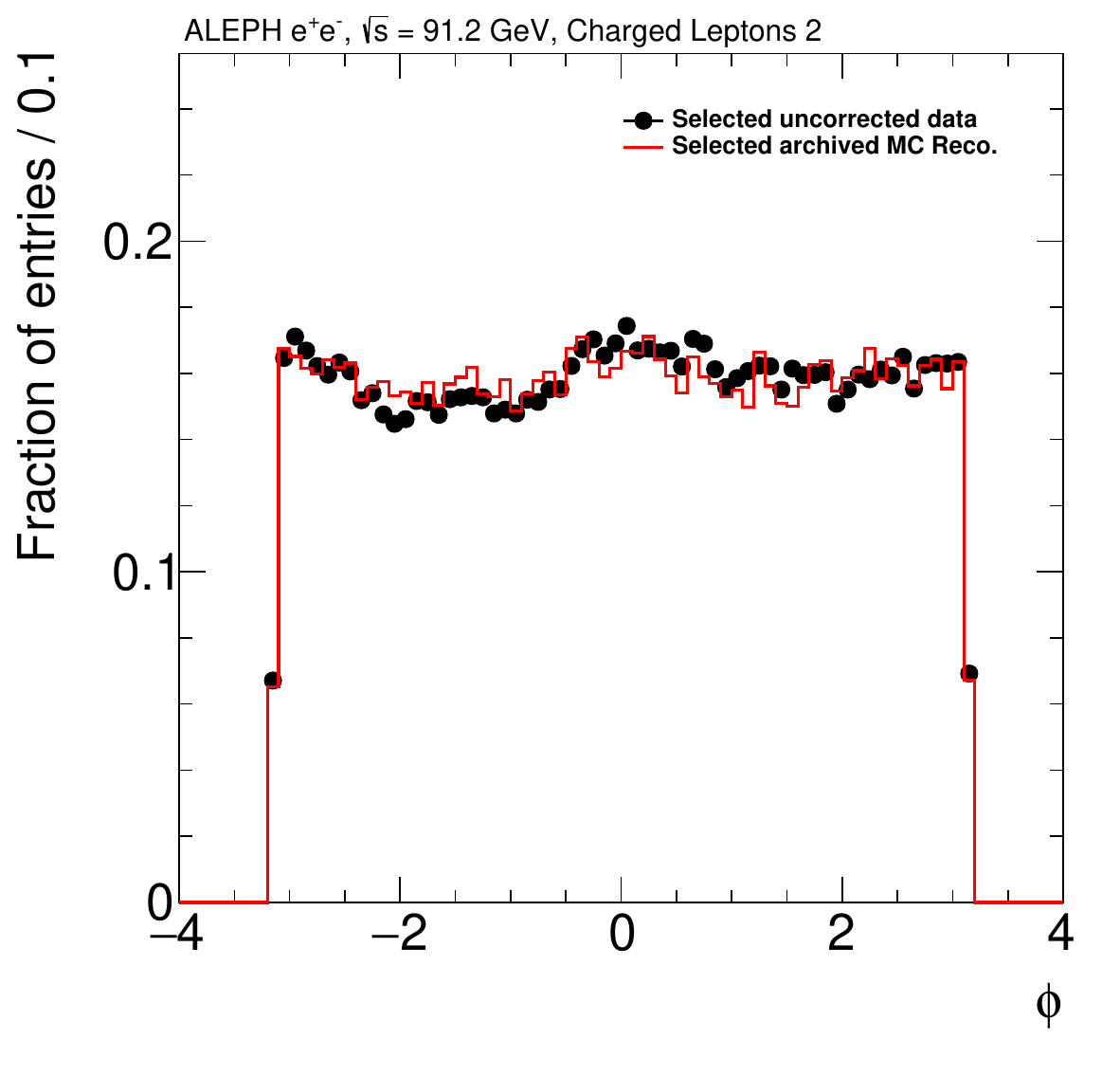}
    \caption{}
\end{subfigure}
\begin{subfigure}[b]{0.32\textwidth}
    \includegraphics[width=\textwidth,angle=0]{figures/nominal/h_pwflag2_pt.pdf}
    \caption{}
\end{subfigure}
\begin{subfigure}[b]{0.32\textwidth}
    \includegraphics[width=\textwidth,angle=0]{figures/nominal/h_pwflag2_ntpc.pdf}
    \caption{}
\end{subfigure}
\begin{subfigure}[b]{0.32\textwidth}
    \includegraphics[width=\textwidth,angle=0]{figures/nominal/h_pwflag2_d0.pdf}
    \caption{}
\end{subfigure}
\begin{subfigure}[b]{0.32\textwidth}
    \includegraphics[width=\textwidth,angle=0]{figures/nominal/h_pwflag2_z0.pdf}
    \caption{}
\end{subfigure}
\begin{subfigure}[b]{0.32\textwidth}
    \includegraphics[width=\textwidth,angle=0]{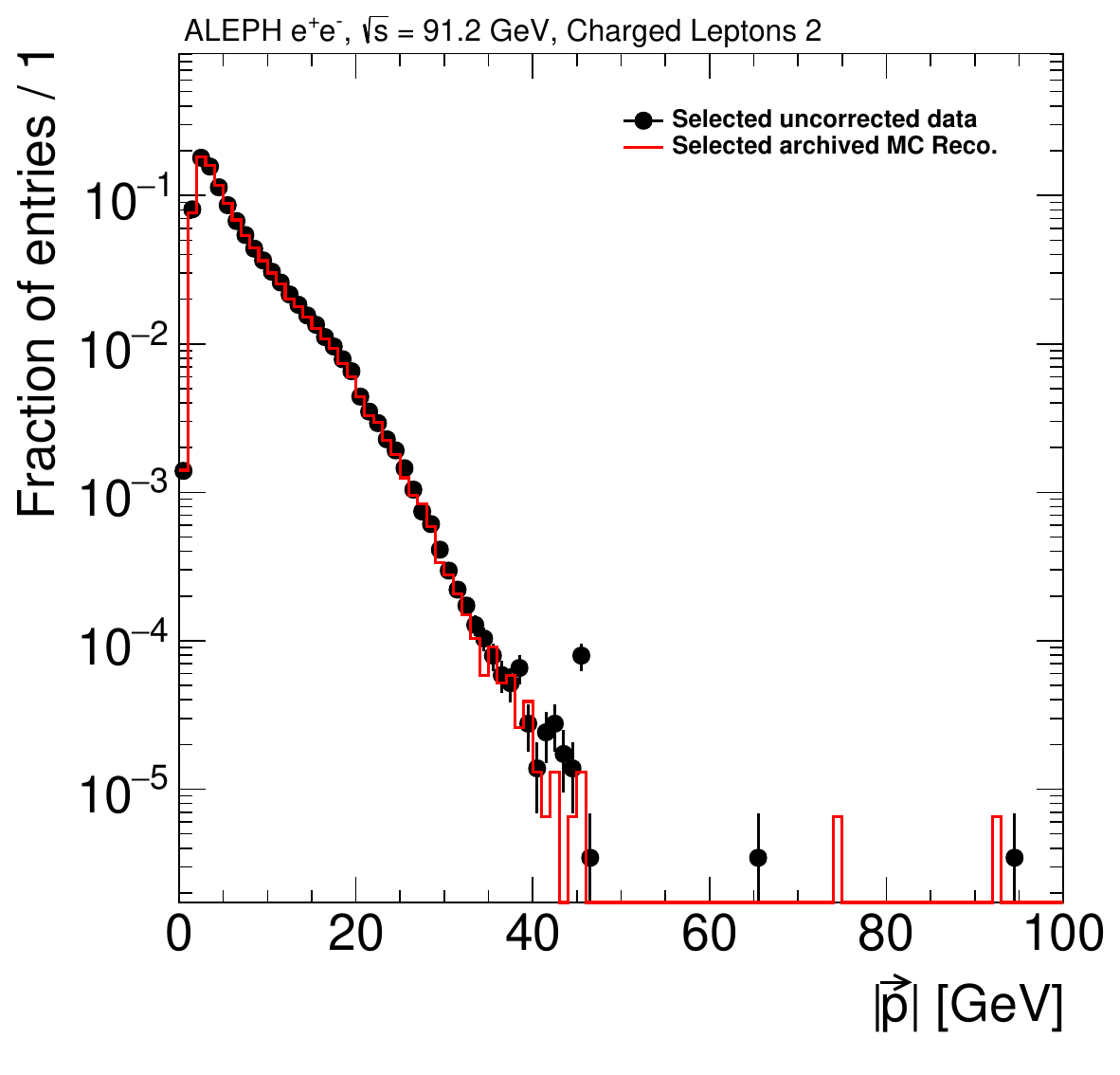}
    \caption{}
\end{subfigure}
\begin{subfigure}[b]{0.32\textwidth}
    \includegraphics[width=\textwidth,angle=0]{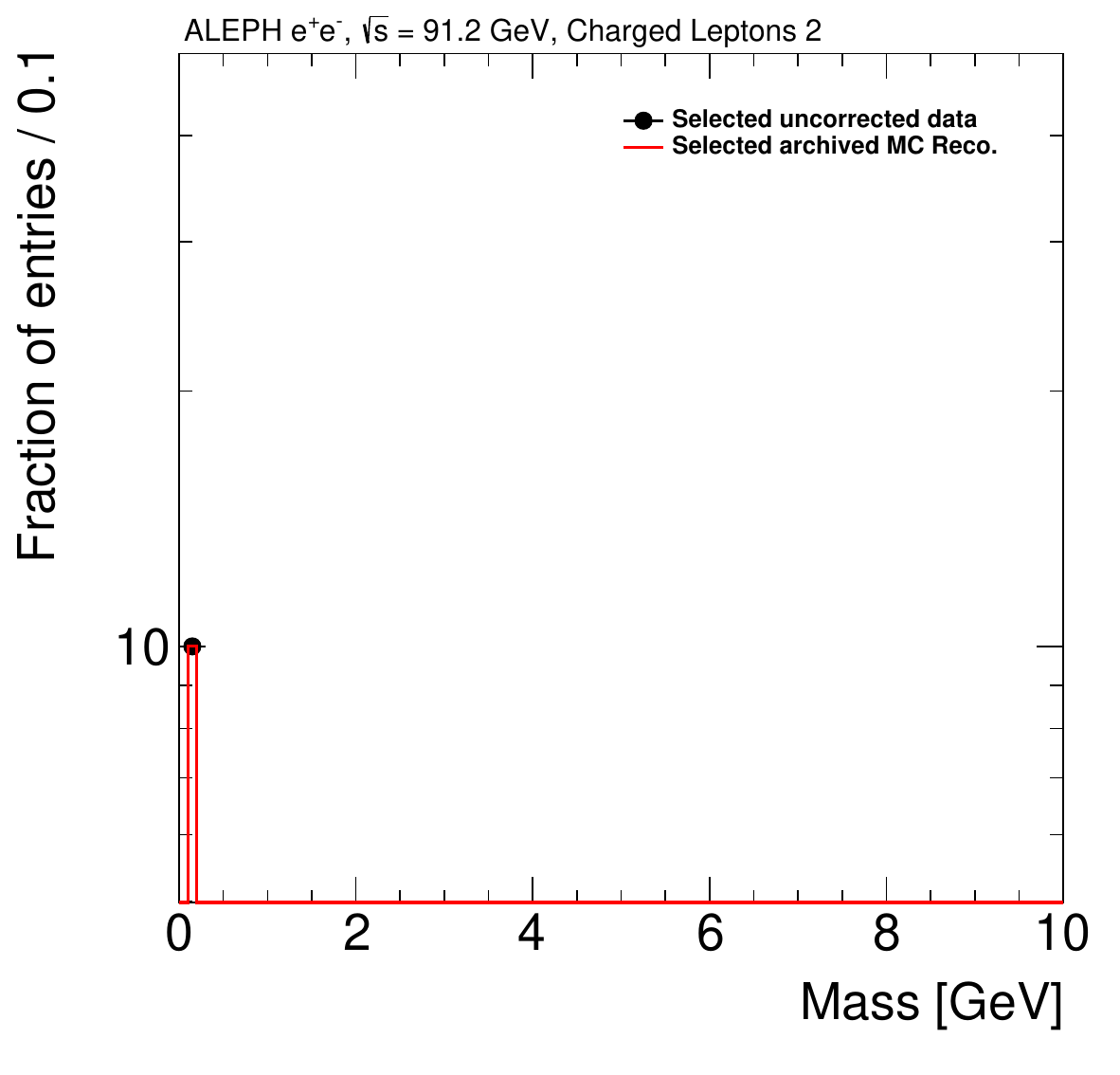}
    \caption{}
\end{subfigure}
\begin{subfigure}[b]{0.32\textwidth}
    \includegraphics[width=\textwidth,angle=0]{figures/nominal/h_pwflag2_energy.pdf}
    \caption{}
\end{subfigure}
\caption{Kinematic distributions for all charged leptons 2.}
\label{fig:kinem_pwflag2}
\end{figure}

\begin{figure}[ht]
\centering
\begin{subfigure}[b]{0.32\textwidth}
    \includegraphics[width=\textwidth,angle=0]{figures/nominal/h_pwflag3_cosTheta.pdf}
    \caption{}
\end{subfigure}
\begin{subfigure}[b]{0.32\textwidth}
    \includegraphics[width=\textwidth,angle=0]{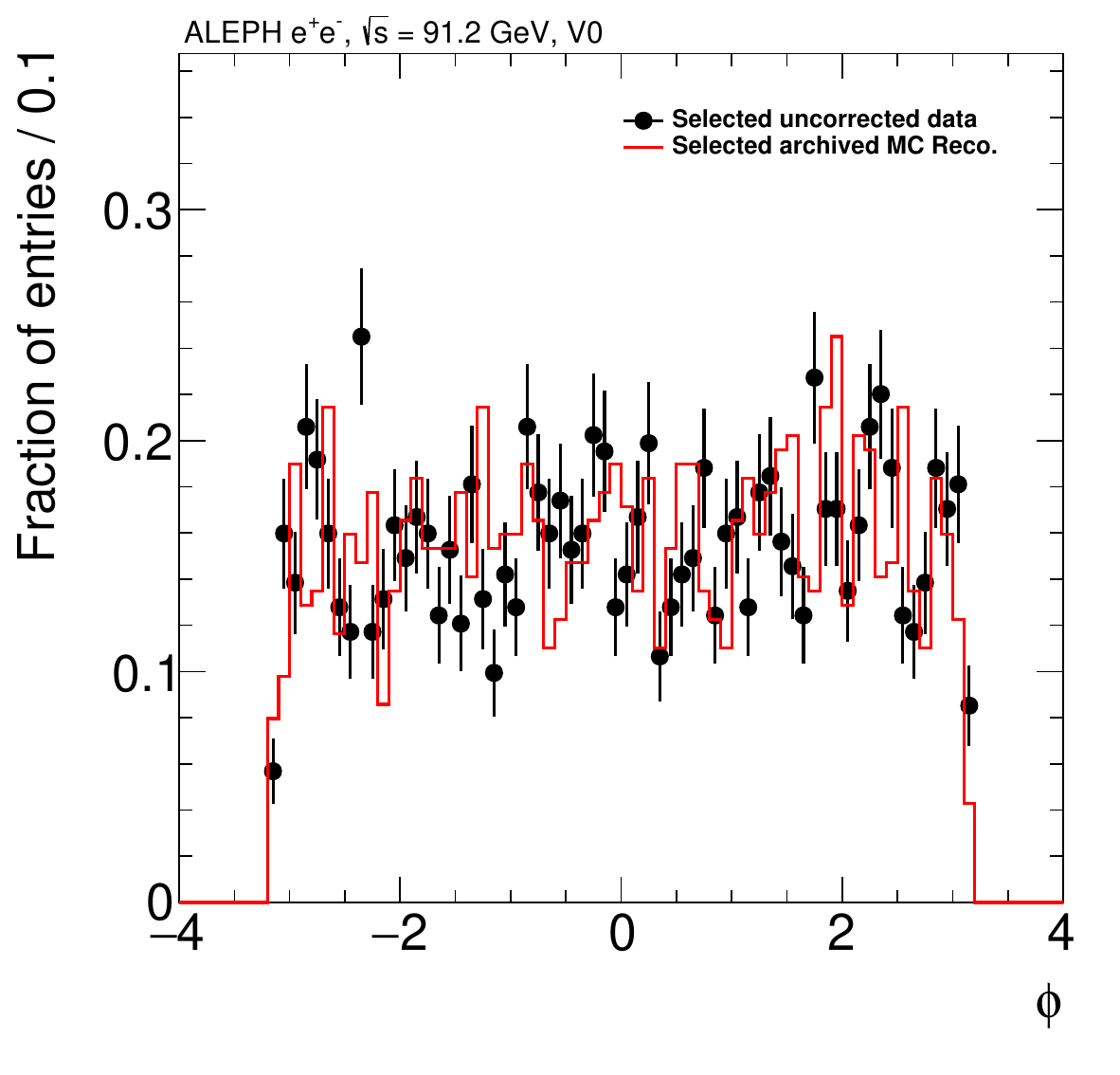}
    \caption{}
\end{subfigure}
\begin{subfigure}[b]{0.32\textwidth}
    \includegraphics[width=\textwidth,angle=0]{figures/nominal/h_pwflag3_pt.pdf}
    \caption{}
\end{subfigure}
\begin{subfigure}[b]{0.32\textwidth}
    \includegraphics[width=\textwidth,angle=0]{figures/nominal/h_pwflag3_ntpc.pdf}
    \caption{}
\end{subfigure}
\begin{subfigure}[b]{0.32\textwidth}
    \includegraphics[width=\textwidth,angle=0]{figures/nominal/h_pwflag3_d0.pdf}
    \caption{}
\end{subfigure}
\begin{subfigure}[b]{0.32\textwidth}
    \includegraphics[width=\textwidth,angle=0]{figures/nominal/h_pwflag3_z0.pdf}
    \caption{}
\end{subfigure}
\begin{subfigure}[b]{0.32\textwidth}
    \includegraphics[width=\textwidth,angle=0]{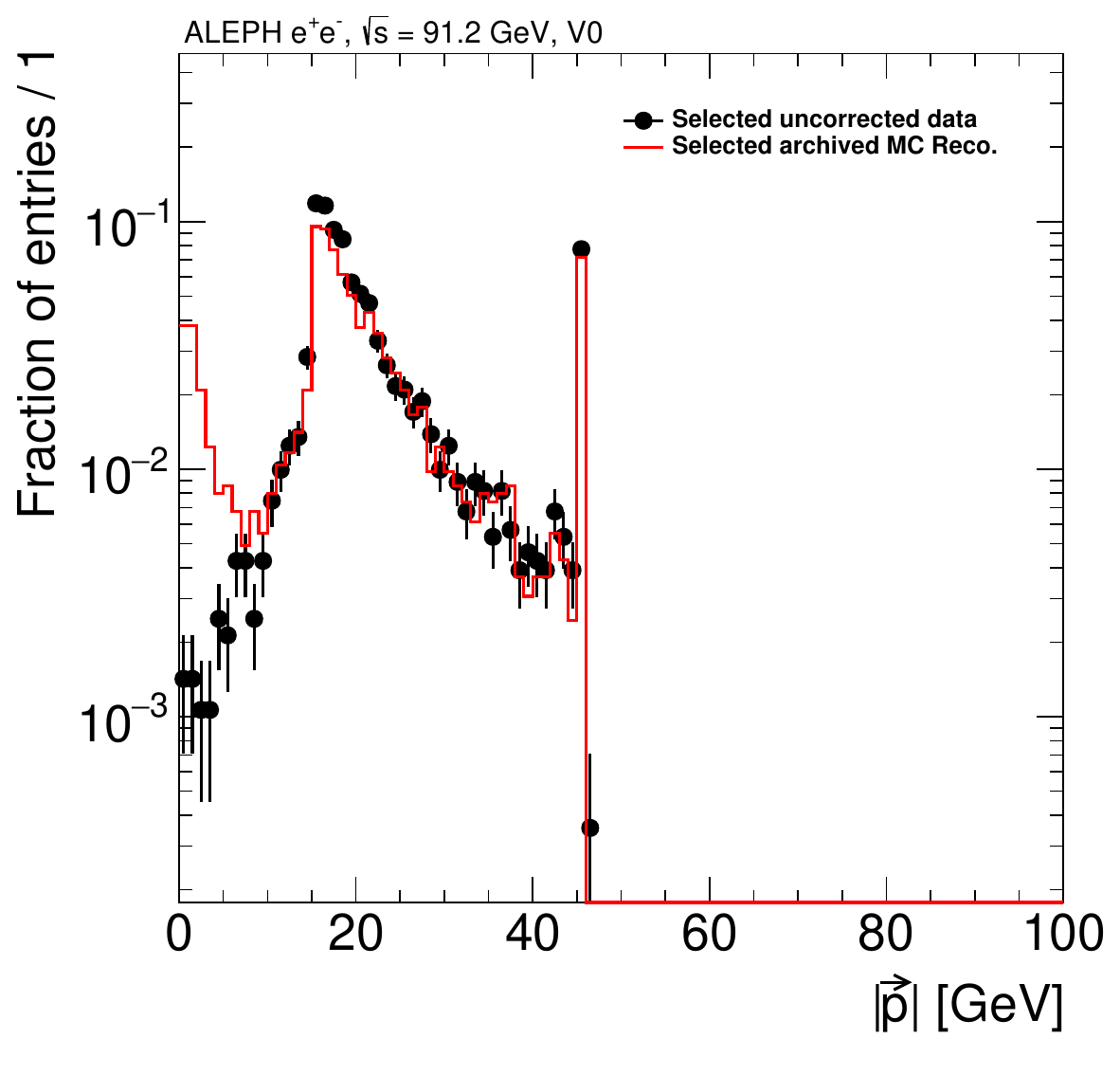}
    \caption{}
\end{subfigure}
\begin{subfigure}[b]{0.32\textwidth}
    \includegraphics[width=\textwidth,angle=0]{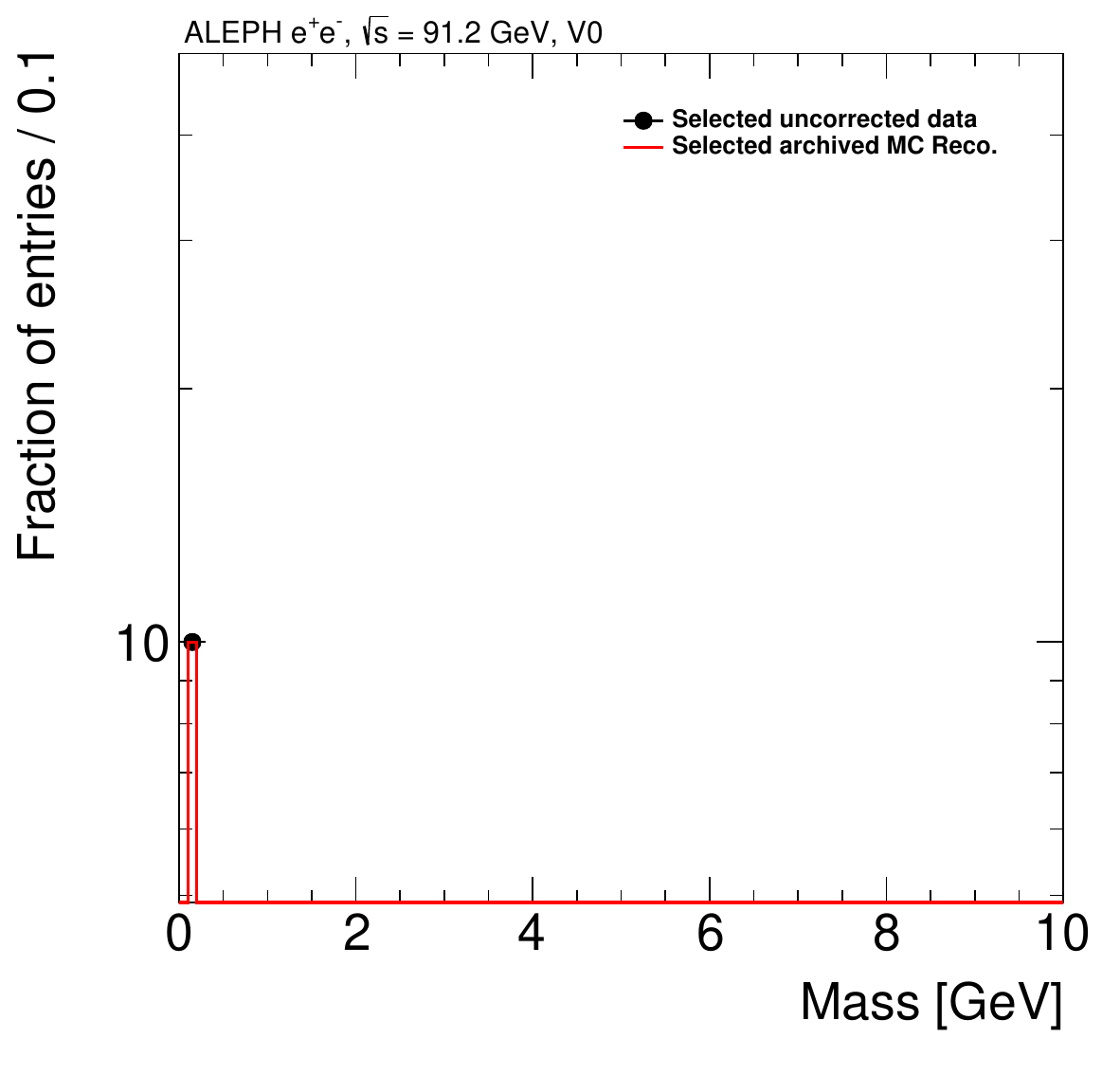}
    \caption{}
\end{subfigure}
\begin{subfigure}[b]{0.32\textwidth}
    \includegraphics[width=\textwidth,angle=0]{figures/nominal/h_pwflag3_energy.pdf}
    \caption{}
\end{subfigure}
\caption{Kinematic distributions for all V0.}
\label{fig:kinem_pwflag3}
\end{figure}

\begin{figure}[ht]
\centering
\begin{subfigure}[b]{0.32\textwidth}
    \includegraphics[width=\textwidth,angle=0]{figures/nominal/h_pwflag4_cosTheta.pdf}
    \caption{}
\end{subfigure}
\begin{subfigure}[b]{0.32\textwidth}
    \includegraphics[width=\textwidth,angle=0]{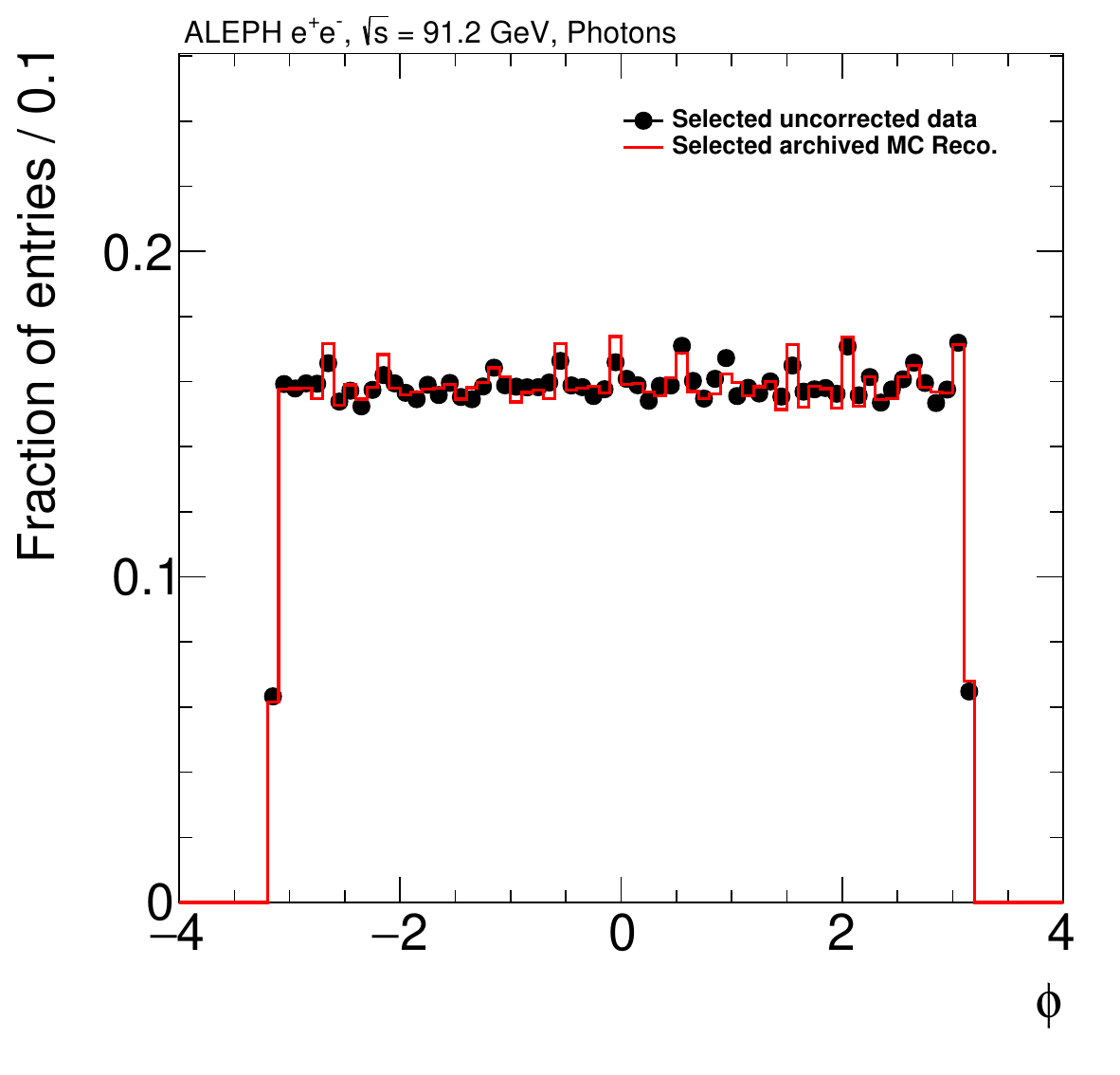}
    \caption{}
\end{subfigure}
\begin{subfigure}[b]{0.32\textwidth}
    \includegraphics[width=\textwidth,angle=0]{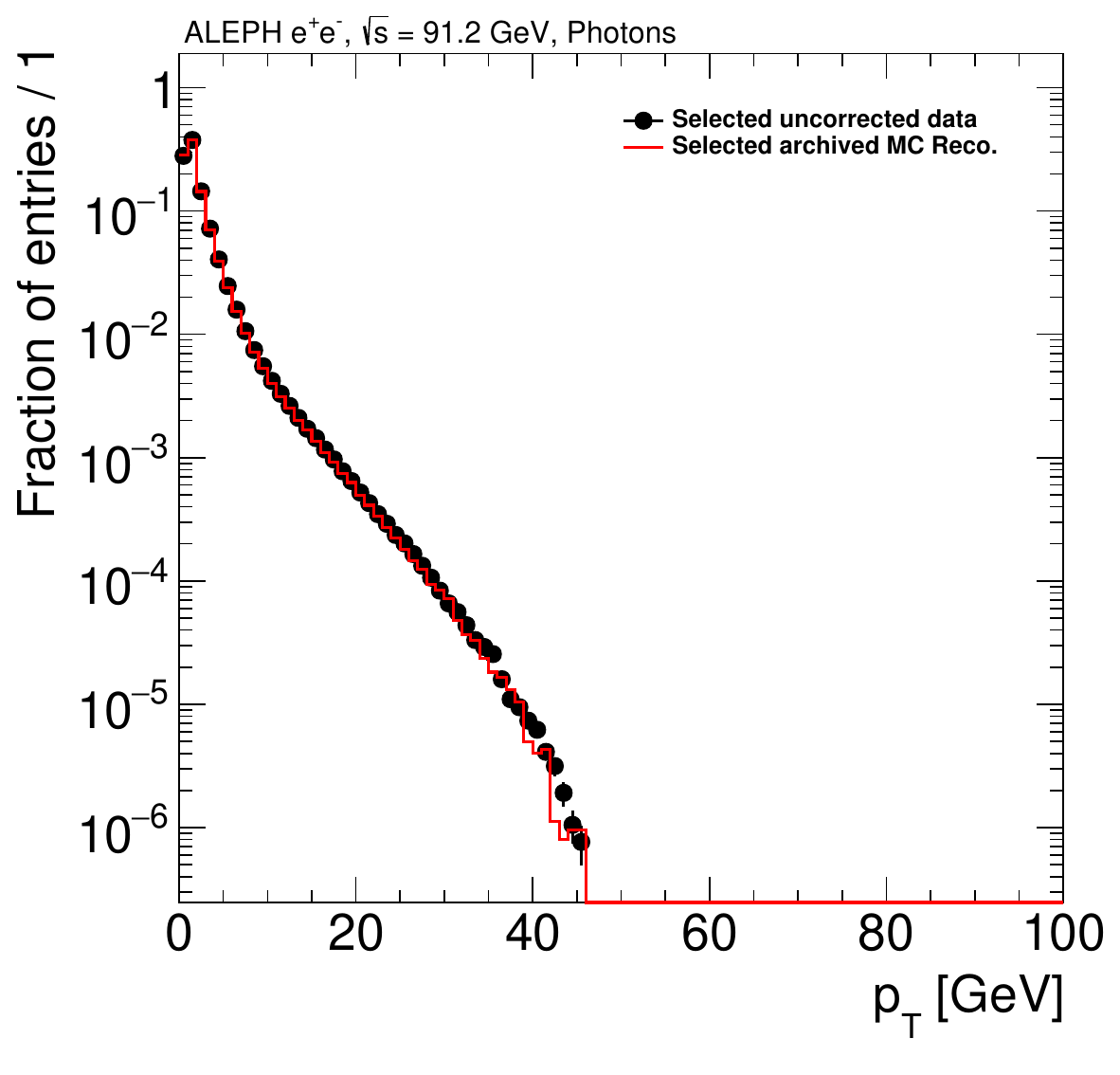}
    \caption{}
\end{subfigure}
\begin{subfigure}[b]{0.32\textwidth}
    \includegraphics[width=\textwidth,angle=0]{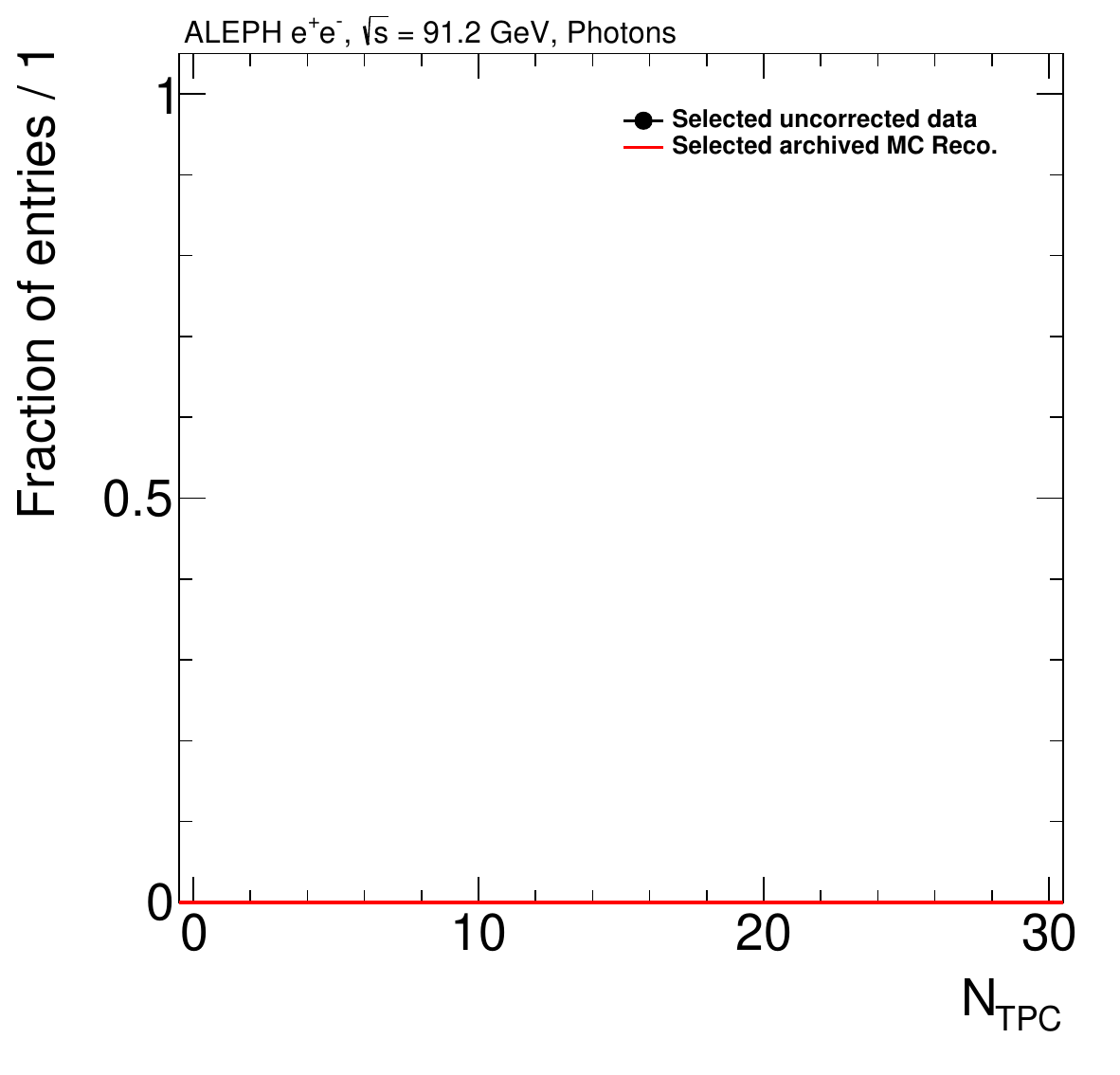}
    \caption{}
\end{subfigure}
\begin{subfigure}[b]{0.32\textwidth}
    \includegraphics[width=\textwidth,angle=0]{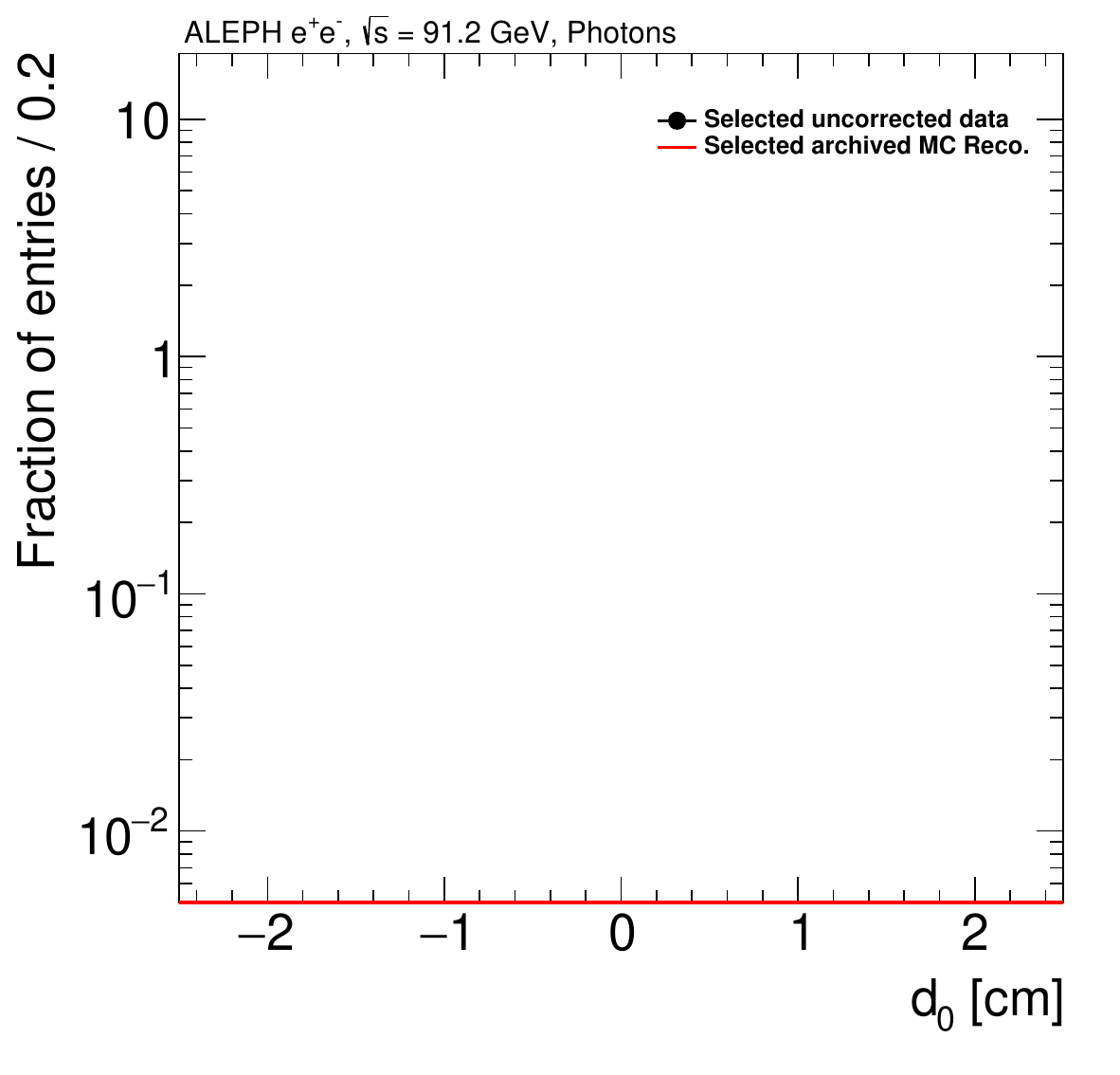}
    \caption{}
\end{subfigure}
\begin{subfigure}[b]{0.32\textwidth}
    \includegraphics[width=\textwidth,angle=0]{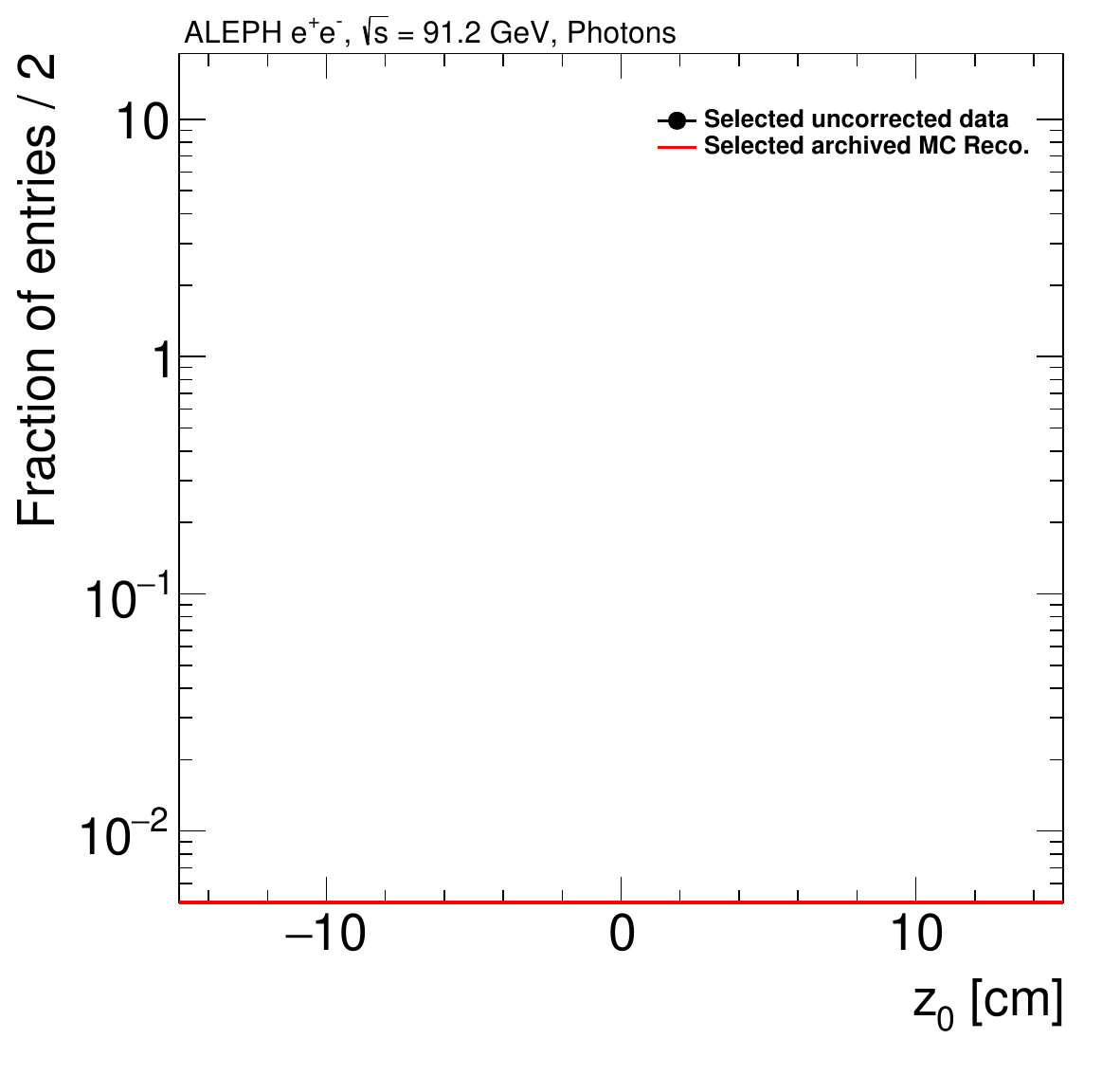}
    \caption{}
\end{subfigure}
\begin{subfigure}[b]{0.32\textwidth}
    \includegraphics[width=\textwidth,angle=0]{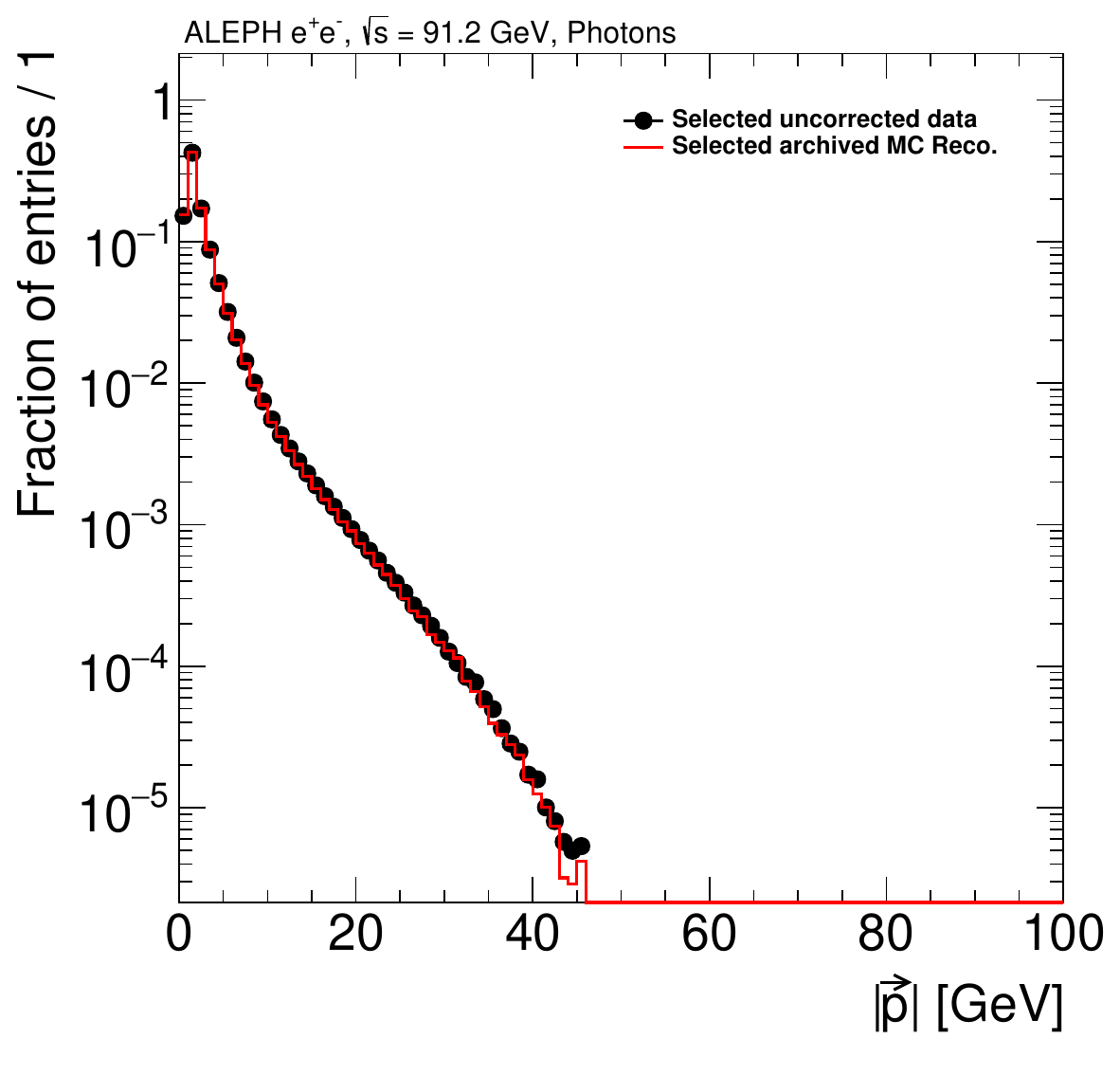}
    \caption{}
\end{subfigure}
\begin{subfigure}[b]{0.32\textwidth}
    \includegraphics[width=\textwidth,angle=0]{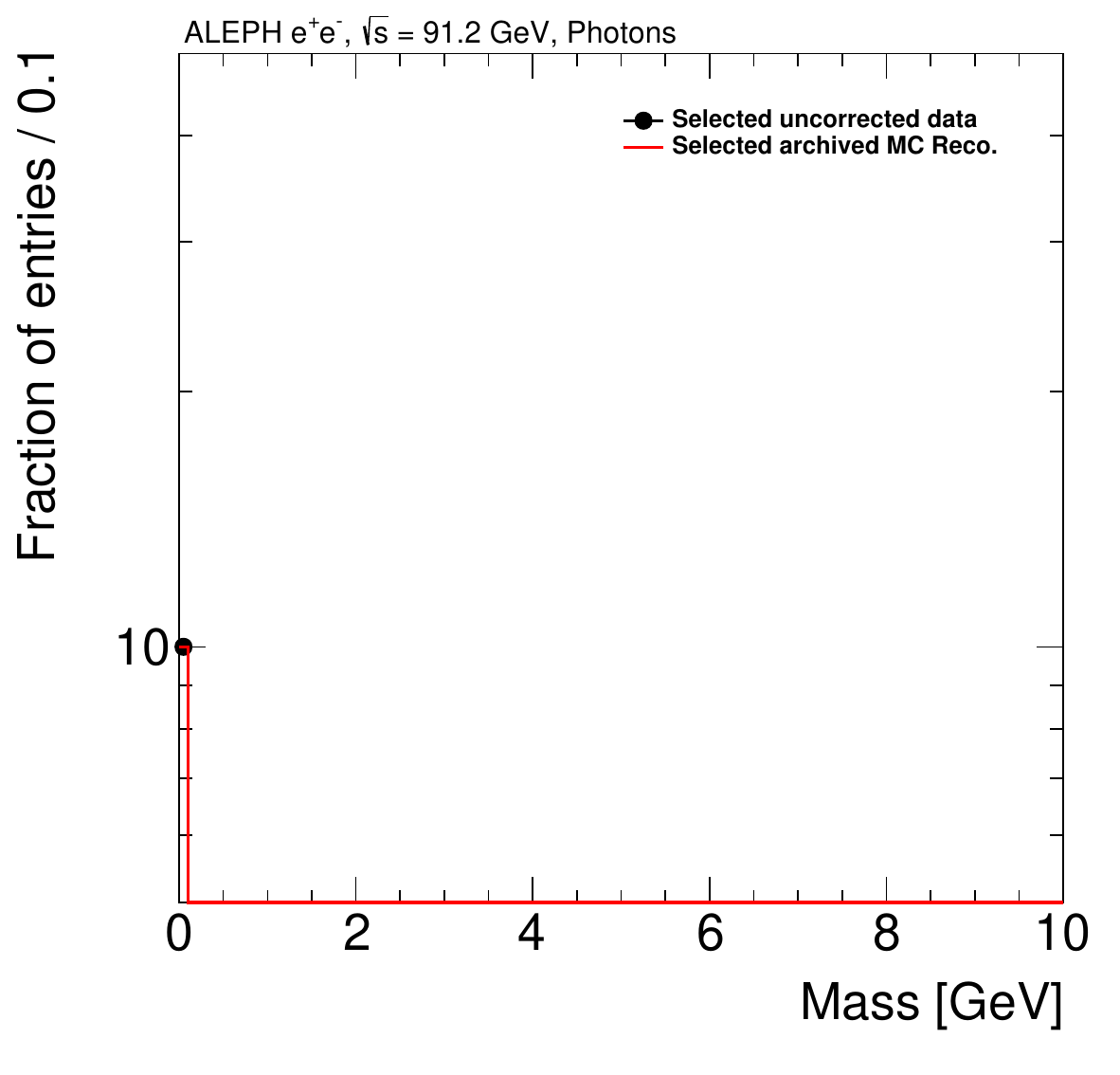}
    \caption{}
\end{subfigure}
\begin{subfigure}[b]{0.32\textwidth}
    \includegraphics[width=\textwidth,angle=0]{figures/nominal/h_pwflag4_energy.pdf}
    \caption{}
\end{subfigure}
\caption{Kinematic distributions for all photons.}
\label{fig:kinem_pwflag4}
\end{figure}

\begin{figure}[ht]
\centering
\begin{subfigure}[b]{0.32\textwidth}
    \includegraphics[width=\textwidth,angle=0]{figures/nominal/h_pwflag5_cosTheta.pdf}
    \caption{}
\end{subfigure}
\begin{subfigure}[b]{0.32\textwidth}
    \includegraphics[width=\textwidth,angle=0]{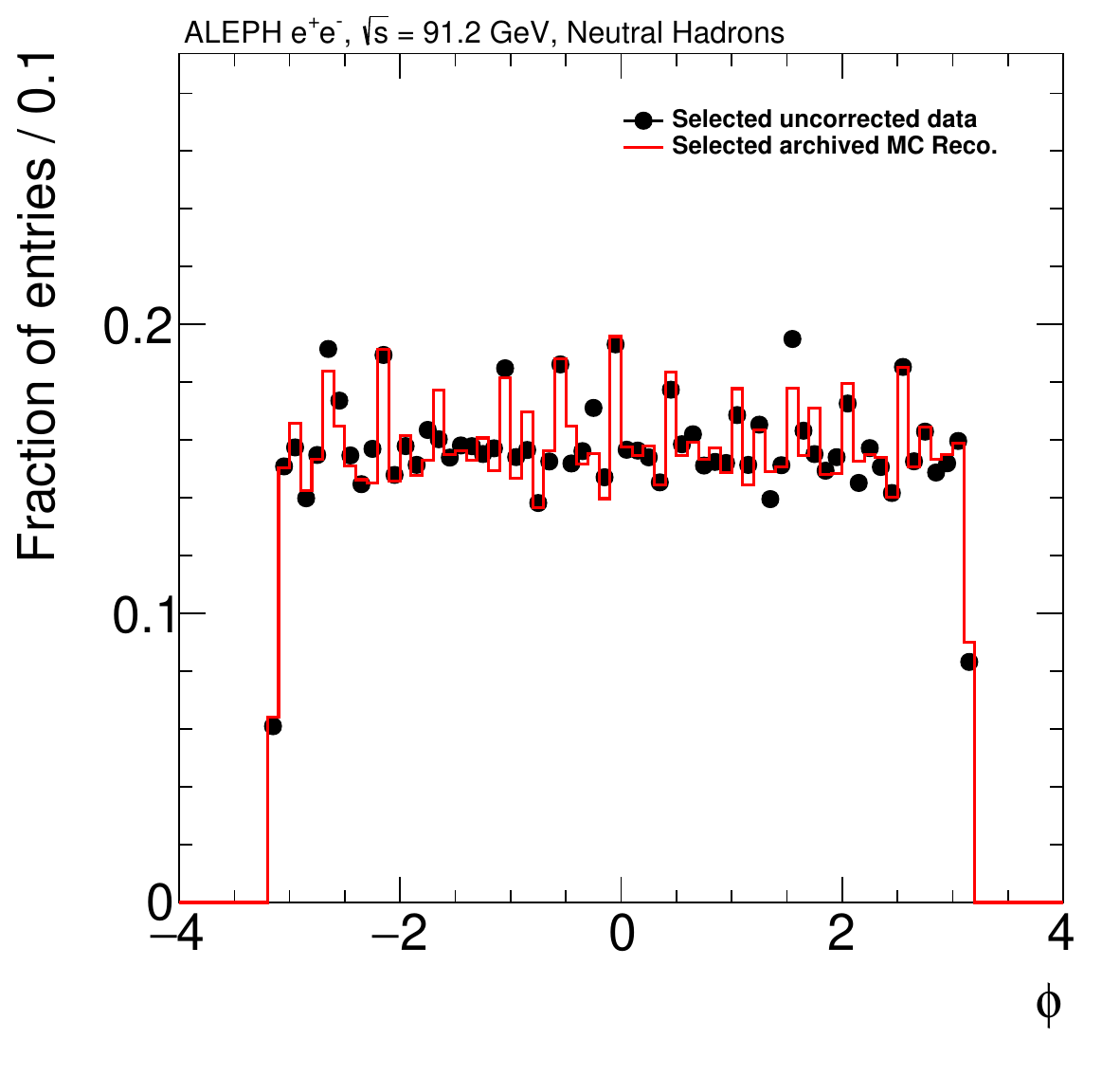}
    \caption{}
\end{subfigure}
\begin{subfigure}[b]{0.32\textwidth}
    \includegraphics[width=\textwidth,angle=0]{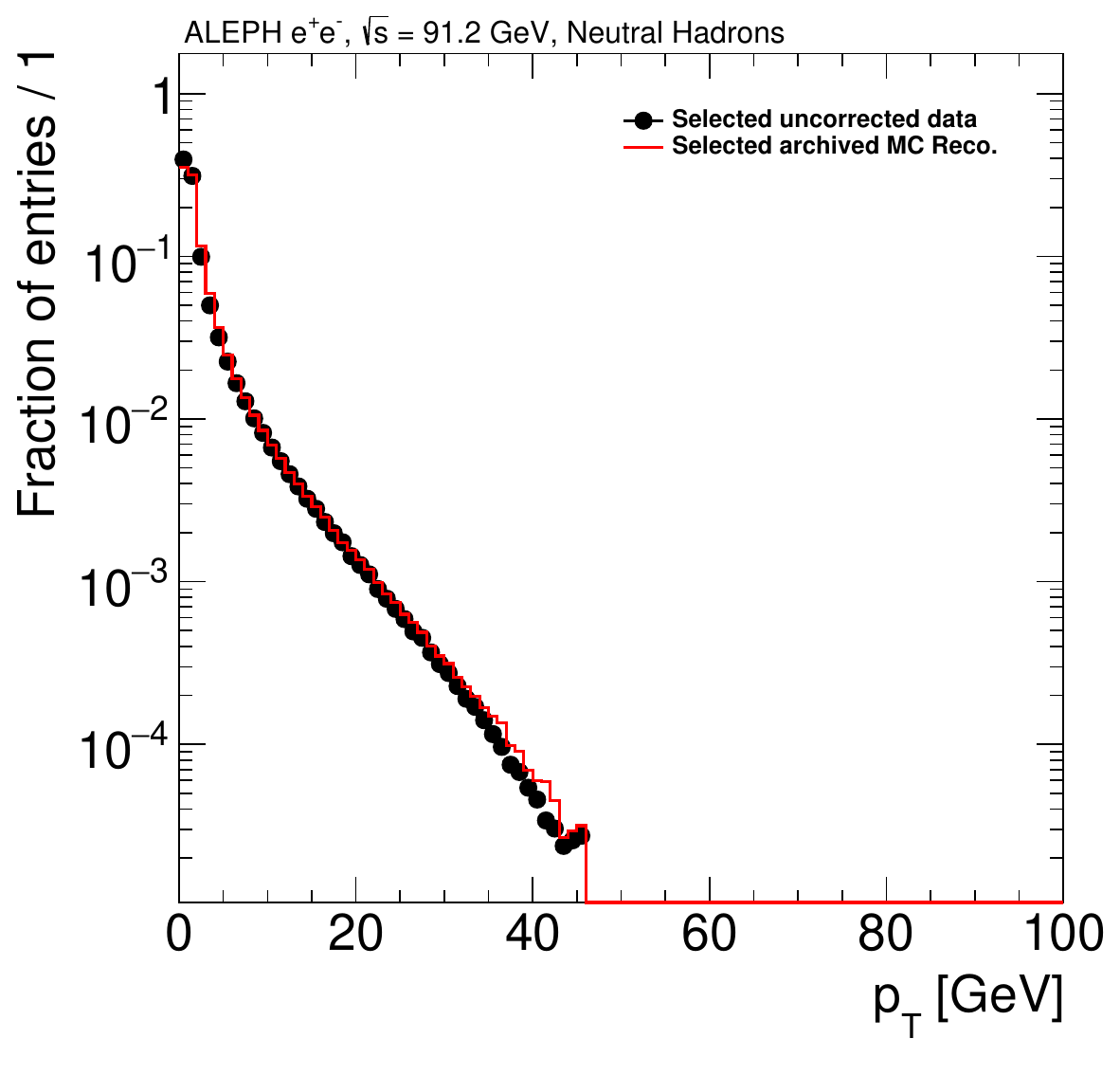}
    \caption{}
\end{subfigure}
\begin{subfigure}[b]{0.32\textwidth}
    \includegraphics[width=\textwidth,angle=0]{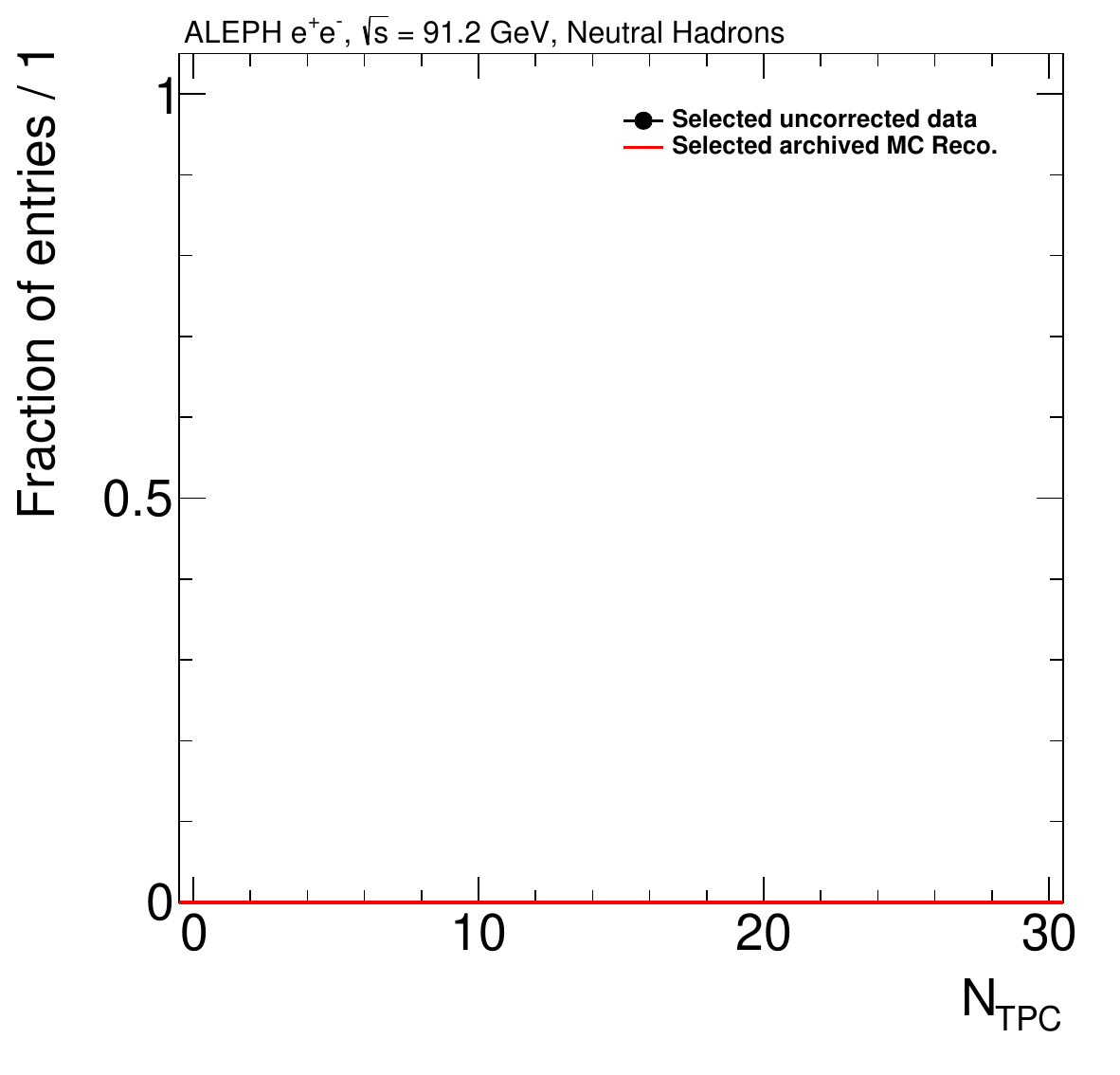}
    \caption{}
\end{subfigure}
\begin{subfigure}[b]{0.32\textwidth}
    \includegraphics[width=\textwidth,angle=0]{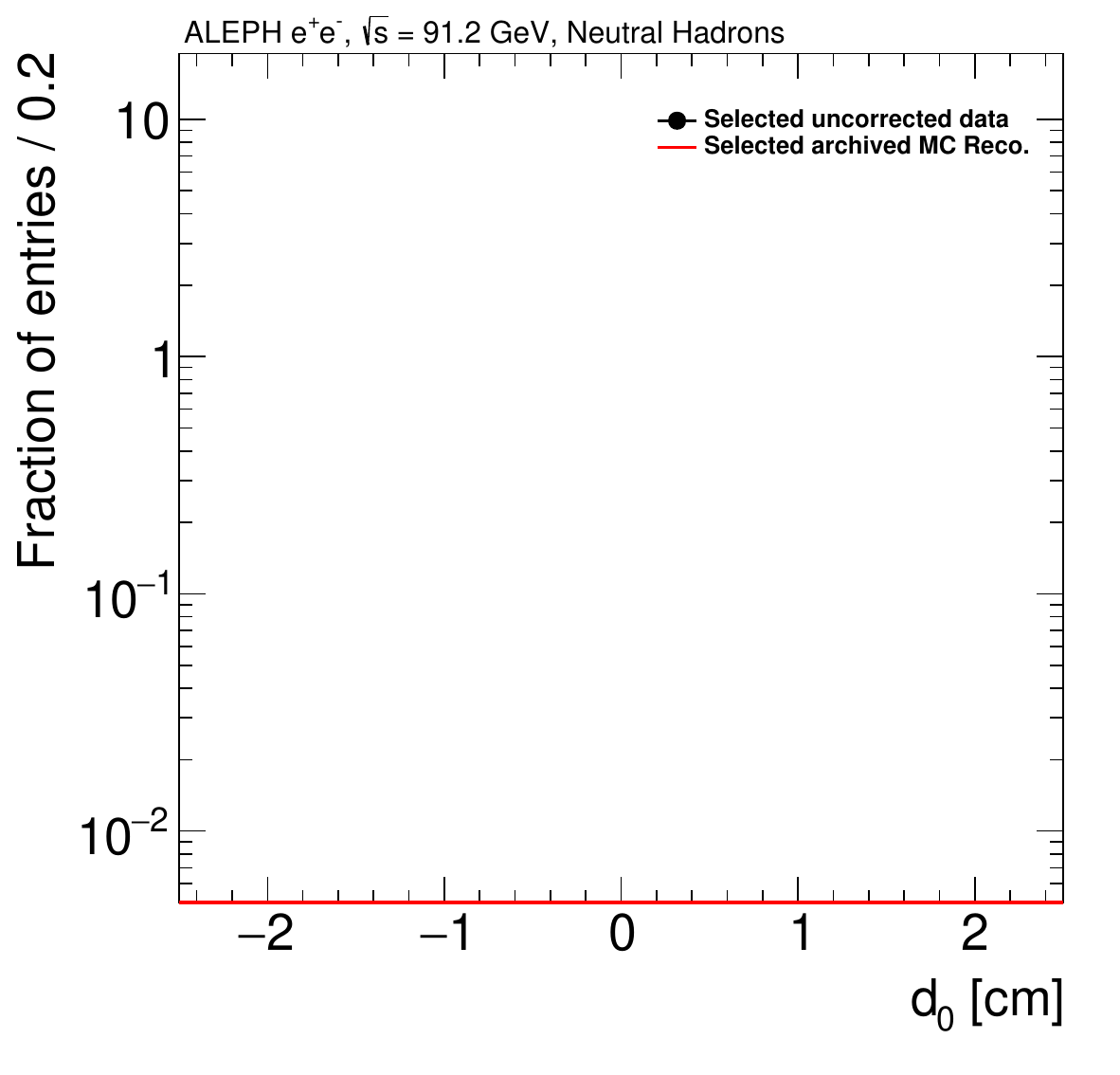}
    \caption{}
\end{subfigure}
\begin{subfigure}[b]{0.32\textwidth}
    \includegraphics[width=\textwidth,angle=0]{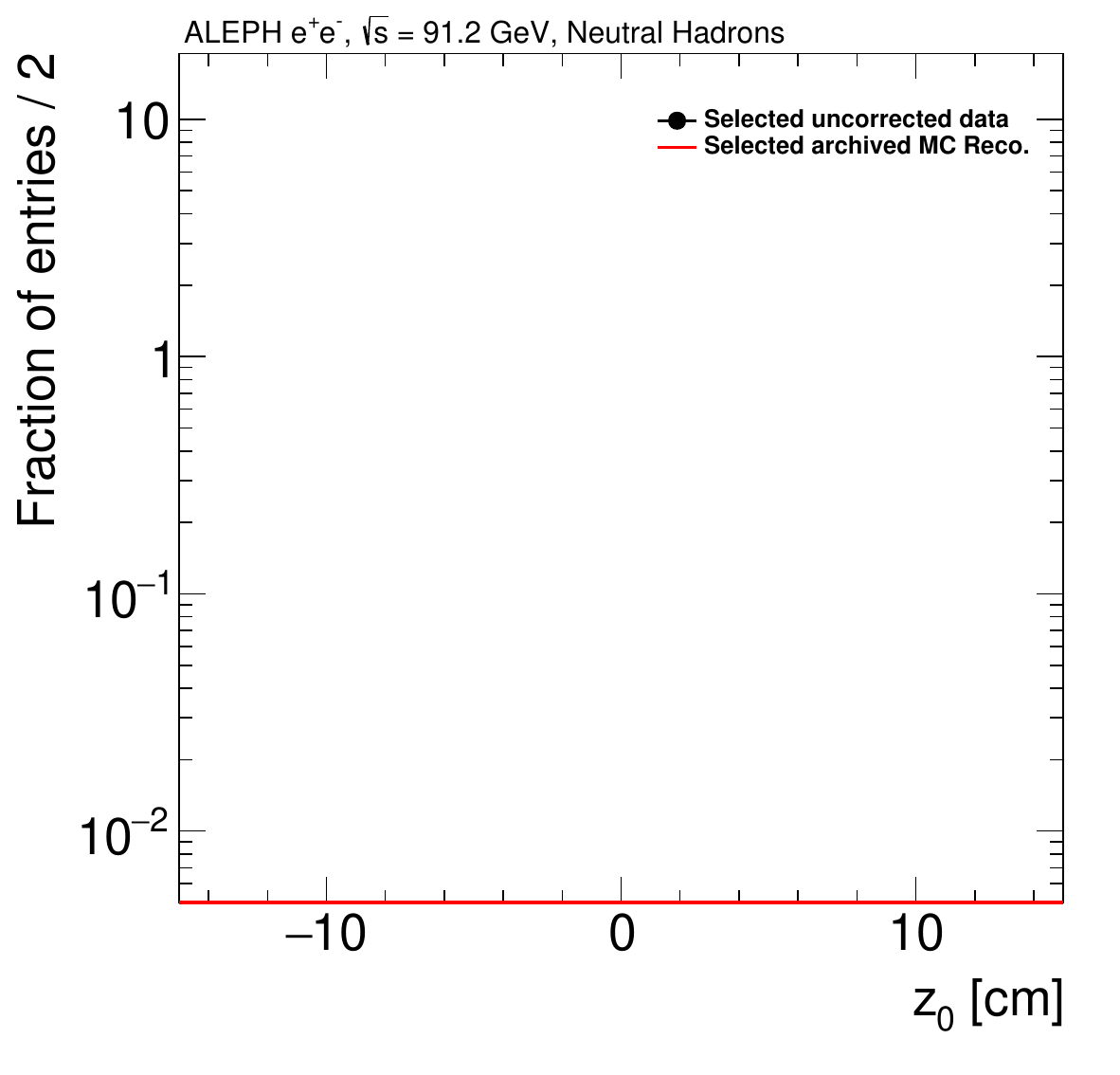}
    \caption{}
\end{subfigure}
\begin{subfigure}[b]{0.32\textwidth}
    \includegraphics[width=\textwidth,angle=0]{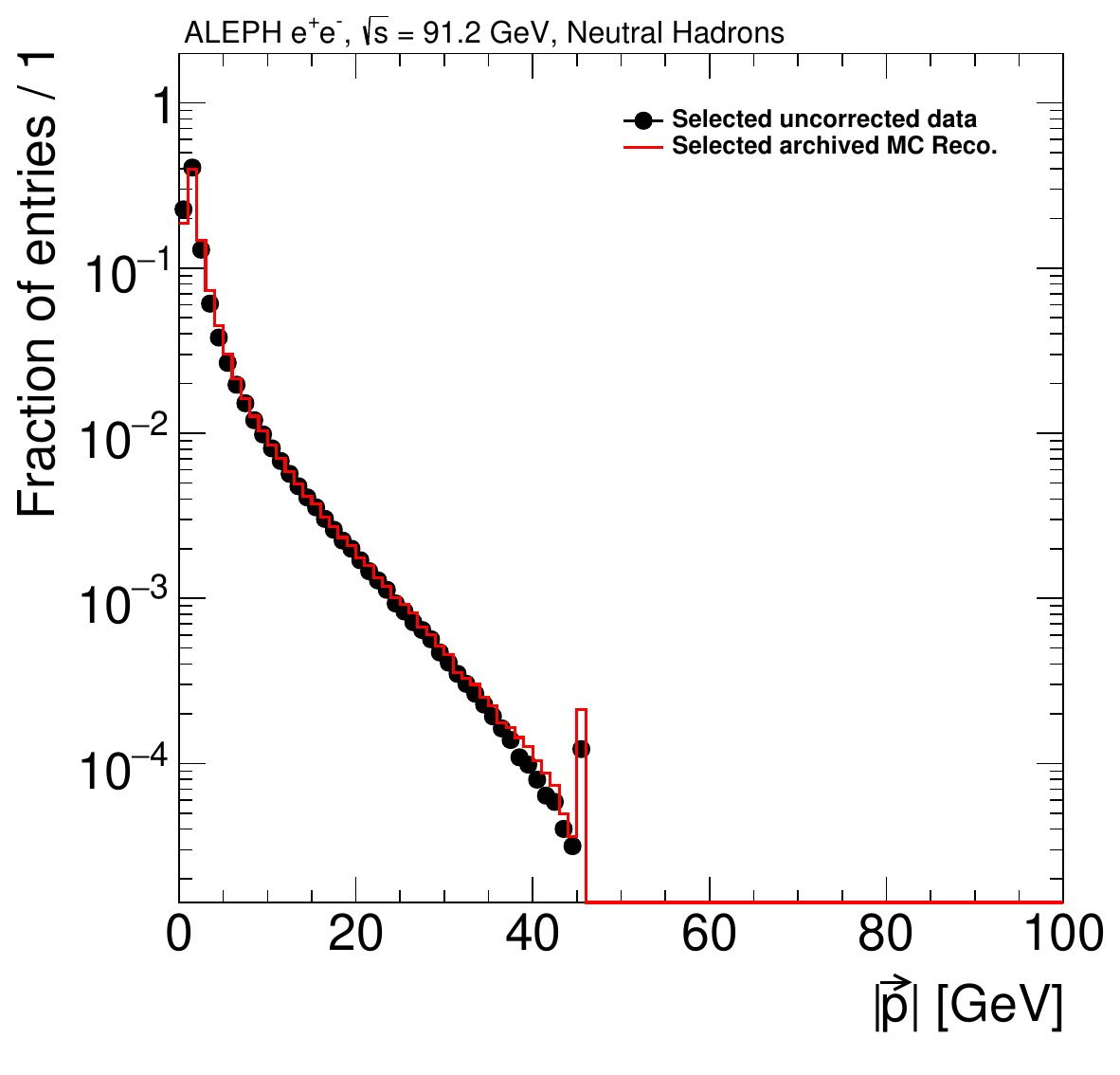}
    \caption{}
\end{subfigure}
\begin{subfigure}[b]{0.32\textwidth}
    \includegraphics[width=\textwidth,angle=0]{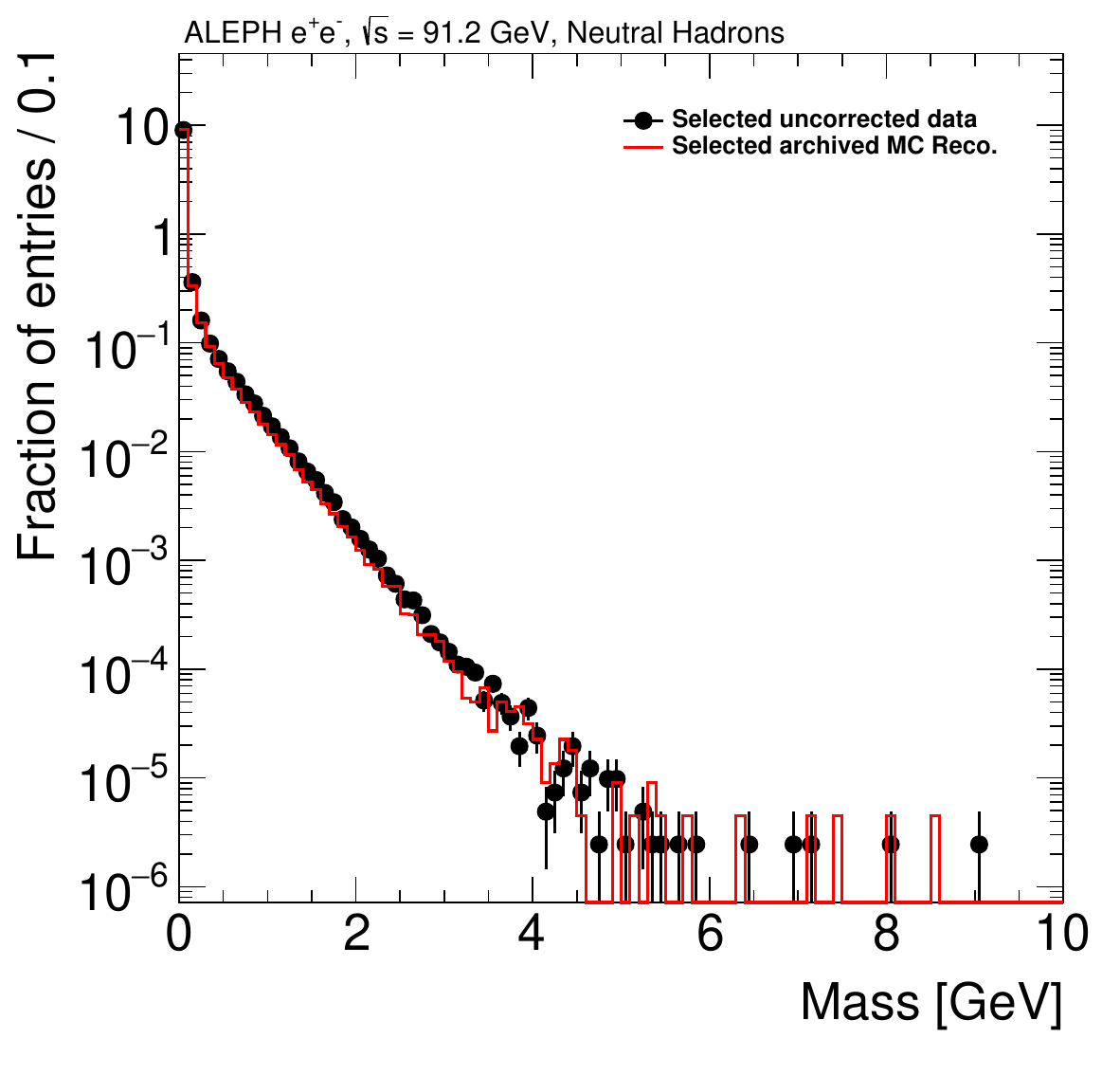}
    \caption{}
\end{subfigure}
\begin{subfigure}[b]{0.32\textwidth}
    \includegraphics[width=\textwidth,angle=0]{figures/nominal/h_pwflag5_energy.pdf}
    \caption{}
\end{subfigure}
\caption{Kinematic distributions for all neutral hadrons.}
\label{fig:kinem_pwflag5}
\end{figure}

\begin{figure}[ht]
\centering
\begin{subfigure}[b]{0.32\textwidth}
    \includegraphics[width=\textwidth,angle=0]{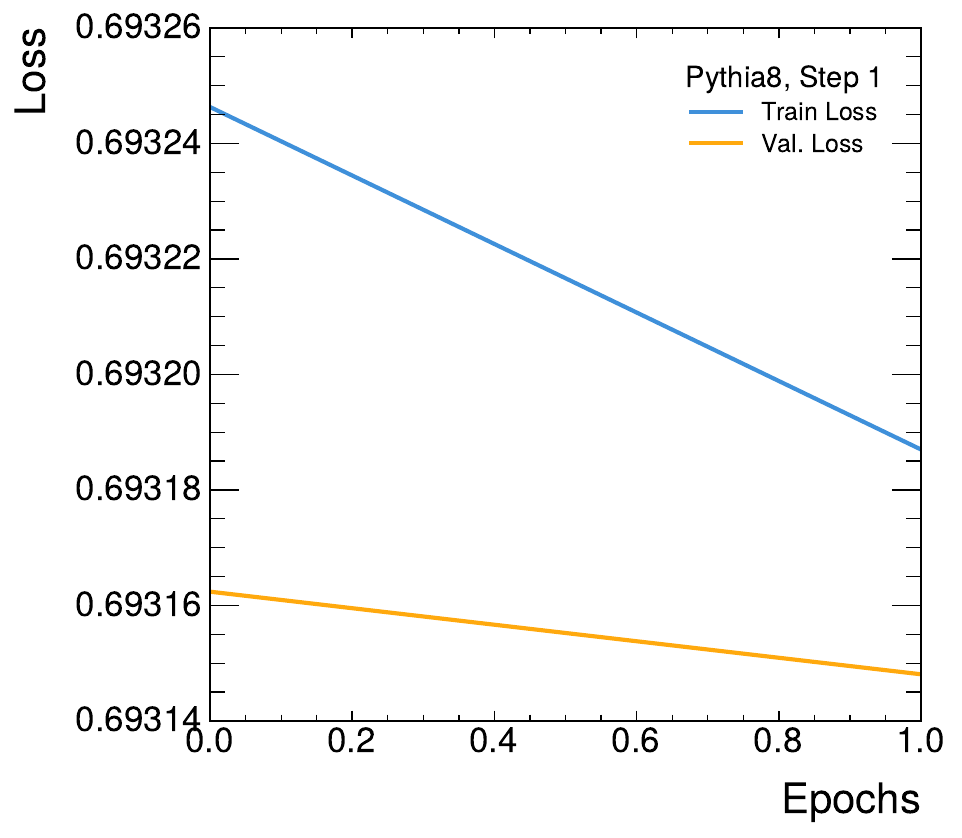}
    \caption{}
    \label{fig:theoryReweighting_lossCurves_step1_pythia8}
\end{subfigure}
\begin{subfigure}[b]{0.32\textwidth}
    \includegraphics[width=\textwidth,angle=0]{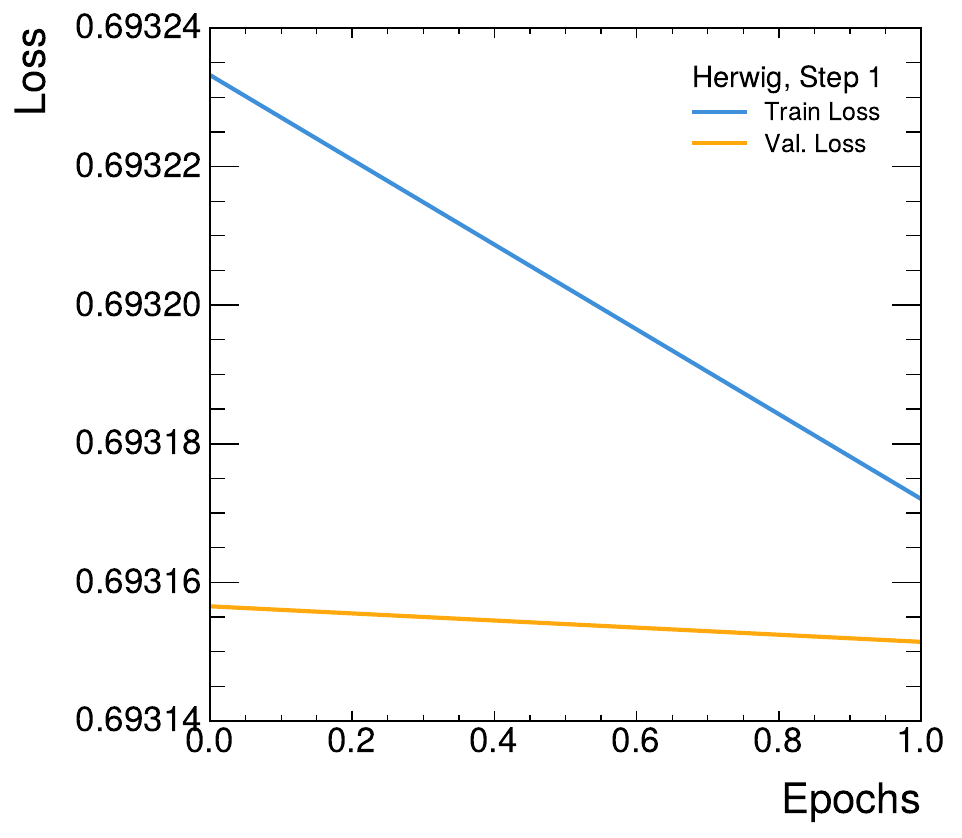}
    \caption{}
    \label{fig:theoryReweighting_lossCurves_step1_herwig}
\end{subfigure}
\begin{subfigure}[b]{0.32\textwidth}
    \includegraphics[width=\textwidth,angle=0]{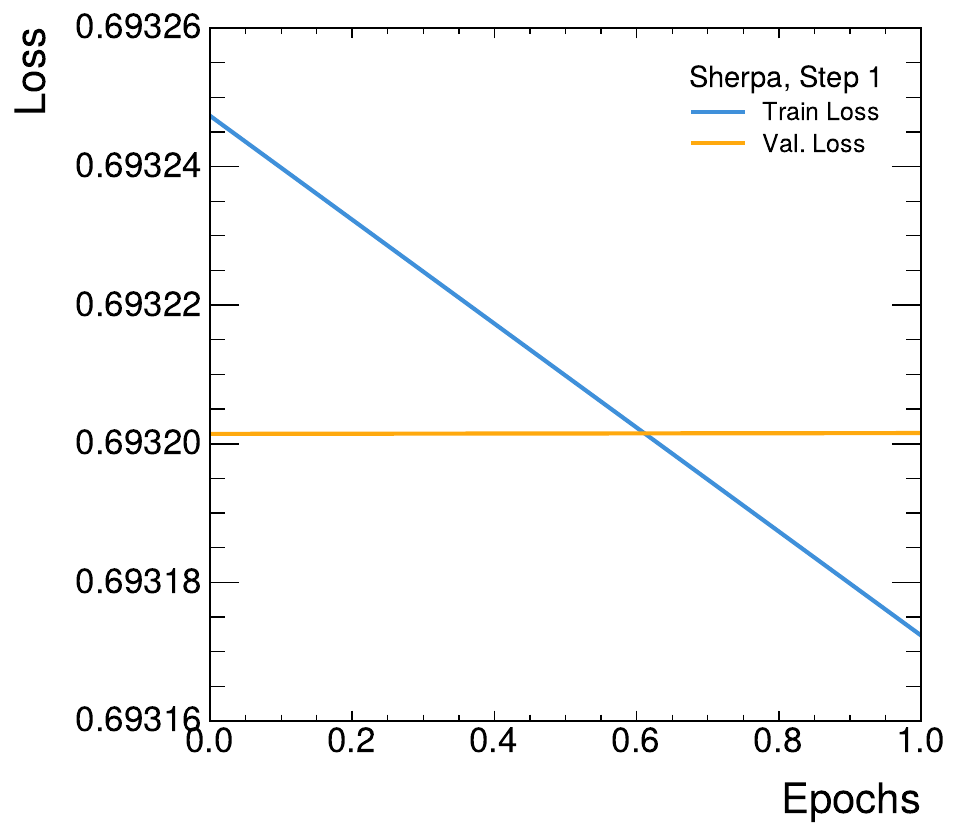}
    \caption{}
    \label{fig:theoryReweighting_lossCurves_step1_sherpa}
\end{subfigure}

\begin{subfigure}[b]{0.32\textwidth}
\includegraphics[width=\textwidth,angle=0]{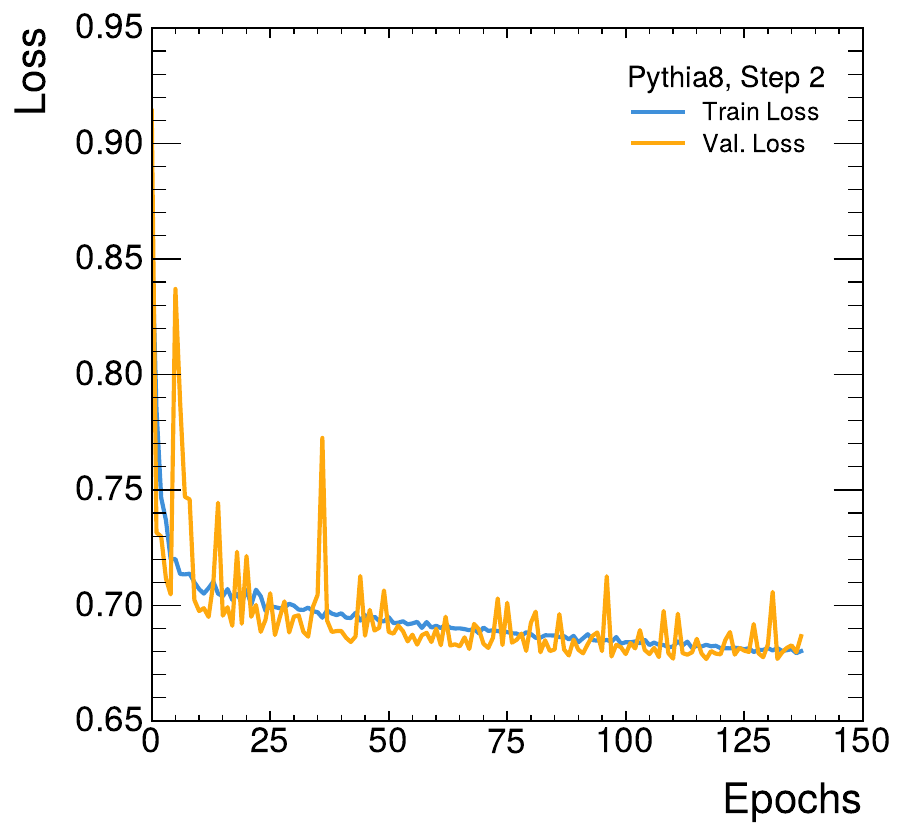}
\caption{}
\label{fig:theoryReweighting_lossCurves_step2_pythia8}
\end{subfigure}
\begin{subfigure}[b]{0.32\textwidth}
\includegraphics[width=\textwidth,angle=0]{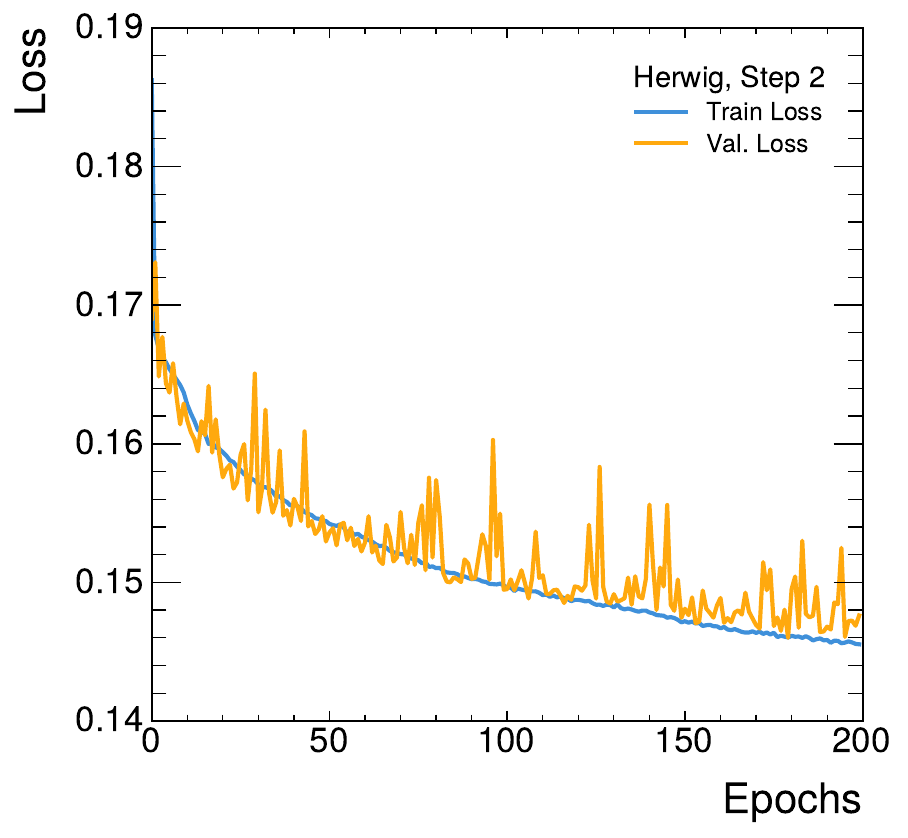}
\caption{}
\label{fig:theoryReweighting_lossCurves_step2_herwig}
\end{subfigure}
\begin{subfigure}[b]{0.32\textwidth}
\includegraphics[width=\textwidth,angle=0]{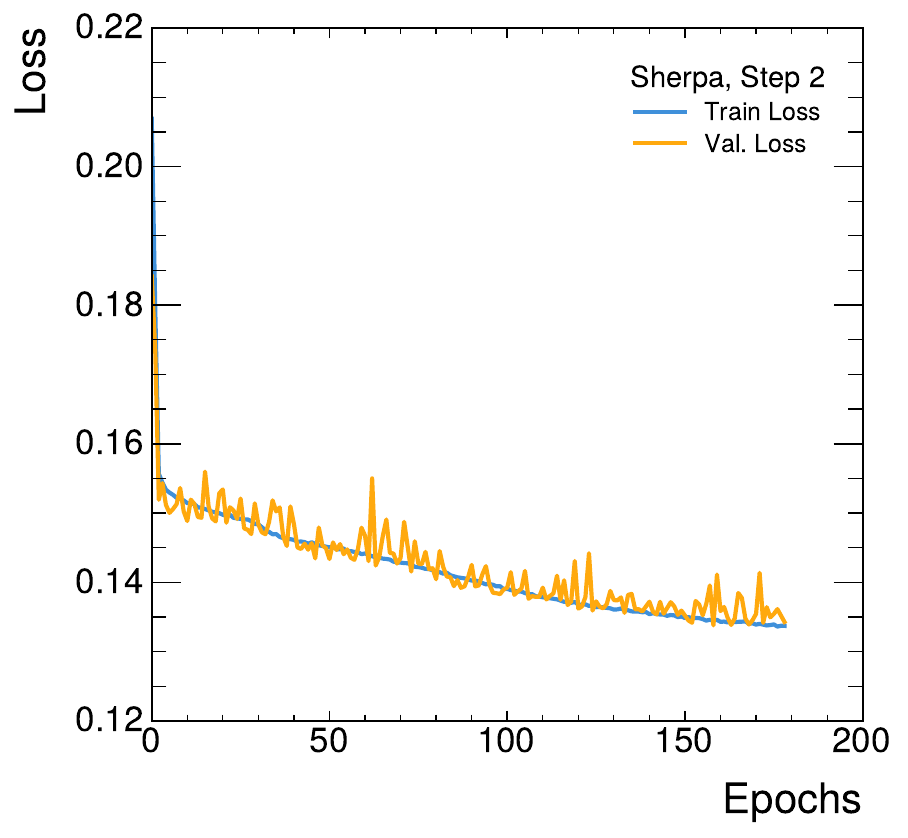}
\caption{}
\label{fig:theoryReweighting_lossCurves_step2_sherpa}
\end{subfigure}
\caption{Training and validation loss curves for the second stage of the particle-level reweighting procedure. The particle edge transformer is trained to reweight the archived $\textsc{Pythia}$ 6.1 MC to match the alternative MC sample. Early stopping is applied based on the validation loss to prevent overfitting. The steady decrease and stabilization of the loss indicate successful learning of the reweighting function.}
\label{fig:theoryReweighting_lossCurves}
\end{figure}

\begin{figure}[t!]
\centering
\begin{subfigure}[b]{0.3\textwidth}
    \includegraphics[width=\textwidth,angle=0]{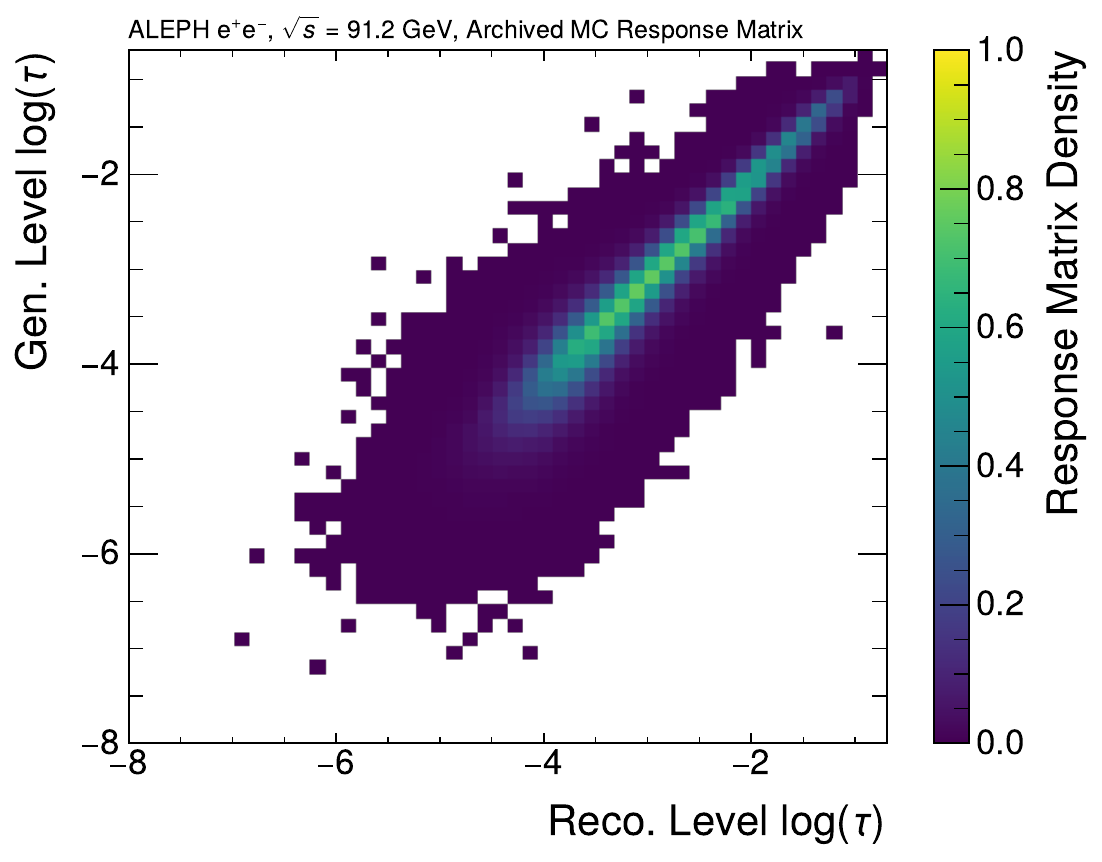}
    \caption{}
    \label{fig:pythia8_responseMatrix_a}
\end{subfigure}
\begin{subfigure}[b]{0.3\textwidth}
    \includegraphics[width=\textwidth,angle=0]{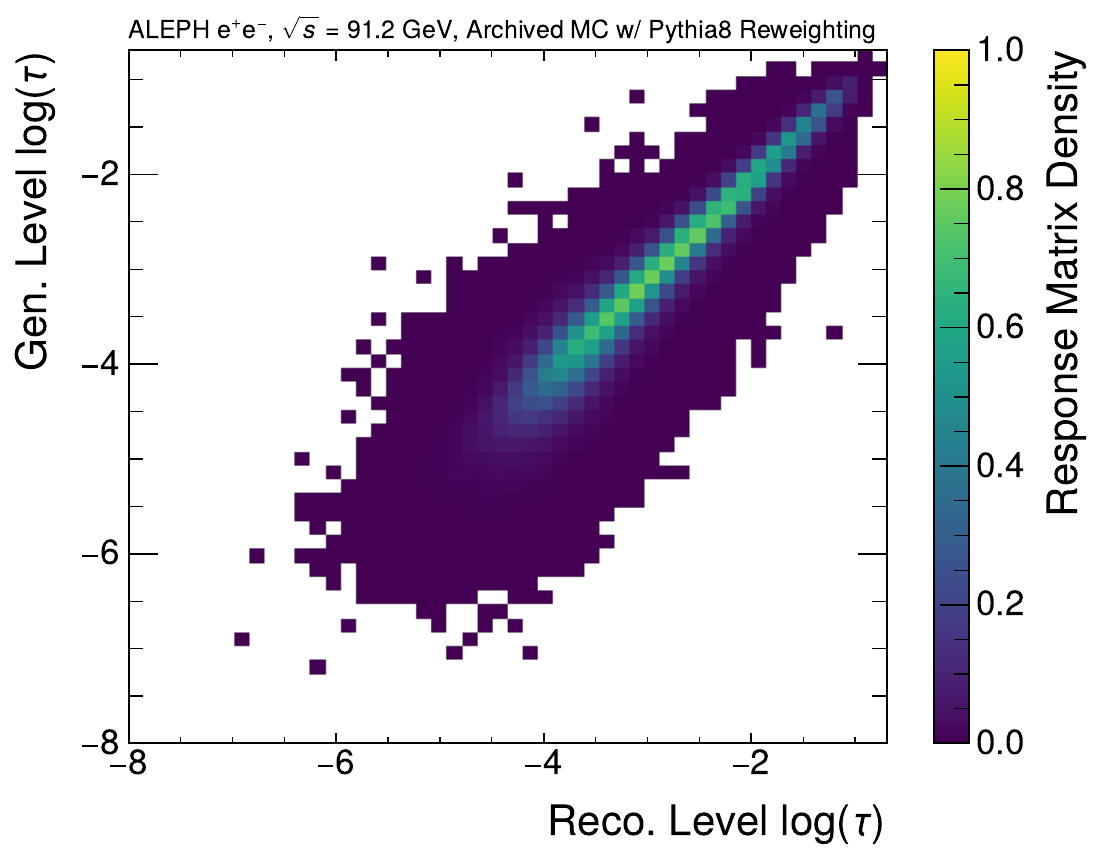}
    \caption{}
    \label{fig:pythia8_responseMatrix_b}
\end{subfigure}
\begin{subfigure}[b]{0.3\textwidth}
    \includegraphics[width=\textwidth,angle=0]{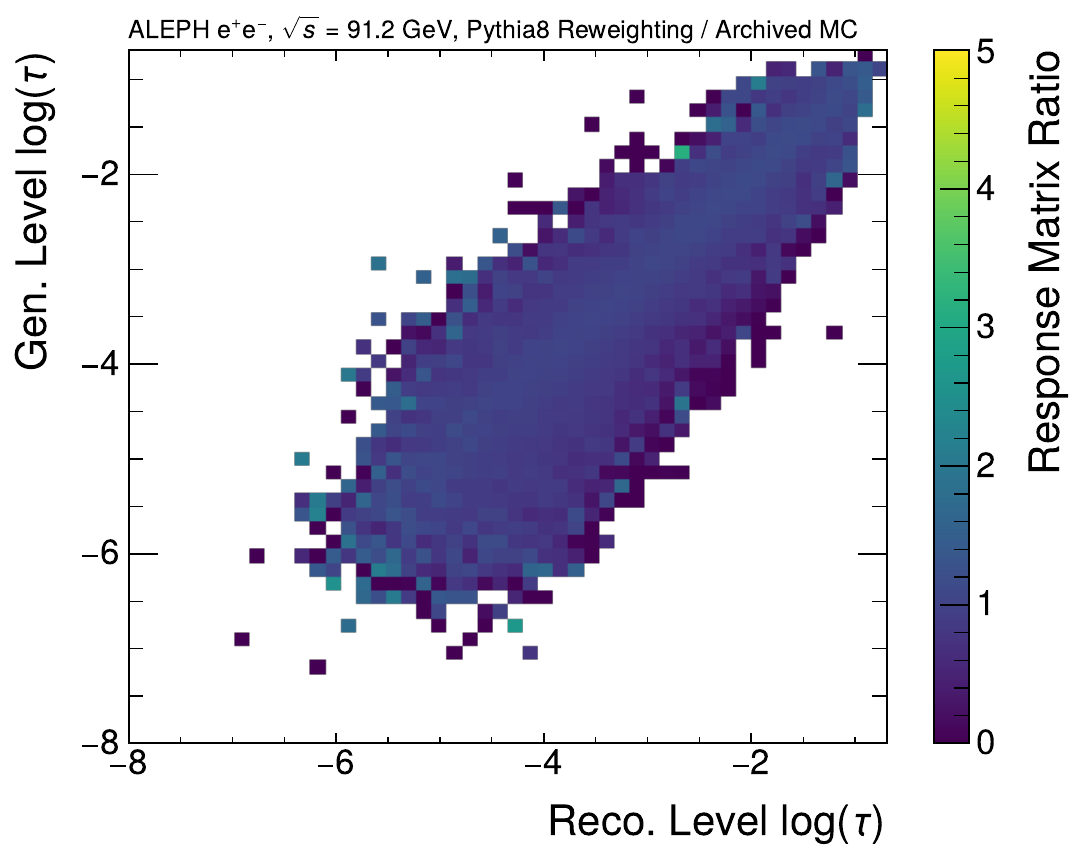}
    \caption{}
    \label{fig:pythia8_responseMatrix_c}
\end{subfigure}
\begin{subfigure}[b]{0.3\textwidth}
    \includegraphics[width=\textwidth,angle=0]{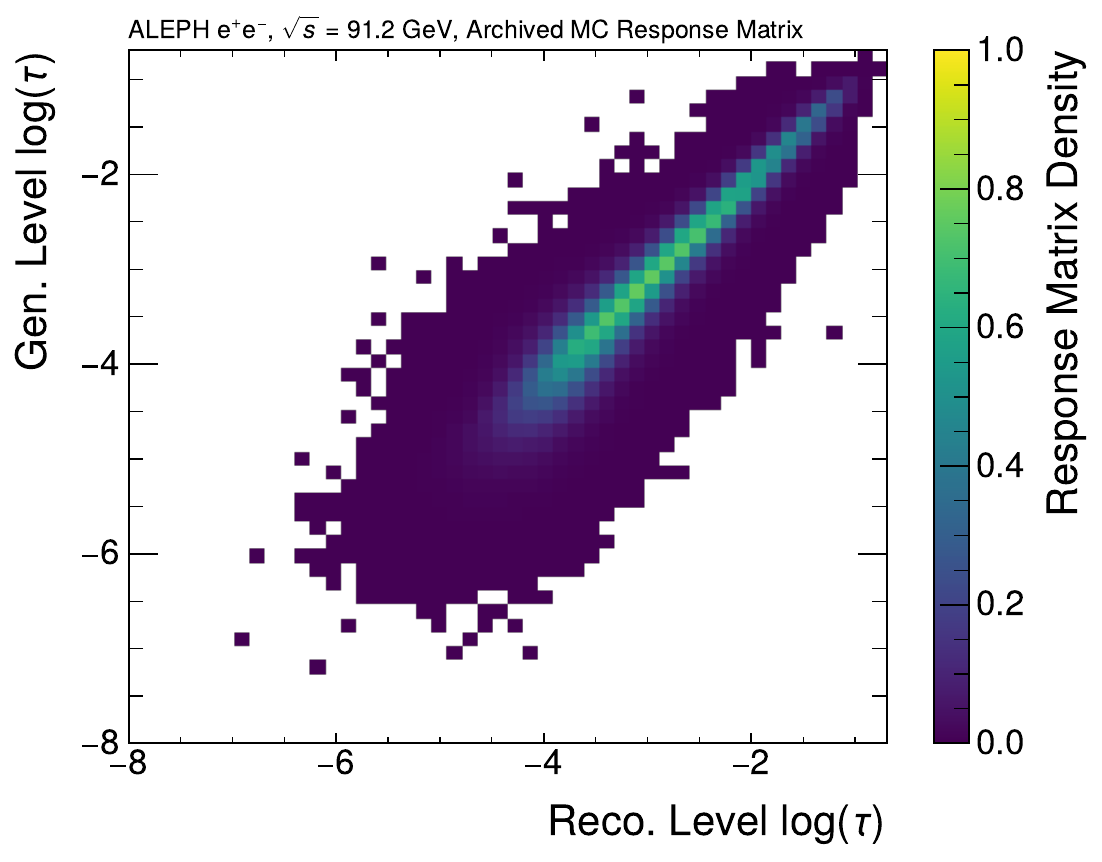}
    \caption{}
    \label{fig:Herwig_responseMatrix_a}
\end{subfigure}
\begin{subfigure}[b]{0.3\textwidth}
    \includegraphics[width=\textwidth,angle=0]{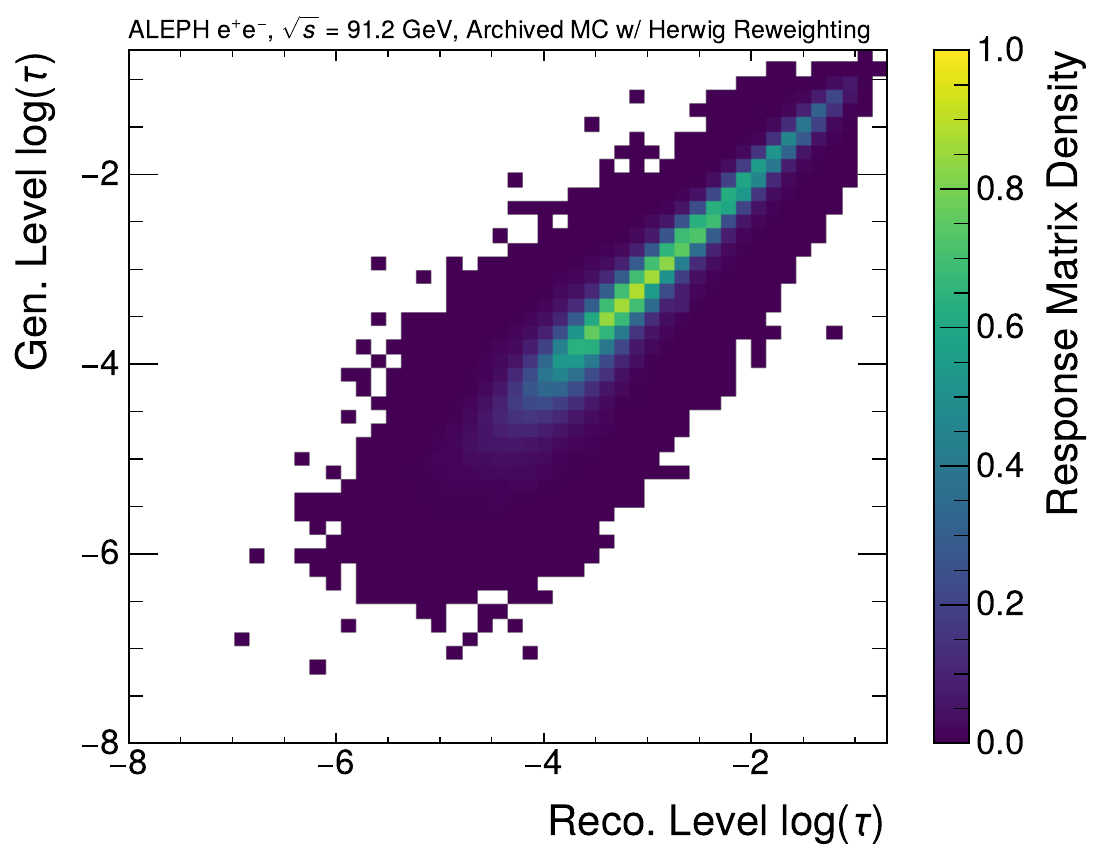}
    \caption{}
    \label{fig:Herwig_responseMatrix_b}
\end{subfigure}
\begin{subfigure}[b]{0.3\textwidth}
    \includegraphics[width=\textwidth,angle=0]{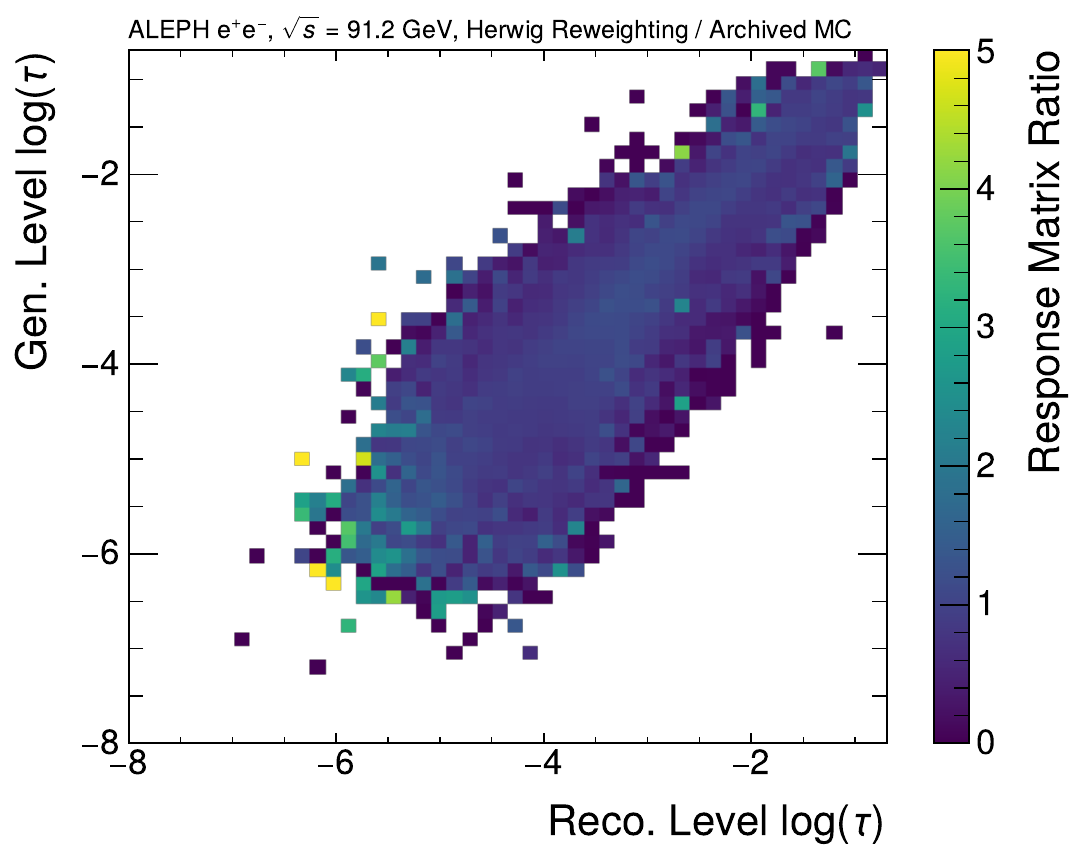}
    \caption{}
    \label{fig:Herwig_responseMatrix_c}
\end{subfigure}
\begin{subfigure}[b]{0.3\textwidth}
    \includegraphics[width=\textwidth,angle=0]{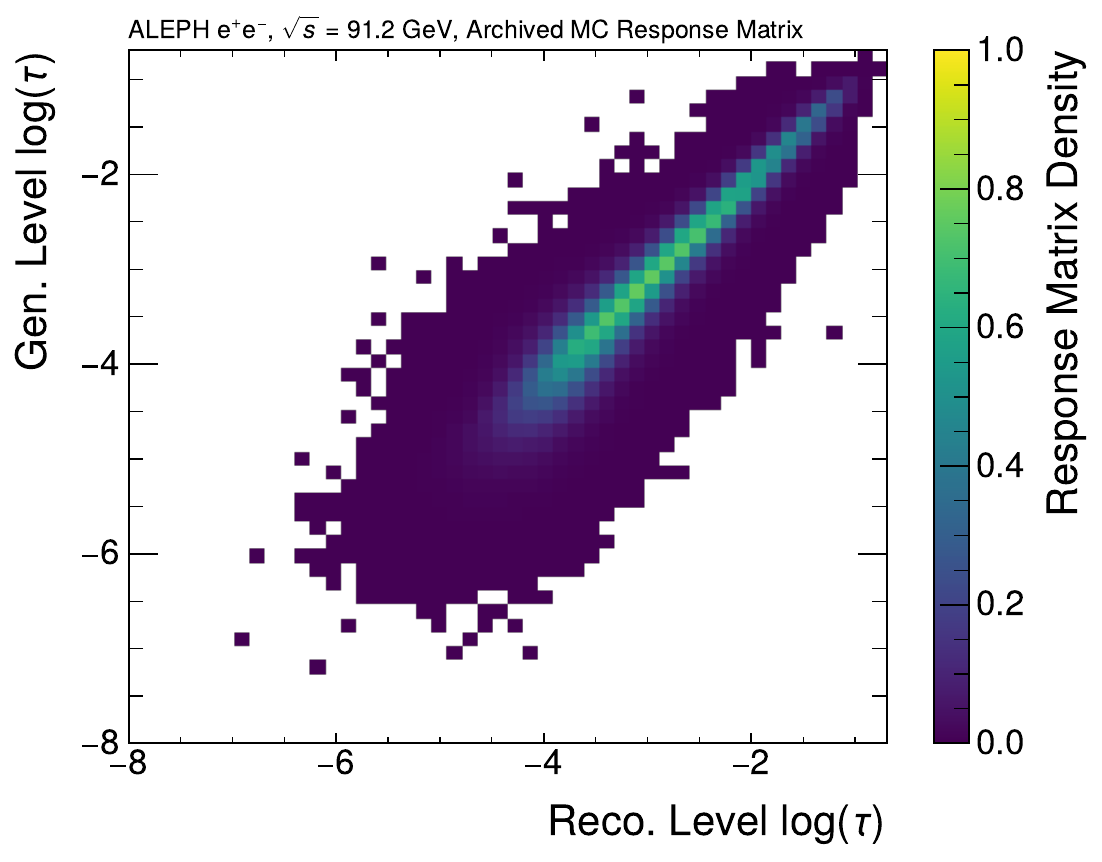}
    \caption{}
    \label{fig:Sherpa_responseMatrix_a}
\end{subfigure}
\begin{subfigure}[b]{0.3\textwidth}
    \includegraphics[width=\textwidth,angle=0]{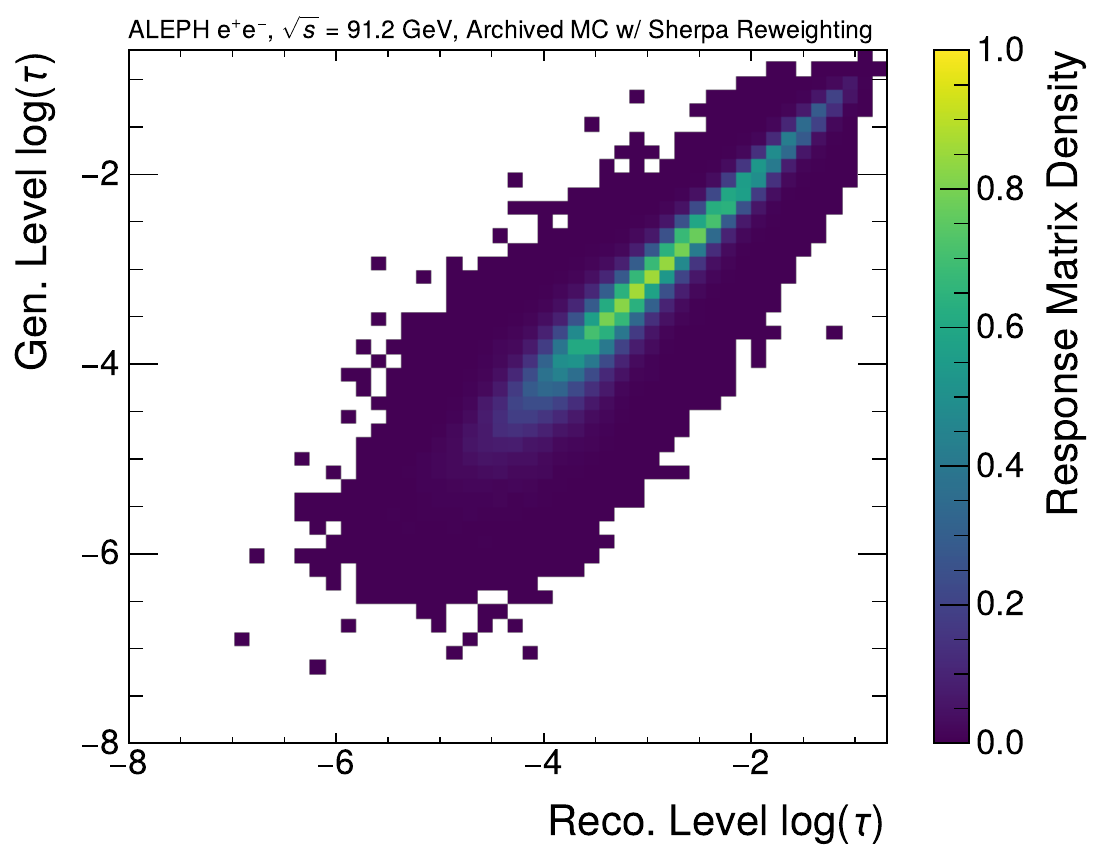}
    \caption{}
    \label{fig:Sherpa_responseMatrix_b}
\end{subfigure}
\begin{subfigure}[b]{0.3\textwidth}
    \includegraphics[width=\textwidth,angle=0]{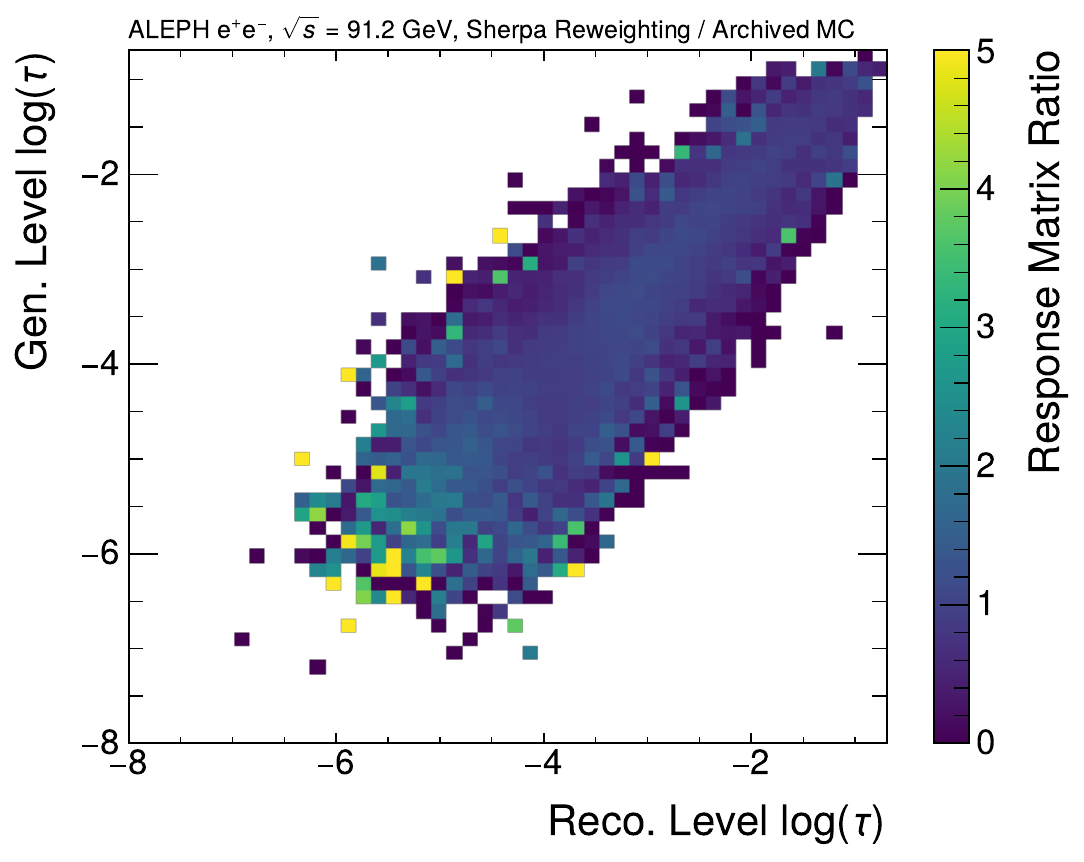}
    \caption{}
    \label{fig:Sherpa_responseMatrix_c}
\end{subfigure}
\caption{Response matrices illustrating detector-level versus generator-level thrust before and after reweighting. Panel (a) shows the nominal response matrix from the archived $\textsc{Pythia}$ 6.1 MC, panel (b) shows the response matrix after applying the theoretical reweighting, and panel (c) displays the ratio of the reweighted to nominal response matrices. The changes highlight the impact of the theoretical variation on the detector response modeling.}
\label{fig:theoryUncert_responseMatrix}
\end{figure}\FloatBarrier\newpage
\section{Derivation of UniFold as unbinned IBU}
\label{sec:app_unifoldAsIBU}

\begin{align}
p_{\text{unfolded}}^{(n)}(t) &= \nu_n(t)p_{\text{Gen.}}(t) \\
\nu_n(t)p_{\text{Gen.}}(t) &= p_{\text{Gen.}}(t)\nu_{n-1}(t) \frac{p_{(w_n^{\text{pull}}, \text{Gen.})}(t)}{p_{(\nu_{n-1}, \text{Gen.})}(t)} \tag{Use Rule 2} \\
&= p_{\text{Gen.}}(t)\nu_{n-1}(t) \frac{\int dm \; p_{\text{Gen.}\mid\text{{Sim.}}}(t\mid m)p_{\text{{Sim.}}}(m)w_{n}(m)}{\nu_{n-1}(t)p_{\text{Gen.}}(t)} \tag{Use Rule 1} \\
&= \int dm \; \nu_{n-1}(t)p_{\text{Gen.}}(t) \frac{p_{\text{Gen.}\mid\text{{Sim.}}}(t\mid m)p_{\text{{Sim.}}}(m)\nu_{n-1}^{\text{push}}(m)}{\nu_{n-1}p_{\text{Gen.}}(t)}\frac{p_{\text{data}}(m)}{p_{(\nu_{n-1}^{\text{push}},\text{Sim.})}(m)} \\
&\rightarrow \text{Use that } \nu_{n-1}^{\text{push}}(m) = \nu_{n-1}(t) \\
&= \int dm \; p_{\text{data}}(m) \nu_{n-1}(t)p_{\text{Gen.}}(t) \frac{p_{\text{Gen.}\mid\text{{Sim.}}}(t\mid m)p_{\text{{Sim.}}}(m)}{p_{\text{Gen.}}(t)}\frac{1}{p_{(\nu_{n-1}^{\text{push}},\text{Sim.})}(m)} \\
&\rightarrow \text{Use Bayes' Rule: } P(m|t) = \frac{P(t|m) P(m)}{P(t)} \\
&= \int dm \; p_{\text{data}}(m) \nu_{n-1}(t)p_{\text{Gen.}}(t) p_{\text{{Sim.}}\mid\text{Gen.}}(m\mid t)\frac{1}{p_{(\nu_{n-1}^{\text{push}},\text{Sim.})}(m)} \\
&\rightarrow \text{Use that } p_{(\nu_{n-1}^{\text{push}},\text{Sim.})}(m) = \int dt' p_{\text{{Sim.}}\mid\text{Gen.}}(m\mid t')\nu_{n-1}(t')p(t') \\
&= \int dm \; p_{\text{data}}(m) \frac{p_{\text{{Sim.}}\mid\text{Gen.}}(m\mid t)\nu_{n-1}(t)p_{\text{Gen.}}(t)}{\int dt' p_{\text{{Sim.}}\mid\text{Gen.}}(m\mid t')\nu_{n-1}(t')p(t')} \\
&\text{or as binned version} \\
&= \sum_i m_i \frac{R_{ij} t_{j}}{\sum_k R_{ik} t_{k}} \tag{IBU!}
\end{align}

\end{document}